\documentclass[fleqn,usenatbib]{mnras}
\usepackage{mathptmx}
\usepackage[T1]{fontenc}
\usepackage{ae,aecompl}

\usepackage{graphicx}	% Including figure files
\usepackage{amsmath}	% Advanced maths commands
\usepackage{amssymb}	% Extra maths symbols
\usepackage{epsfig}
\pdfoutput=1
\title[Long-term emission behaviour of Mrk~421]{Dissecting the long-term emission behaviour of the BL Lac object Mrk~421}

\author[M. I. Carnerero et al.]
{M.~I.~Carnerero             $^{ 1}$\thanks{E-mail:maribel@oato.inaf.it},
C.~M.~Raiteri               $^{ 1}$,
M.~Villata                  $^{ 1}$,
J.~A.~Acosta-Pulido         $^{ 2,3}$,
\newauthor
V.~M.~Larionov              $^{ 4,5}$,
P.~S.~Smith                 $^{ 6}$,
F.~D'Ammando                $^{ 7,8}$,
I.~Agudo                    $^{ 9}$,
\newauthor
M.~J.~Ar\'evalo             $^{ 2,3}$,
R.~Bachev                   $^{10}$,
J.~Barnes                   $^{11}$,
S.~Boeva                    $^{10}$,
V.~Bozhilov                 $^{12}$,
\newauthor
D.~Carosati                 $^{13,14}$,
C.~Casadio                  $^{15}$,
W.~P.~Chen                  $^{16}$,
G.~Damljanovic              $^{17}$,
\newauthor
E.~Eswaraiah                $^{16}$,
E.~Forn\'e                  $^{18}$,
G.~Gantchev                 $^{12}$,
J.~L.~G\'omez               $^{ 9}$,
\newauthor
P.~A.~Gonz\'alez-Morales    $^{ 2,3}$,
A.~B.~Gri\~n\'on-Mar\'in      $^{ 2,3}$,
T.~S.~Grishina              $^{ 4}$,
\newauthor
M.~Holden                   $^{19}$,
S.~Ibryamov                 $^{10,20}$,
M.~D.~Joner                 $^{19}$,
B.~Jordan                   $^{21}$,
S.~G.~Jorstad               $^{4,22}$,
\newauthor
M.~Joshi                    $^{22}$,
E.~N.~Kopatskaya            $^{ 4}$,
E.~Koptelova                $^{16}$,
O.~M.~Kurtanidze            $^{23,24,25}$,
\newauthor
S.~O.~Kurtanidze            $^{23}$,
E.~G.~Larionova             $^{ 4}$,
L.~V.~Larionova             $^{ 4}$,
G.~Latev                    $^{10}$,
\newauthor
C.~L\'azaro                 $^{ 2,3}$,
R.~Ligustri                 $^{26}$,
H.~C.~Lin                   $^{16}$,
A.~P.~Marscher              $^{22}$,
\newauthor
C.~Mart\'inez-Lombilla      $^{ 2,3}$,
B.~McBreen                  $^{27}$,
B.~Mihov                    $^{10}$,
S.~N.~Molina                $^{ 9}$,
\newauthor
J.~W.~Moody                 $^{19}$,
D.~A.~Morozova              $^{ 4}$,
M.~G.~Nikolashvili           $^{23}$,
K.~Nilsson                  $^{28}$,
\newauthor
E.~Ovcharov                 $^{12}$,
C.~Pace                     $^{29}$,
N.~Panwar                   $^{16}$,
A.~Pastor~Yabar             $^{ 2,3}$,
R.~L.~Pearson               $^{19}$,
\newauthor
F.~Pinna                    $^{ 2,3}$,
C.~Protasio                 $^{ 2,3}$,
N.~Rizzi                    $^{30}$,
F.~J.~Redondo-Lorenzo       $^{ 2,3}$,
\newauthor
G.~Rodr\'iguez-Coira        $^{ 2,3}$,
J.~A.~Ros                   $^{18}$,
A.~C.~Sadun                 $^{31}$,
S.~S.~Savchenko             $^{ 4}$,
\newauthor
E.~Semkov                   $^{10}$,
L.~Slavcheva-Mihova         $^{10}$,
N.~Smith                    $^{32}$,
A.~Strigachev               $^{10}$,
\newauthor
Yu.~V.~Troitskaya           $^{ 4}$,
I.~S.~Troitsky              $^{ 4}$,
A.~A.~Vasilyev              $^{ 4}$,
and O.~Vince                $^{17}$
}

\date{Accepted XXX. Received YYY; in original form ZZZ}

\pubyear{2015}

\begin{document}
\label{firstpage}
\pagerange{\pageref{firstpage}--\pageref{lastpage}}
\maketitle

\begin{abstract}
We report on long-term multi-wavelength monitoring of blazar Mrk~421 by the GASP-WEBT collaboration and Steward Observatory, and by the {\em Swift} and {\em Fermi} satellites. We study the source behaviour in the period 2007--2015, characterized by several extreme flares. The ratio between the optical, X-ray, and $\gamma$-ray fluxes is very variable. The $\gamma$-ray flux variations show fair correlation with the optical ones starting from 2012.
We analyse spectropolarimetric data and find wavelength-dependence of the polarisation degree ($P$), which is compatible with the presence of the host galaxy, and no wavelength-dependence of the electric vector polarisation angle (EVPA). Optical polarimetry shows a lack of simple correlation between $P$ and flux and wide rotations of the EVPA.
We build broad-band spectral energy distributions with simultaneous near-infrared and optical data from the GASP-WEBT and UV and X-ray data from the \textit{Swift} satellite. 
They show strong variability in both flux and X-ray spectral shape and suggest a shift of the synchrotron peak up to a factor $\sim 50$ in frequency. 
The interpretation of the flux and spectral variability is compatible with jet models including at least two emitting regions that can change their orientation with respect to the line of sight.
\end{abstract}

\begin{keywords}
galaxies: active - BL Lacertae objects: general - BL Lacertae objects: individual: Mrk 421 - galaxies: jets
\end{keywords}

%%%%%%%%%%%%%%%%%%%%%%%%%%%%%%%%%%%%%%%%%%%%%%%%%%

%%%%%%%%%%%%%%%%% BODY OF PAPER %%%%%%%%%%%%%%%%%%

\section{Introduction}

The active galactic nuclei (AGNs) known as ``Blazars" are the ideal sources to study extragalactic jets, since in these objects one of the two jets coming out from the central black hole points toward us and its emission is thus enhanced by Doppler beaming.
The low-energy radiation that we observe from the radio to the optical--X-ray frequencies is ascribed to synchrotron radiation from relativistic electrons, while the highest-energy radiation is most likely produced by inverse-Compton scattering on the same relativistic electrons.
The photon seeds for the latter process can come either from the jet itself (Synchrotron Self Compton - or SSC - models) or from the disk, broad line region, torus (External Compton - or EC - models). High-energy radiation can also be produced by hadronic processes \citep{bot13}.
At low frequencies we can measure the degree and angle of polarisation of blazar emission. 
The study of their variability and possible wavelength dependence is important to infer the jet properties, because the polarisation is tied to the jet magnetic field structure \citep[e.g.][]{smi96,vis98}.

Mrk~421 at $z=0.031$ \citep{ulr75} is one of the best monitored blazar over the entire electromagnetic spectrum. It is classified as a high-energy-peaked BL Lac (HBL), which means that the synchrotron peak (and usually also the inverse-Compton peak) in its spectral energy distribution (SED) is positioned at relatively high frequencies.  

It was the first blazar that was detected at energies $E>500 \rm \, GeV$ \citep{pun92} and many observing campaigns have recently been organized to analyse the source behaviour at TeV frequencies, usually including multiwavelength data \citep[e.g.][]{aha05,alb07,don09,abd11,ale12,ale15a,ale15b,ahn16b,bal16}.
It was observed by the EGRET instrument onboard the \textit{Compton Gamma-Ray Observatory} with an average flux at $E>100 \rm \, MeV$ of $(13.9 \pm 1.8) \times \rm 10^{-8} \, ph \, cm^{-2} \, s^{-1}$ and a photon index $\Gamma=1.57 \pm 0.15$ \citep{har99}. The \textit{Fermi} Large Area Telescope (LAT) Third Source Catalog \citep{ace15} reports a $\gamma$-ray flux of $\sim 18 \times \rm 10^{-8} \, ph \, cm^{-2} \, s^{-1}$ between 100 MeV and 100 GeV, and a photon index $\Gamma=1.77 \pm 0.08$. 
The source is very bright and variable at X-rays. In particular, a strong X-ray flare was observed in 2013 \citep{pia14,pal15,sin15,kap16}. 
Optical observations are available since 1899 \citep{mil75} and show that large-amplitude, rapid variability is a distinctive feature also in the optical band.
The radio morphology reveals a bright nucleus and a one-sided jet with stationary or subluminal components \citep{pin10,bla13}.
An extreme radio flare was observed in 2012, possibly connected with preceding $\gamma$-ray flares \citep{hov15}. Possible radio--$\gamma$ correlation was also found in 2011 by \citet{lic14}. 

In this paper we analyse the Mrk~421 long-term flux and polarisation behaviour. Very preliminary results were reported in \citet{car16}. We present the optical and near-infrared data obtained by the GLAST-AGILE Support Program (GASP) of the Whole Earth Blazar Telescope (WEBT) collaboration\footnote{http://www.oato.inaf.it/blazars/webt/}. We compare the optical and near-infrared flux variations with the X-ray and UV light curves obtained by the \textit{Swift} satellite and with the $\gamma$-ray light curve from the \textit{Fermi} satellite, by means of a cross-correlation analysis on the full dataset available from 2007 to 2015. Moreover, we analyse the photopolarimetric behaviour and the spectropolarimetric data acquired at the Steward Observatory in the framework of the monitoring program in support to the \textit{Fermi} mission\footnote{http://james.as.arizona.edu/$\sim$psmith/Fermi/}. Finally, broad-band SEDs are built from the near-infrared to the X-ray energies to investigate the source spectral variability in the synchrotron part of the spectrum.

\section{Optical photometry}

Optical observations in $R$ band for the GASP-WEBT were performed with 34 telescopes in 26 observatories around the world: Abastumani (Georgia), AstroCamp (Spain), Belogradchik (Bulgaria), Calar Alto\footnote{Calar Alto data was acquired as part of the MAPCAT project: http://www.iaa.es/$\sim$iagudo/$\_$iagudo/MAPCAT.html} (Spain), Castelgrande (Italy), Crimean (Russia), L'Ampolla (Spain), Lowell (Perkins, USA), Lulin (Taiwan), New Mexico Skies (USA), Pulkovo (Russia), ROVOR (USA), Roque de los Muchachos (KVA and Liverpool, Spain), Rozhen (Bulgaria), SAI Crimean (Russia), Sabadell (Spain), Sirio (Italy), Skinakas (Greece), St.\ Petersburg (Russia), Talmassons (Italy), Teide (BRT, IAC80 and STELLA-I, Spain), Tijarafe (Spain), Torino (Italy), Tuorla (Finland), Astronomical Station Vidojevica - ASV (Serbia) and West Mountain (USA).
Further $R$-band data were provided by the Steward Observatory (USA).
Calibration of the source magnitude was obtained with respect to the reference stars 1, 2 and 3 by \citet{vil98}.

\begin{figure*}
\centering
\includegraphics[width=13cm]{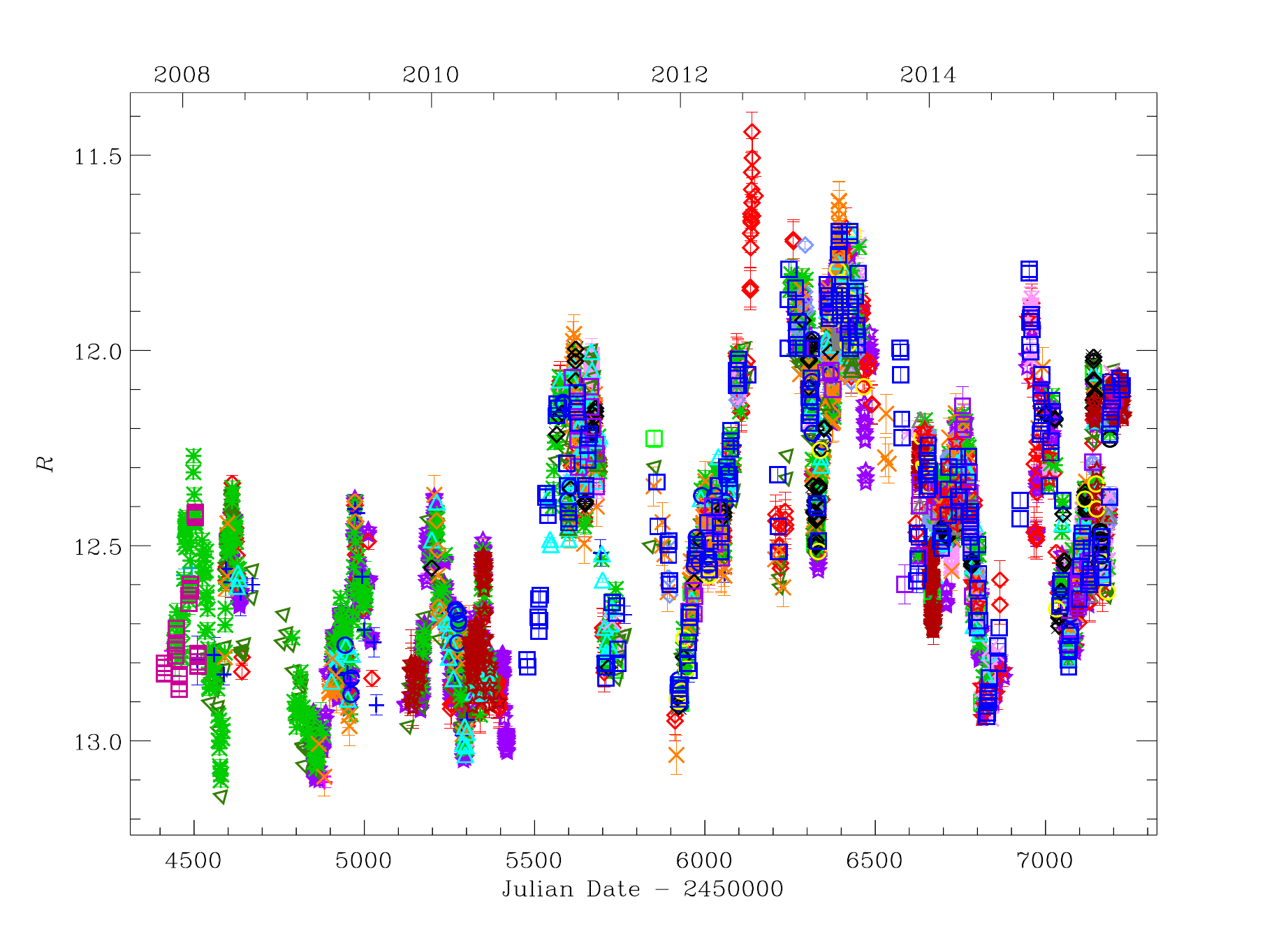}
  \caption{Optical light curve of Mrk~421 built with data from the GASP-WEBT collaboration and Steward Observatory in $R$ band. Different colours and symbols highlight data from different telescopes. No correction for the host-galaxy contribution and Galactic extinction has been applied.}
\label{fi:mk421_optico}
\end{figure*}

The light curve in $R$ band was built by carefully assembling the datasets coming from the different telescopes. Moreover, binning was used to reduce the noise of data acquired close in time by the same telescope. Offsets among different GASP datasets caused by partial inclusion of the host galaxy were minimized by adopting the same prescriptions for the photometry, i.e.\ an aperture radius of 7.5 arcsec. 
The Steward photometry was obtained with an extraction aperture of $\rm 7.6 \, arcsec \times 10 \, arcsec$, so that we had to add 2 mJy to the source flux density to make the Steward data match the GASP data.

The final light curve is shown in Fig.\ \ref{fi:mk421_optico}, where different symbols and colours highlight data from the various telescopes. It includes 5591 data points in the period 2007 November 8 ($\rm JD=2454412.7$) to 2015 July 23 ($\rm JD=2457227.4$). They represent observed magnitudes, with no correction for the Galactic extinction and host-galaxy contribution. Strong variability characterises the entire period on a large variety of time scales.

\begin{figure}
\centering
\includegraphics[width=8cm]{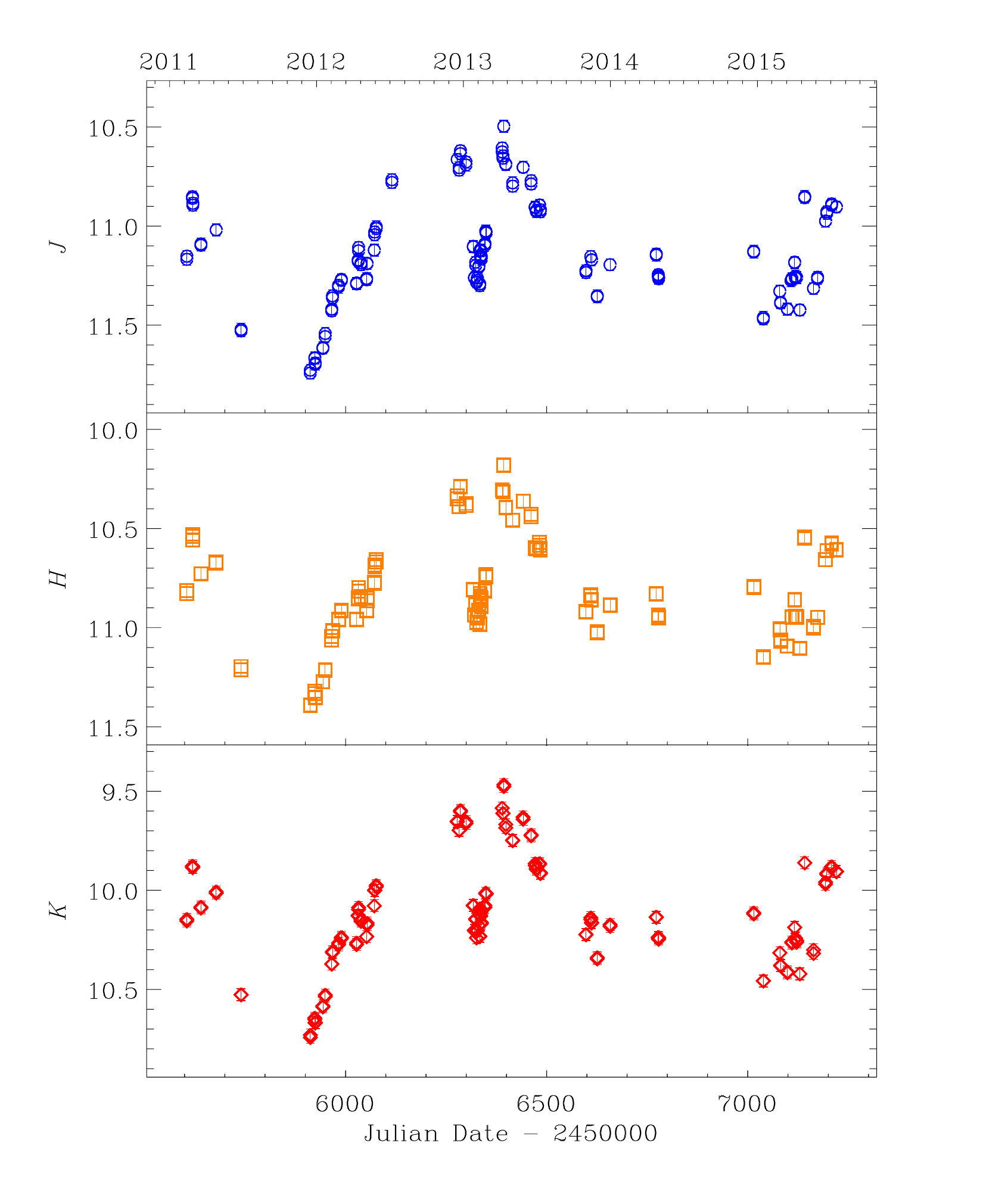}
  \caption{Near-infrared light curves of Mrk~421 built with data from the GASP-WEBT collaboration in $J$, $H$ and $K$ bands. No correction for the host-galaxy contribution and Galactic extinction has been applied.}
\label{fi:1101_ir}
\end{figure}

\section{Near-infrared photometry}

The near-infrared light curves of Mrk~421 in the period 2011--2015 are shown in Fig.\ \ref{fi:1101_ir} in $J$, $H$ and $K$ bands. The GASP-WEBT observations were performed by the 1.5 m TCS telescope in the Teide observatory. 

The source calibration was obtained with respect to stars in the source field of view (FOV), whose magnitudes were adopted from the Two Micron All-Sky Data Release\footnote{http://www.ipac.caltech.edu/2mass/} \citep[][2MASS]{skr06}. Because of the small FOV ($\rm 4 \, arcsec \, \times \, 4 \, arcsec$) and consequent small number of reliable reference stars, we found a systematic offset between the zero-points of the Mrk~421 images with respect to the zero-points derived from the other fields images. We corrected for this difference and 
further checked that the inferred colour indices of the source were consistent with those reported in the 2MASS-point source catalog.

As in the case of the optical data, the near-infrared light curves were carefully checked and cleaned by reducing the data scattering through the binning of data close in time. These light curves are generally under-sampled for a detailed comparison with the optical data, but where sufficient sampling has been achieved, they indicate that a close correspondence exists, as expected if the near-infrared and optical emissions are produced by the same mechanism in the same jet region. We note that the amount of variability is nearly the same in the three near-infrared bands. 

\section{Removal of the host galaxy contribution}

The Mrk~421 host galaxy is relatively bright in optical bands, very strong in the near-infrared, while its flux is small in UV bands.
To remove the host-galaxy contribution, we used a de Vaucouleurs profile, as done by \citet{rai10} for BL Lacertae:

\begin{equation*}\label{vacu2}
I(r)=I_{\rm e} \, {\rm e}^{-7.67[(r/r_{\rm e})^{0.25}-1]}
\end{equation*}

where $r_{\rm e}$ is the effective radius, i.e.\ the radius of the isophote containing half of the total luminosity and $I_{\rm e}$ is the surface brightness at the effective radius. 
We used $r_{\rm e}=8.2 \pm 0.2$ arcsec and $R_{\rm host}=13.18$ mag \citep{nil07} to estimate that the host galaxy contribution to the observed fluxes is $p = 48$ per cent of the whole galaxy flux with an aperture radius $r_{\rm a}=7.5$ arcsec, as used by the WEBT observers. We also estimated 
$p = 37$ per cent for $r_{\rm a}=5.0$ arcsec, the value that we will use in Section\ \ref{uvot} for analysing the UV data (see Table\ \ref{ta:vacou}). 

In the $R$ band, we found that 7.86 mJy must be subtracted from the observed photometric flux densities to isolate the non-thermal continuum of the active nucleus.
We then calculated the host-galaxy contribution in the other bands by applying the colour indices determined by \cite{man01} for elliptical galaxies to the de-reddened $R$-band magnitude. We adopted a Galactic extinction value of $A_R = 0.041$ mag from \cite{sch98} and derived extinction in the other bands through the \citet{car89} laws, setting $R_V=A_V/E(B-V)=3.1$, the mean value for the interstellar medium.

The optical and near-infrared host magnitudes were converted into flux densities using the zero-mag fluxes given by \cite{bes98}. The whole galaxy flux densities were multiplied by the $p(r_{\rm a})$ values to derive the contribution to the source photometry within the aperture radius. The results are shown in Table\ \ref{ta:vacou}.

In the UV case, we used the template of a 13 Gyr elliptical galaxy that is available from the SWIRE project\footnote{http://www.iasf-milano.inaf.it/polletta/templates/ swire$\_$templates.html}
\citep{polletta2007}. We scaled the template in order to have the host-galaxy flux expected in the $U$ filter. Galactic extinction in the UV bands was estimated by convolving the \cite{car89} laws with the filter effective area and source spectrum.
The results are in Table\ \ref{ta:vacou}. As can be seen, the host galaxy contribution is relevant in the near-infrared, whereas it is negligible in the UV.

\begin{table}
 \centering
  \caption{UV, optical and near-infrared observing bands with the corresponding Galactic extinction values $A_\lambda$, photometry aperture radius $r_{\rm a}$, percentage of the host-galaxy flux included in the given aperture $p(r_{\rm a})$, host-galaxy flux density contribution to the source photometry $F_{\rm gal}$, and median observed flux density $\langle F_{\nu}^{\rm obs} \rangle$, including both Mrk~421 and the host galaxy.}
   \label{ta:vacou}
  \begin{tabular}{@{}cccccc@{}}
  \hline
Filters & $A_\lambda$  & $r_{\rm a}$  & $p(r_{\rm a})$   &$F_{\rm gal}$ & $\langle F_{\nu}^{\rm obs} \rangle$\\
        & [mag]        & [arcsec]  & [per cent] &   [mJy]      &   [mJy]          \\
 \hline
$w2$& 0.112 &5.0&37 &0.039&12.011\\
$m2$&0.118  &5.0&37 &0.067&14.305\\
$w1$&0.095  &5.0&37 &0.122&13.922\\
$U$ &0.083  &5.0&37 &0.497&-\\
%$V$ &0.051  &3.0&25 &2.794&26.584\\
$R$ &0.041  &7.5&48 &7.862&31.615\\
$J$ &0.014  &7.5&48 &22.613&54.241\\
$H$ &0.009  &7.5&48 &26.786&46.071\\
$K$ &0.006  &7.5&48 &20.409&56.102\\
\hline
\end{tabular}
\end{table}

We checked that using the SWIRE Template method to calculate the host galaxy contribution at lower frequencies gives the same results obtained by the colour indices method within a few mJy.

\section{Colour analysis}

Analysis of colour variations is an important tool to investigate the spectral behaviour of the source and, in turn, the nature of its emission.

\begin{figure}
\centering
\includegraphics[width=8.5cm]{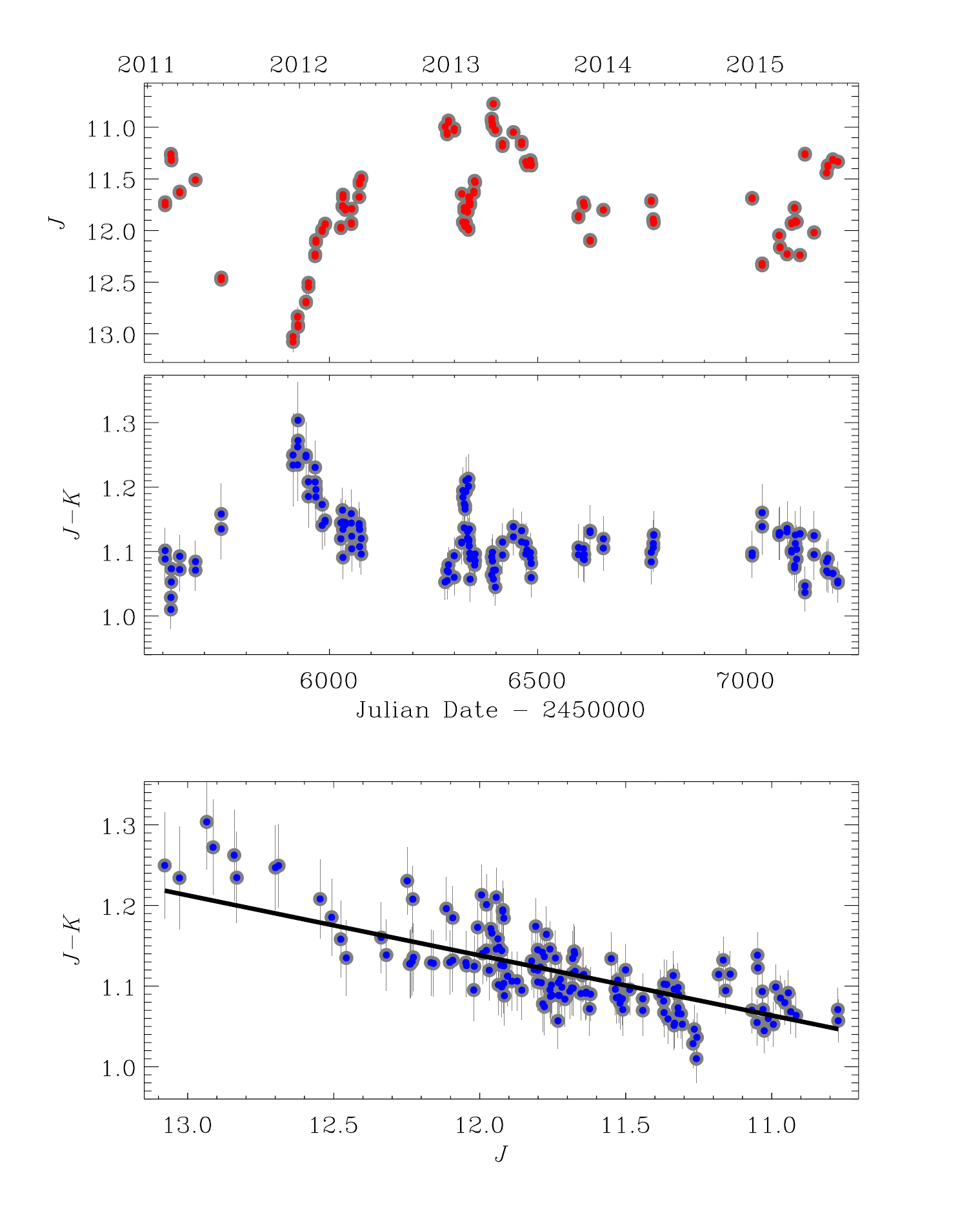}
  \caption{The $J$-band light curve in the 2011--2015 period (top panel); the corresponding $J-K$ colour index as a function of time (middle panel) and brightness level (bottom panel). In the bottom panel the solid line represents a linear fit to the data. The data have been corrected for the host galaxy contribution.}
\label{fi:colour_1101}
\end{figure}

In Fig.\ \ref{fi:colour_1101} we show the $J$-band light curve (top panel) together with the corresponding $J-K$ colour indices as a function of time (middle panel) and brightness level (bottom panel). The colour indices were calculated by selecting $J$ and $K$ data points with small errors and taken within at most 15 min. We obtained that the average $J-K$ value is 1.12, with a standard deviation of 0.05. The data were corrected for the host galaxy contribution as explained in Section 4.

It can be seen that, in general, the values of $J-K$ decrease with increasing flux, which is more evident in the bottom panel, where the behaviour of the colour index with brightness is displayed. We observe a bluer-when-brighter trend, as expected for a BL Lac object \citep[e.g.][]{ike11}, with a linear Pearson's correlation coefficient of 0.79 and Spearman's rank correlation coefficient of 0.77.

\section{Observations by \textit{Swift}}

In this section, we analyse the \textit{Swift} satellite data on Mrk~421 obtained with the UV/Optical Telescope \citep[UVOT;][]{rom05} and X-ray Telescope \citep[XRT;][]{bur05} instruments. During the 2007--2015 period, the source was observed by \textit{Swift} in 727 epochs.

\subsection{UVOT}
\label{uvot}

The UVOT instrument on board \textit{Swift} observed Mrk~421 mostly in the UV bands $w1$, $m2$ and $w2$, and sometimes also in the optical bands $v$, $b$, $u$. We downloaded these data from the NASA's High Energy Astrophysics Science Archive Research Center (HEASARC)\footnote{http://heasarc.nasa.gov} and reduced them with the {\sevensize \bf HEAsoft} package version 6.17 and the calibration release 20150717 of the CALDB database available at HEASARC. For each epoch, multiple images in the same filter were first summed with the task {\tt uvotimsum} and then aperture photometry was performed with {\tt uvotsource}.
We extracted source counts from a circular region with 5 arcsec radius centred on the source and background counts from a circle with 15 arcsec radius in a source-free field region.

\begin{figure}
\centering
\includegraphics[width=8cm]{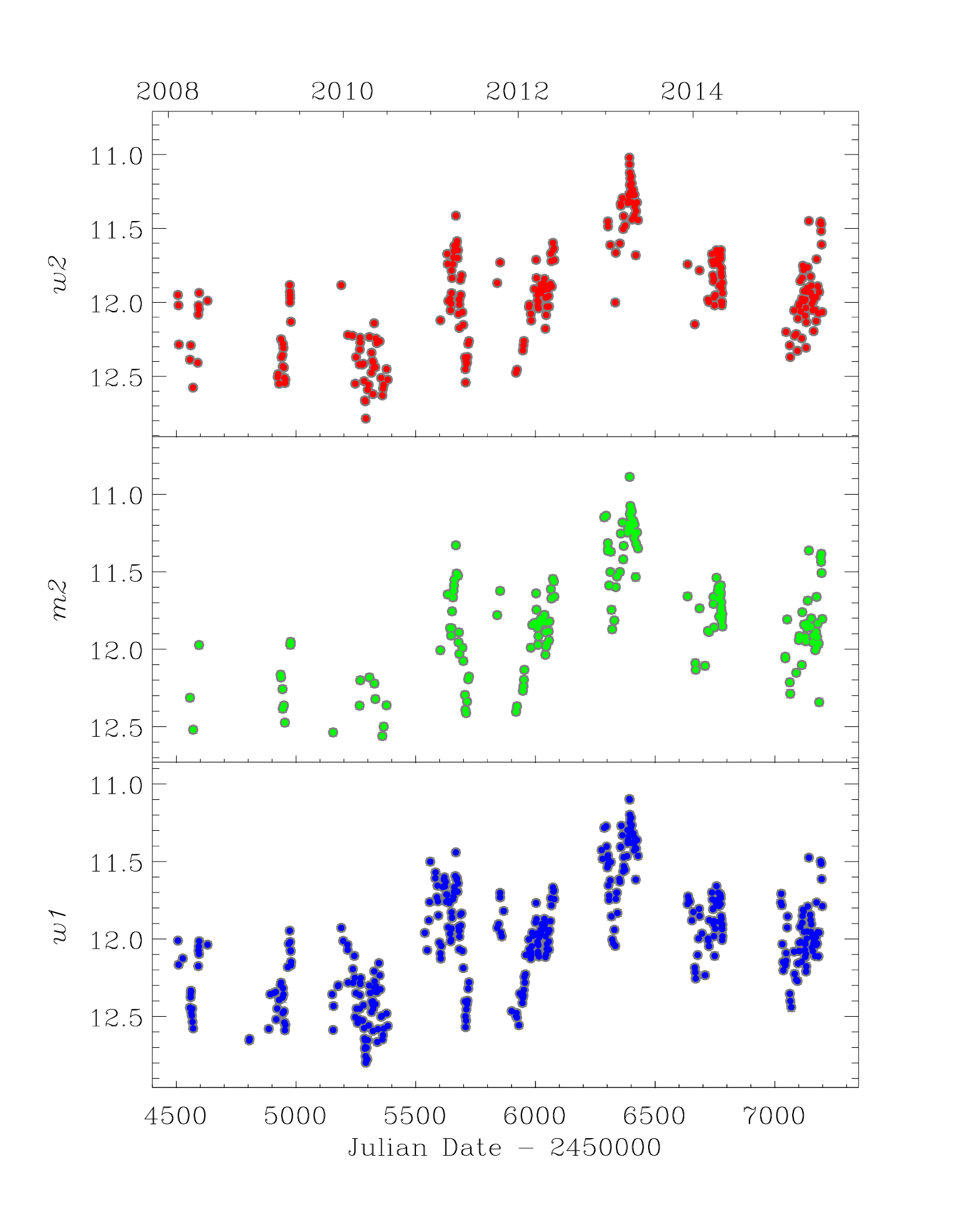}
  \caption{UV light curves of Mrk 421 built with \textit{Swift}-UVOT data.}
\label{fi:1101_uv}
\end{figure}

The UVOT light curves are shown in Fig.\ \ref{fi:1101_uv}. They confirm the general behaviour shown by the ground-based optical and near-infrared curves in Figs.\ \ref{fi:mk421_optico} and \ref{fi:1101_ir}.

\subsection{XRT}

We processed the XRT data with the {\sevensize \bf HEAsoft} package version 6.17 and the CALDB calibration files updated 20150721. The task {\tt xrtpipeline} was executed with standard screening criteria. Only observations performed in pointing mode and with more than 50 counts were selected for further analysis. In the 2007--2015 period, we were left with 710 observations in Windowed timing (WT) mode and only 16 in photon counting (PC) mode, so that we concentrated on the former.

We selected event grades 0-2 and used a circular region with 70 arcsec radius centred on the source to extract the source counts, and a similar region shifted away from the source along the window to extract the background counts. We verified that the background is negligible, as background counts are in average 1.5 per cent and at maximum 5 per cent of the source counts, so we did not correct for it.
Only three observations in WT mode have a mean rate greater than 100 cts s$^{-1}$, implying pile-up.
To correct for pile-up we discarded the inner 3-pixel radius circle in the source extraction region.

\begin{figure}
\centering
\includegraphics[width=8cm]{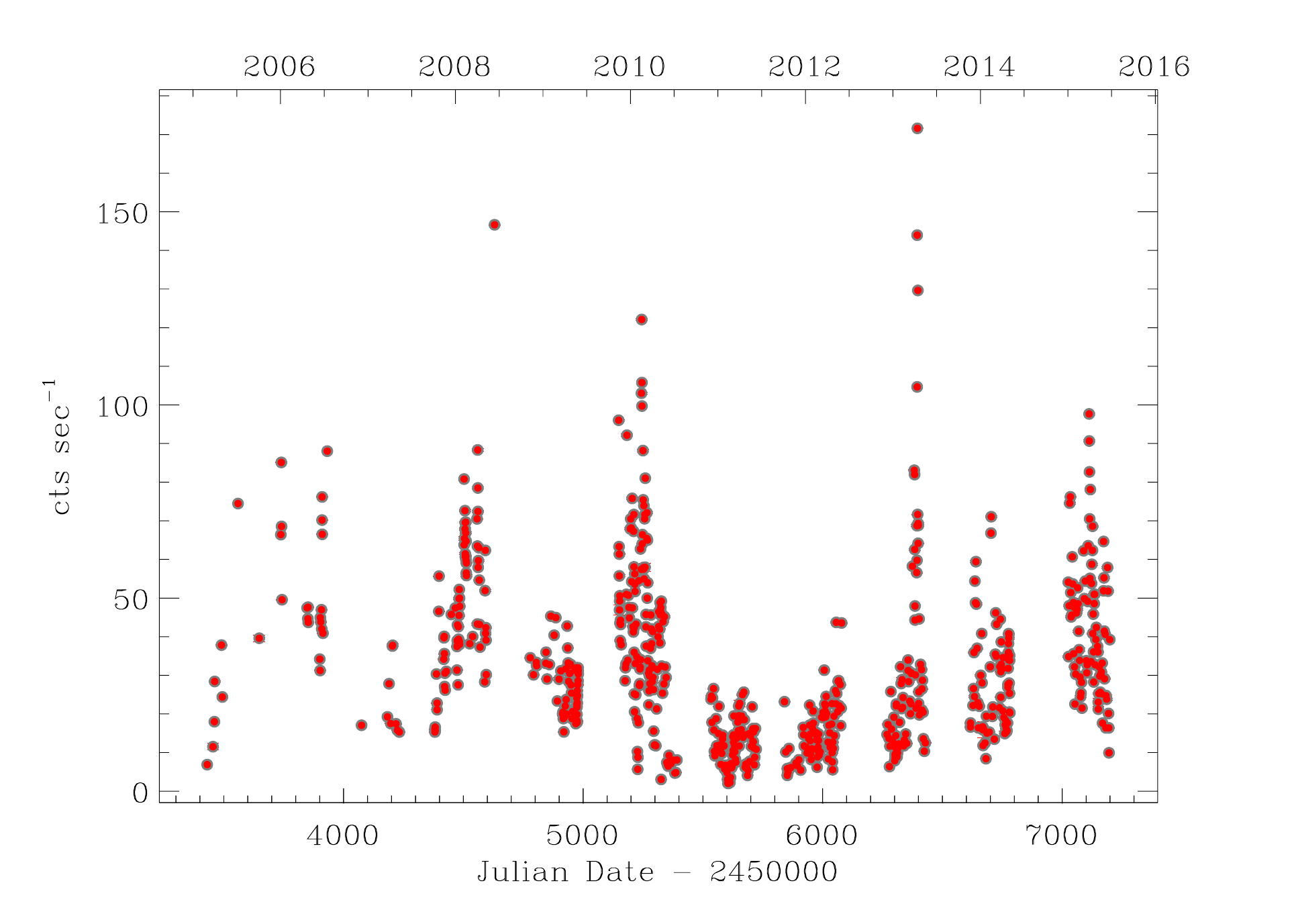}
  \caption{X-ray light curve of Mrk~421 obtained from data collected by the XRT instrument onboard the \textit{Swift} satellite in the period 2007--2015.}
\label{fi:1101_xray}
\end{figure}

We used the {\tt xrtmkarf} task to generate ancillary response files (ARF), which account for different extraction regions, vignetting, and PSF corrections. The X-ray light curve is shown in Fig.\ \ref{fi:1101_xray} and is discussed in the next Section.

By means of the task {\tt grppha} we associated the source spectra with the ARF and CALDB redistribution matrix function (RMF) files, and binned the source spectra in order to have a minimum of 20 counts in each bin for the $\chi^2$ statistics. These grouped spectra were then analysed with the {\sevensize \bf Xspec} package, using the energy channels greater than 0.35 keV to improve the goodness of the fit.

\begin{figure}
\centering
\includegraphics[width=8cm]{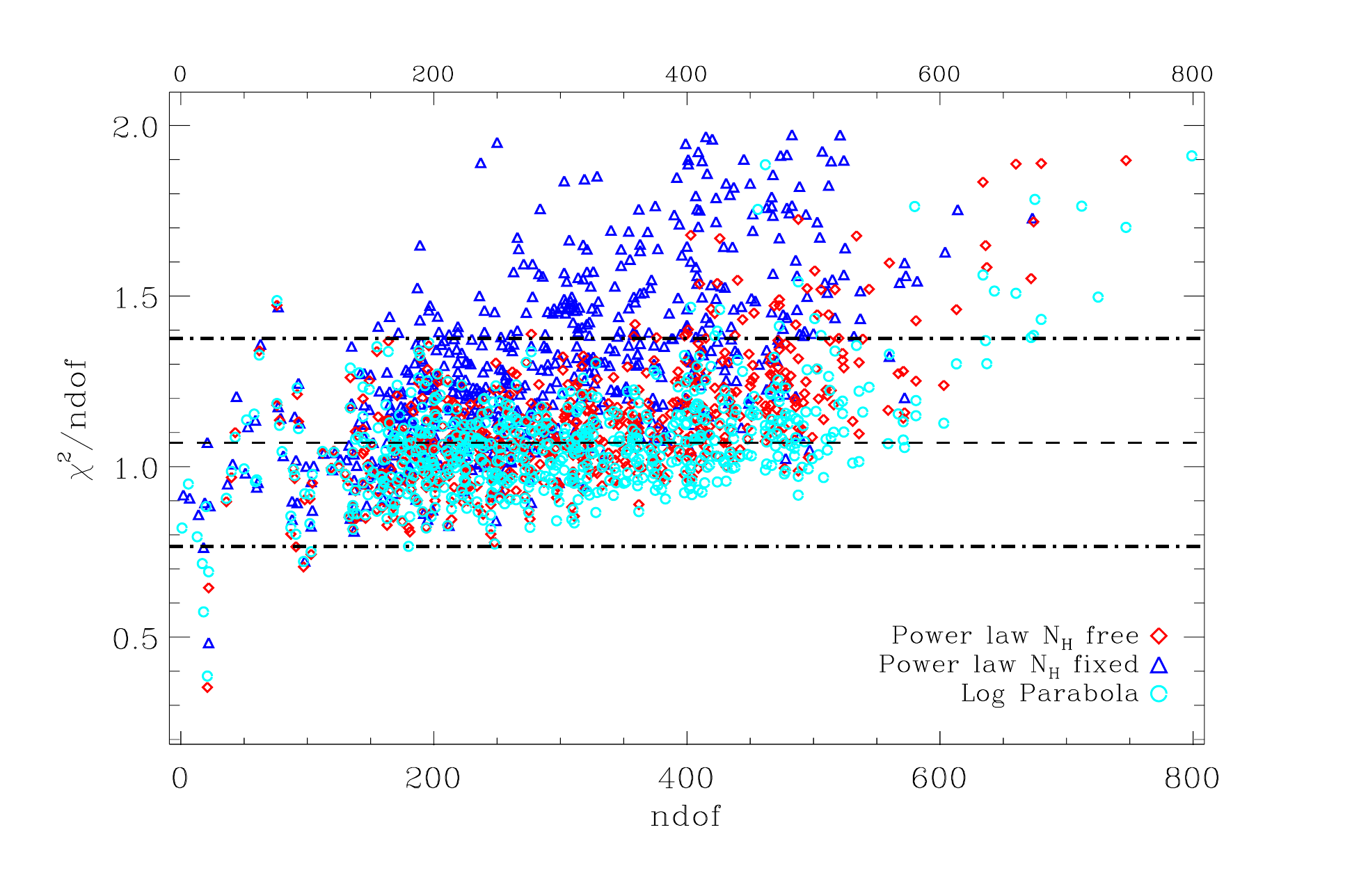}
  \caption{Reduced $\chi^2$ versus the number of degrees of freedom for the different models applied to the XRT spectra of Mrk~421. Blue triangles represent the results of the power-law model with $N_{\rm H}$ fixed to the Galactic value, red diamonds those of the power-law model with $N_{\rm H}$ free, and the cyan circles those of the log-parabola model with Galactic $N_{\rm H}$.}
\label{fi:1101_chi2}
\end{figure}

\begin{figure}
\centering
\includegraphics[width=6cm,angle=-90]{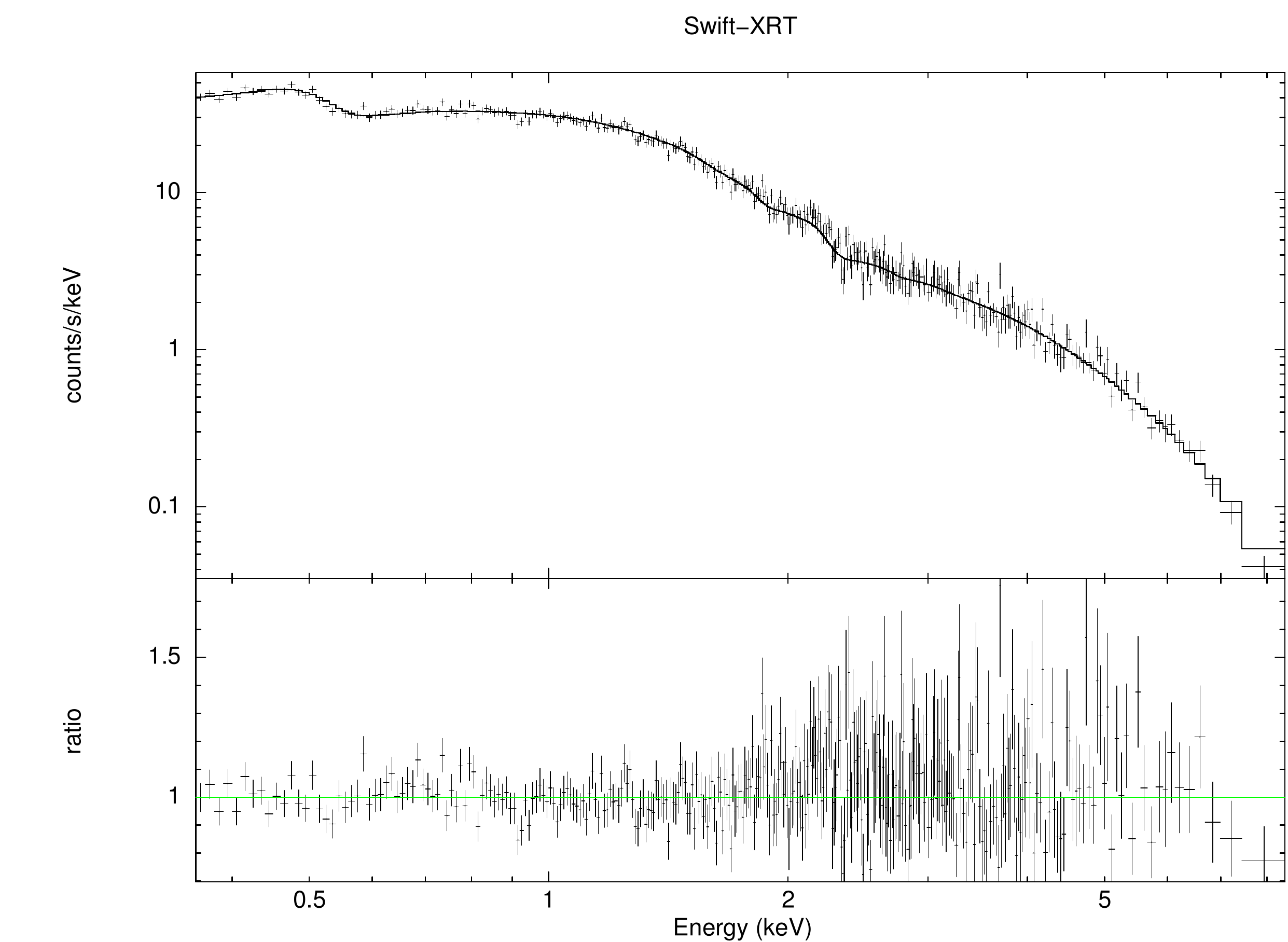}
  \caption{The XRT spectrum of Mrk~421 on 2008 February 13. The best fit was obtained with a log-parabola model. The bottom panel shows the ratio of the data to the folded model.}
\label{fi:1101_spectrum_xrt}
\end{figure}

We applied three different models for the spectral fitting. 1) An absorbed power-law model, where absorption is modelled according to \citet{wil00} and the Hydrogen column is fixed to the Galactic value $N_{\rm H}=1.61 \times 10 ^{20} \rm cm^{-2}$, as derived from the 21 cm measure by \cite{loc95}; 2) an absorbed power-law model with $N_{\rm H}$ free; 3) an absorbed log-parabola model with $N_{\rm H}$ fixed to the Galactic value. We favoured the third model, whose $\chi^2$ is usually smaller than that of the other models, and that produces results with smaller errors. In Fig.\ \ref{fi:1101_chi2} we show the $\chi^2$ versus the number of degrees of freedom (ndof). The $\chi^2$ is more stable when the log-parabola model is applied, but it increases with ndof. This is possibly due to a pronounced curvature. Fig.\ \ref{fi:1101_spectrum_xrt} shows an example of XRT spectrum. It was best-fitted with a log parabola.

\begin{figure}
\centering
\includegraphics[width=8cm]{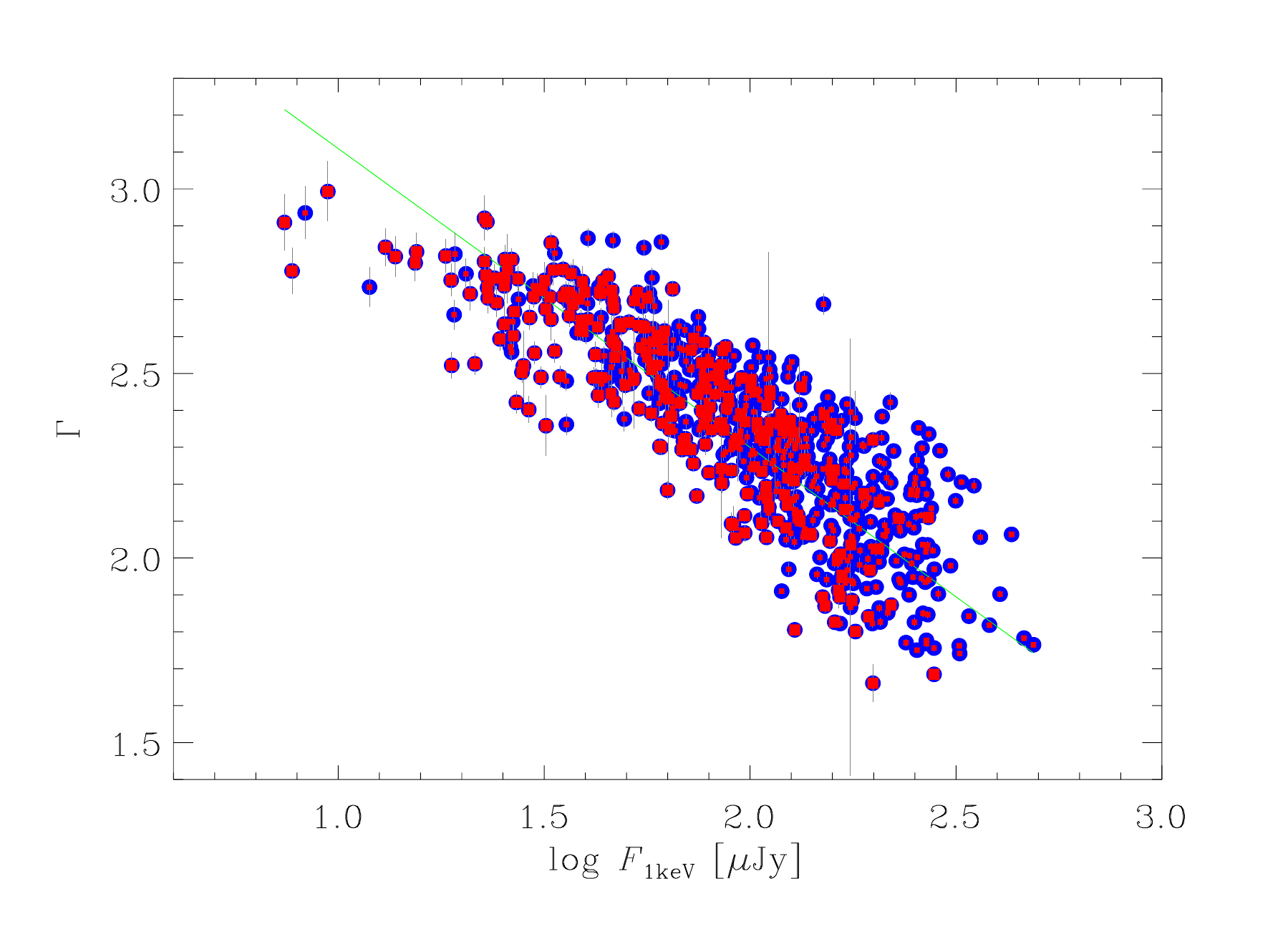}
  \caption{The X-ray photon index $\Gamma$ as a function of the unabsorbed flux density at 1 keV. Data with error less than 30 per cent of the flux are shown. Red squares refer to the best-fitted cases, where the reduced $\chi^2$ is in the range 0.8--1.2 and the number of degrees of freedom is $> 10$. The solid line represents a linear fit to the data.}
\label{fi:1101_gamma_fix}
\end{figure}

In the case of a power-law model with fixed $N_{\rm H}$, the photon index $\Gamma$ ranges from 1.66 to 2.99, indicating a spectrum that oscillates from hard to soft. The average value is 2.34, with standard deviation of 0.26. To understand whether these spectral changes correspond to real variations or are due to noise, we recall the definition of the mean fractional variation $F_{\rm var} = \sqrt{\sigma^2-\delta^2}/\langle f \rangle$ \citep{pet01}, which is commonly used to characterise variability. Here $\langle f \rangle$ is the mean value of the variable we are analysing, $\sigma^2$ its variance and $\delta^2$ the mean square uncertainly. In our case, $F_{\rm var}=0.11$, so we conclude that the variations reflect genuine source variability rather than noise. Figure \ref{fi:1101_gamma_fix} displays the photon index $\Gamma$ as a function of the flux density at 1 keV. We note that the lowest $\Gamma$ values correspond to the highest fluxes, in agreement with the harder-when-brighter trend often observed in blazars.
However, this model produces statistically unacceptable fits (see Fig.\ \ref{fi:1101_chi2}).

\begin{figure}
\centering
\includegraphics[width=8cm]{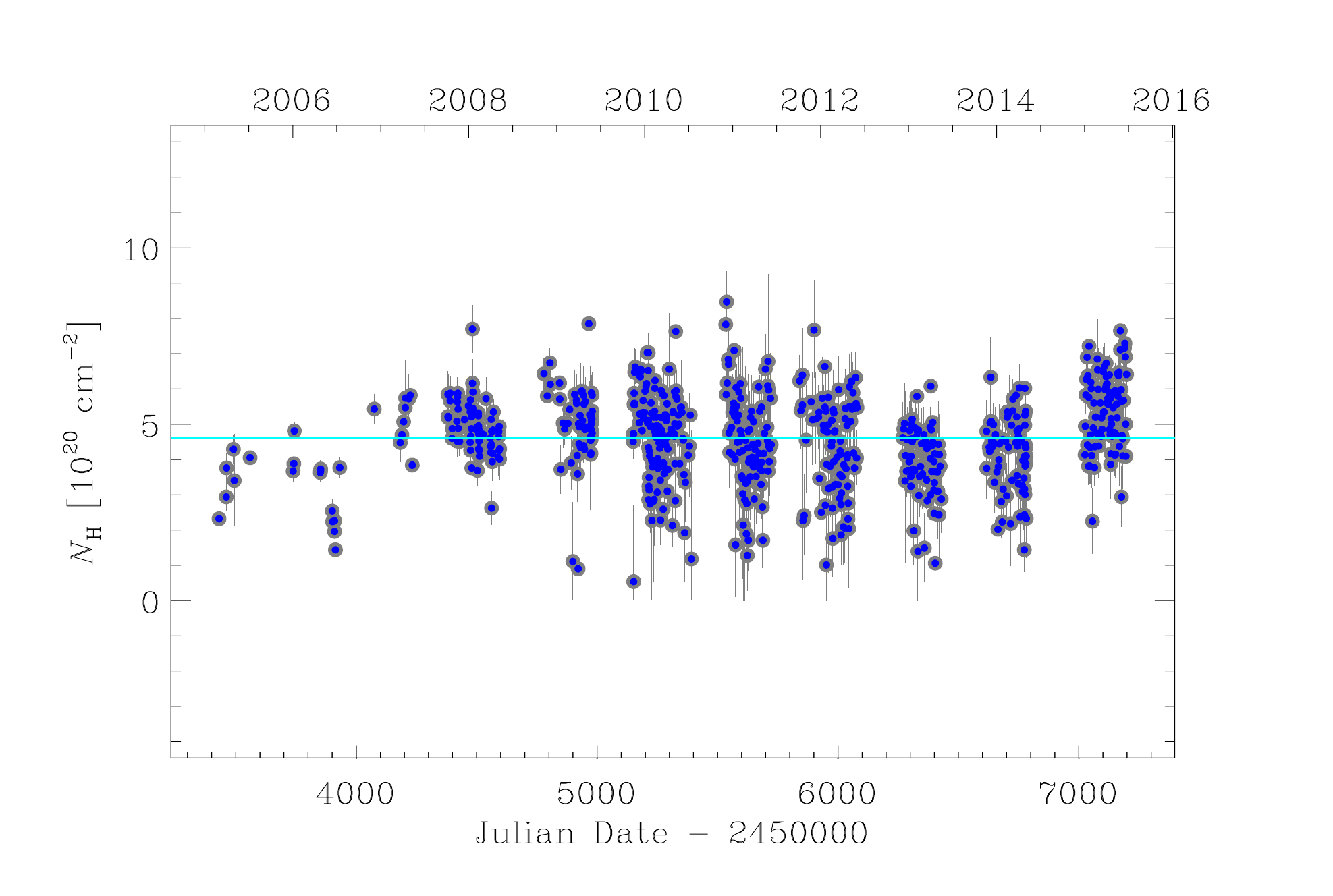}
  \caption{The Galactic Hydrogen column $N_{\rm H}$ as a function of time when a power-law model with free absorption is applied to the XRT spectra. The cyan line marks the average value.}
\label{fi:1101_nh}
\end{figure}

On the other hand, the power-law model with $N_{\rm H}$ free implies a large spread of $N_{\rm H}$ values, which very unlikely corresponds to a physical scenario (see Fig.\ \ref{fi:1101_nh}). We note that in this case the mean value of $N_{\rm H}$ exceeds the Galactic value by a factor $\sim 3$, suggesting that the spectrum is curved.

The log-parabola model has largely been used to fit the X-ray spectrum of this source \citep[e.g.][]{mas04,sin15}. It offers a statistically better fit to the data in case of a curved spectrum. In this model the photon index $\Gamma$ is replaced by two parameters: $\alpha$, the photon index, and $\beta$, the spectral curvature. We obtained $\alpha$ values in the range from 1.58 to 2.99, similar to that found for $\Gamma$ in the power-law case. The average value is 2.27, with standard deviation of 0.28. The mean fractional variation is $F_{\rm var}=0.12$. The $\beta$ parameter goes from $-0.09$ to 0.66, with an average value of 0.24 and standard deviation of 0.10. The mean fractional variation is $F_{\rm var}=0.29$. The large range of $\beta$ values indicates strong curvature changes. However, large uncertainties affect the most extreme $\beta$ values, demanding caution.

\begin{figure*}
\centerline{
    \epsfig{figure=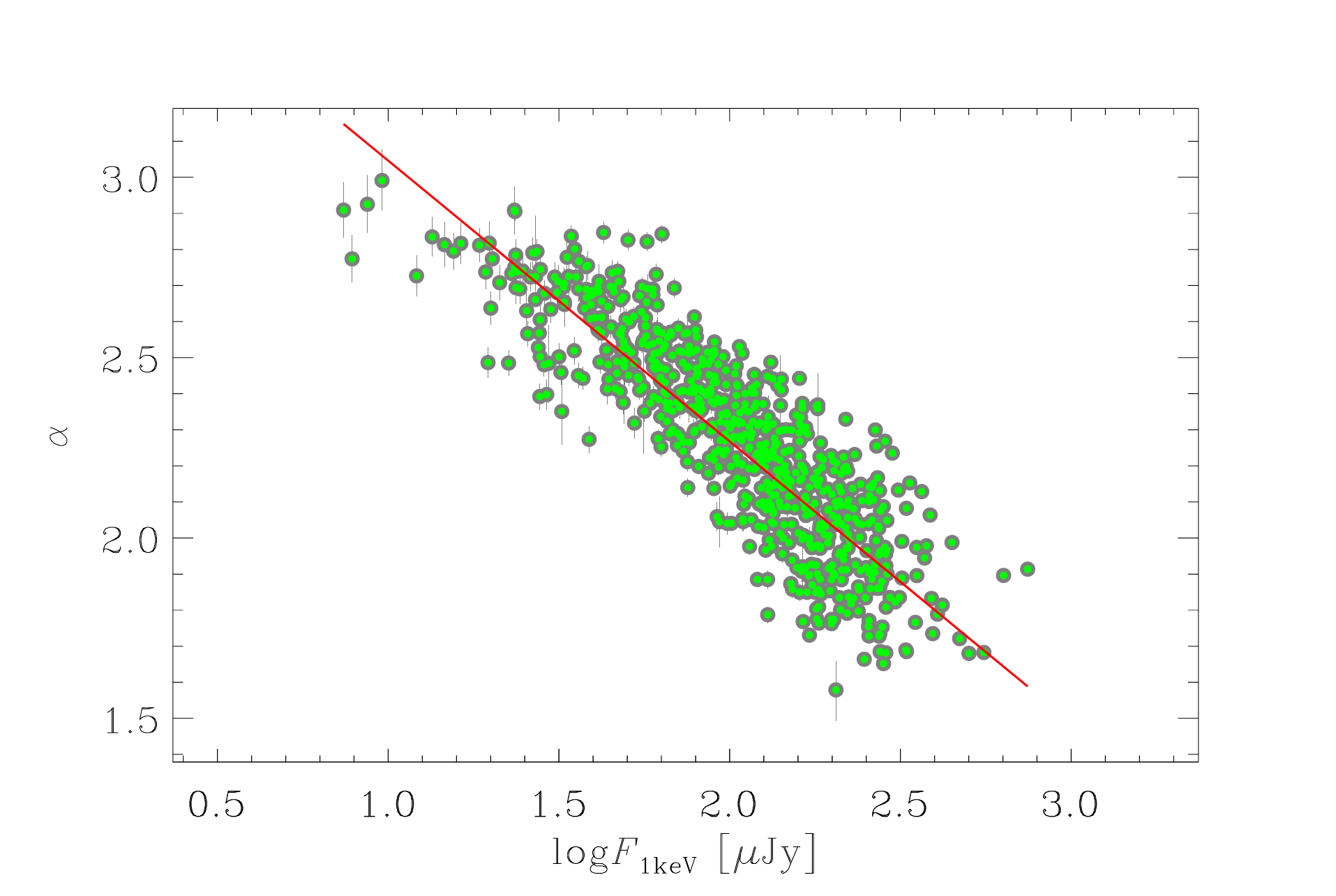,width=7.5cm}
    \epsfig{figure=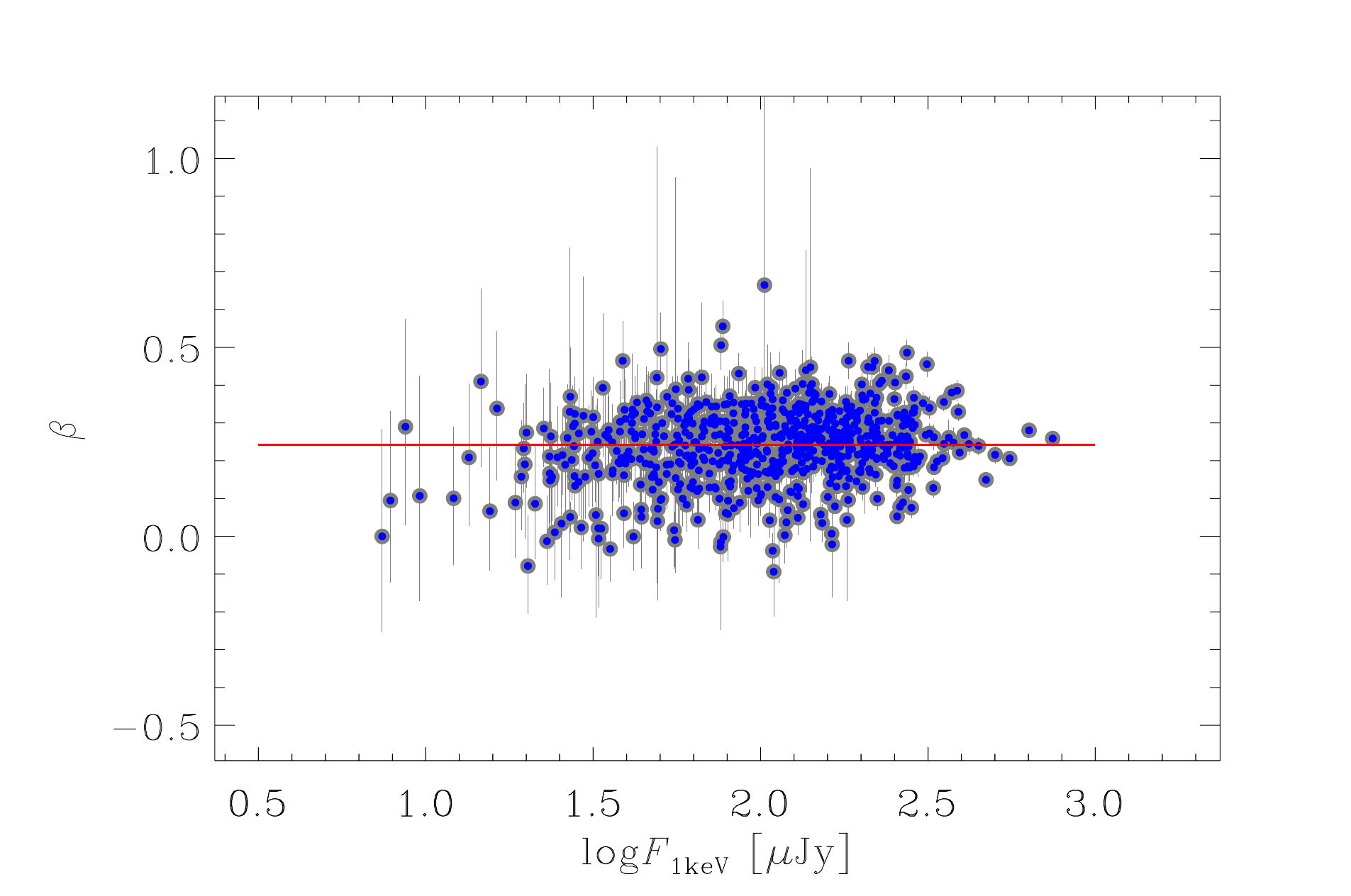,width=7.5cm}
    }
\caption{The behaviour of the $\alpha$ (left) and $\beta$ (right) parameters of the log-parabola model applied to the XRT spectra of Mrk~421 as a function of the unabsorbed flux density at 1 keV. The solid lines represent linear fits to the data.}\label{xrt_flux}
\end{figure*}

In Fig.\ \ref{xrt_flux} we show the trend of the $\alpha$ and $\beta$ parameters of the log-parabola model applied to the X-ray spectra of Mrk~421 as a function of the source flux. While $\alpha$ behaves similarly to $\Gamma$ (Fig.\ \ref{fi:1101_gamma_fix}), confirming the harder-when-brighter spectral property, no clear correlation between $\beta$ and flux is recognizable. We obtained a linear Pearson's correlation coefficient of 0.86/0.20 and Spearman's rank correlation coefficient of 0.88/0.16 for the  $\alpha$/$\beta$ cases.

\section{Observations by {\em Fermi}}
The Large Area Telescope \citep[LAT;][]{atw09} instrument onboard the \textit{Fermi} satellite observes in the 20 MeV--300 GeV energy range. 
In this paper we considered data between 2008 August 4 ($\rm JD=2454683.15$) and 2015 September 10 ($\rm JD=2457275.50$). We used Pass 8 data \citep{atw13}, based on a complete revision of the entire LAT event-level analysis. We adopted the {\sevensize \bf SCIENCETOOLS} software package version v10r0p5 and followed the standard reduction procedure, as done in \citet{car15}. 
We considered a Region of Interest (ROI) of 30\degr\ radius, a maximum zenith angle of 90\degr, and only ``Source" class events (evclass=128, evtype=3). 
The spectral analysis was performed with the science tool {\tt gtlike} and the response function P8R2\textunderscore SOURCE\textunderscore V6. Background was modelled with isotropic (iso\textunderscore source\textunderscore v06.txt) and Galactic diffuse emission (gll\textunderscore iem\textunderscore v06.fit) components. 

As in the 3FGL catalogue, we used a power-law model for the Mrk~421 spectrum. A first maximum likelihood analysis was performed over the whole period to remove from the model the sources having Test Statistics\footnote{This is defined as: ${\rm TS}=2(\log L_1 - \log L_0)$, where $L_1$ and $L_0$ are the likelihood of the data when the model includes or excludes the source, respectively \citep{mat96}.} less than 10. A second maximum likelihood was run on the updated source model.

\begin{figure}
\centering
\includegraphics[width=8cm]{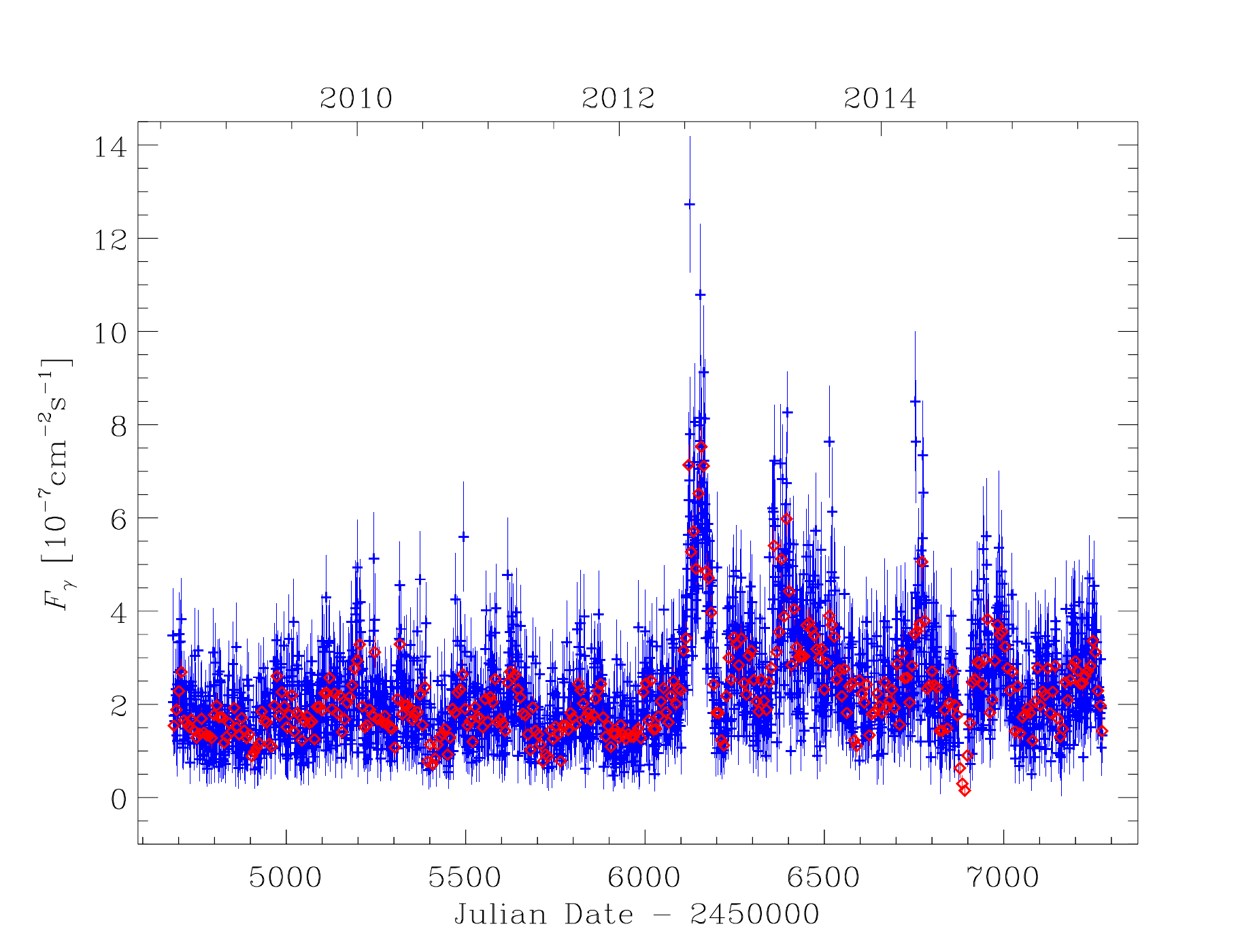}
  \caption{The \textit{Fermi}-LAT 0.1--300 GeV fluxes ($10^{-7} \rm \, ph \, cm^{-2} \, s^{-1}$) derived with different time bins in the 2008--2015 period (red symbols refer to weekly-binned, blue symbols to daily-binned data).}
\label{fi:gamma_curve}
\end{figure}

Integrating over the whole period, the fit gives TS=111714 in the 0.1--300 GeV energy range, with an integrated average flux of $\rm (2.18 \pm 0.02) \times 10^{-7} \, ph \, cm^{-2} \, s^{-1}$ and a photon index $\Gamma=1.77 \pm 0.01$.
The high statistical significance allowed us to obtain weekly-binned and even daily-binned light curves, which are displayed in Fig.\ \ref{fi:gamma_curve}. The spectral index of Mrk~421 and all sources within 10\degr\ were frozen to the values resulting from the likelihood analysis over the entire period.

\section{Multiwavelength behaviour}

\begin{figure*}
\centering
\includegraphics[width=10cm]{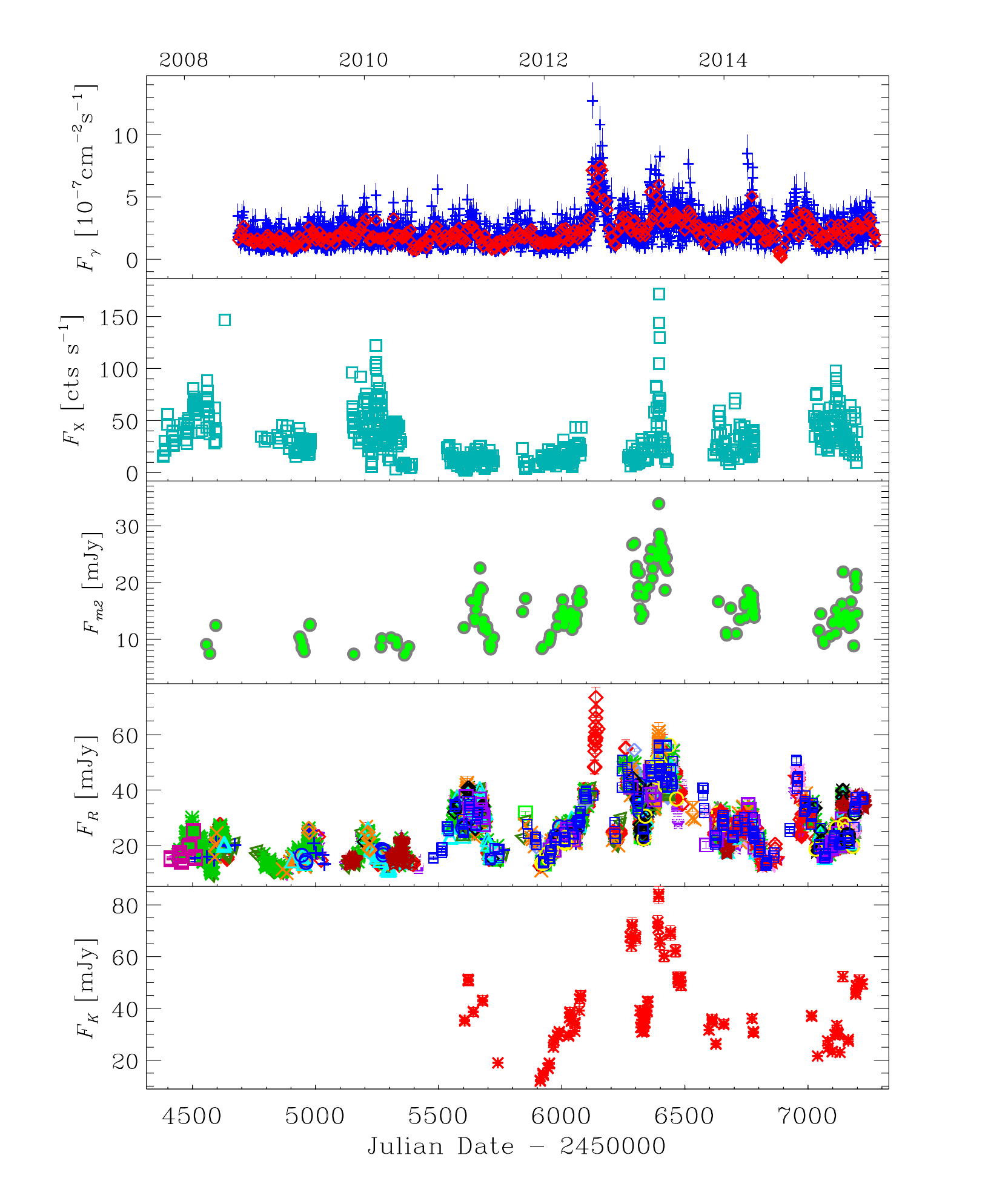}
  \caption{Multiwavelength emission behaviour of Mrk~421 as a function of time. From top to bottom: the $\gamma$-ray fluxes in the 100 MeV -- 300 GeV energy range from {\em Fermi}, red diamonds/blue plus signs refer to weekly/daily-binned data; the X-ray count rate from \textit{Swift}-XRT; the \textit{Swift}-UVOT observed flux densities in the $m2$ band (mJy); the $R$-band observed flux densities (mJy); the $K$-band observed flux densities (mJy). Data in the $m2$, $R$ and $K$ filters were cleaned from the host-galaxy light contamination.}
\label{fi:1101_multi}
\end{figure*}

Figure \ref{fi:1101_multi} compares the time evolution of the Mrk~421 flux at different frequencies in the 2007--2015 period.
The 2013 outburst was observed at all frequencies, while the X-ray outbursts in 2008 and 2010 lack a major optical counterpart (and the latter also a $\gamma$ counterpart) and are difficult to identify in UV because of sparse sampling. In contrast, in 2011 we notice a flare in UV, optical and NIR, but not in X-rays and in $\gamma$-rays. In $\gamma$-rays a major outburst is observed in 2012, at the same time of the strongest optical event. Other noticeable $\gamma$-ray flares were detected in 2013 and 2014. In general, the source behaviour at $\gamma$ energies appears similar to that observed in the optical band, while the X-ray light curve seems quite different \citep[see also][]{don09}.

\begin{figure}
\centering
\includegraphics[width=9cm]{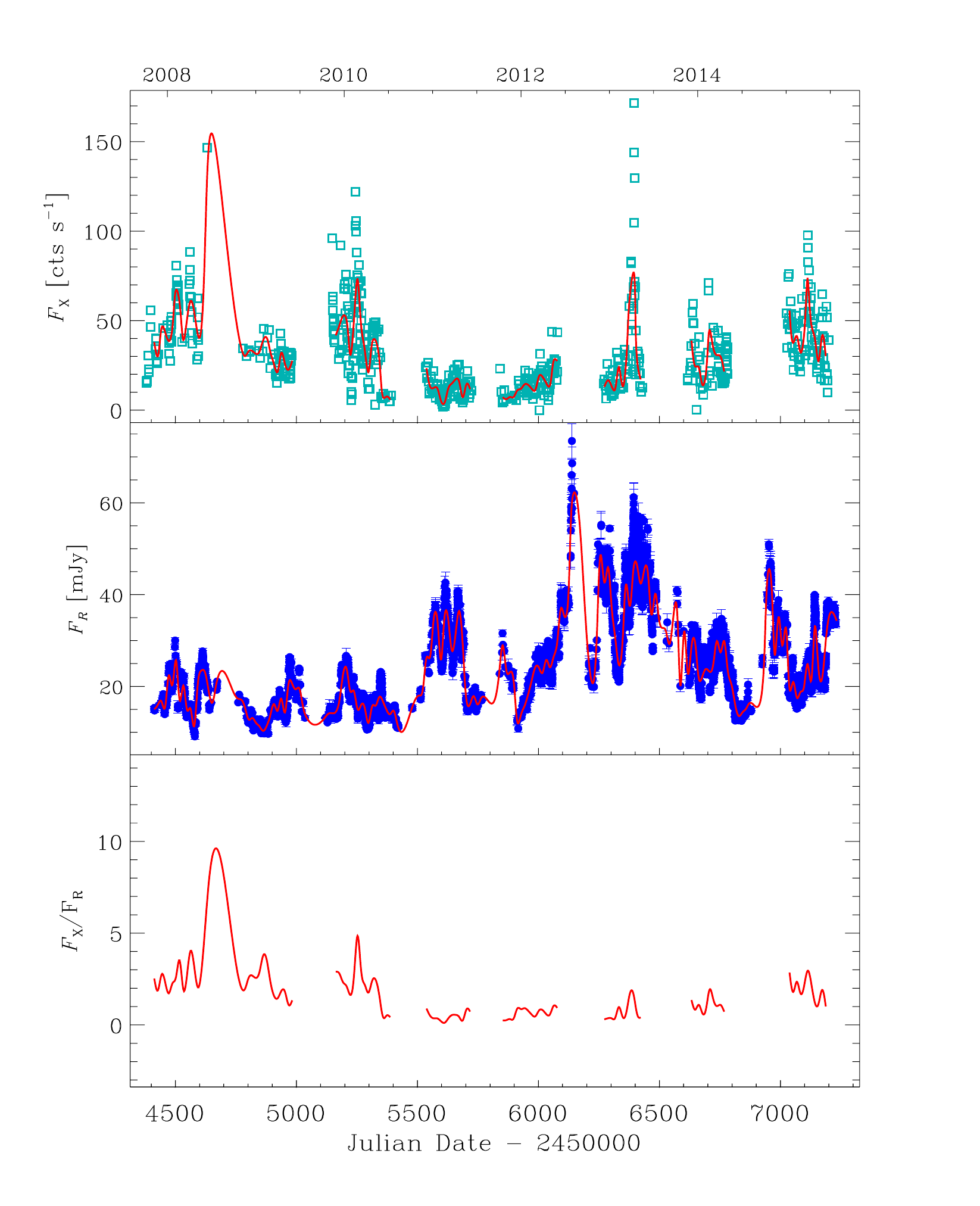}
  \caption{Top: the X-ray count rate from \textit{Swift}-XRT. Middle: the $R$-band flux densities (mJy). In both panels cubic spline interpolations through the 15-d binned light curves are shown. Bottom: the ratio between the X-ray and the optical spline fits.}
\label{fi:1101_spline_15}
\end{figure}

In Fig.\ \ref{fi:1101_spline_15} we compare the source behaviour in X-ray and $R$ band. The long-term trend is traced by means of cubic spline interpolations through the 15-d binned light curves. The ratio between the X-ray and optical splines is displayed in the bottom panel and highlights that the X-ray emission strongly dominates from 2007 June to 2009 June, its importance decreases from 2009 June to 2010 July, and reaches a minimum in 2010 October--2012 June, when the source is very active in the optical band. Starting from 2013, the ratio appears to moderately increase again.

In HBLs the optical and X-ray emissions are thought to be both produced by synchrotron process.
The variability of the X-ray to optical flux ratio in Mrk~421 then may indicate that the jet zones from where the X-ray and optical radiations are emitted do not coincide and are characterised by their own short-term variability.
Moreover, the fact that periods of X-ray flux dominance alternate with periods of optical flux dominance suggests that the corresponding emitting regions belong to a curved jet whose orientation changes may alternatively favour the Doppler enhancement of one region with respect to the other \citep[e.g.][]{vil09a,vil09b}.
An alternative explanation would be that of a one-zone model where the jet parameters change so that the synchrotron peak frequency shifts, modifying the ratio between the X-ray and optical fluxes. However, this kind of model met some difficulties in explaining the behaviour of Mrk~421 during the 2008 active state \citep[e.g.][]{ale12}.

\section{Variability of the optical polarisation}

As mentioned in the Introduction, the polarised blazar emission shows variable degree of linear polarisation ($P$) and electric vector polarisation angle \citep[EVPA; e.g.][]{smi96}. We analyse the polarimetric behaviour of Mrk~421 by means of 1430 optical data acquired as $R$-band photo-polarimetry by the Lowell (Perkins), Crimean, Calar Alto observatories, and as spectropolarimetry by the Steward observatory. In the latter case, the values of $P$ and EVPA are derived from the median of the normalized Stokes' parameters $q=Q/I$ and $u=U/I$ in the 5000--7000 \AA\ bandpass, whose effective wavelength is close to the Cousins' $R$ band.
A description of the data acquisition and reduction procedures is given in \cite{jor10}, \cite{lar08} and \cite{smi03}.

\begin{figure}
\centering
\includegraphics[width=8.5cm]{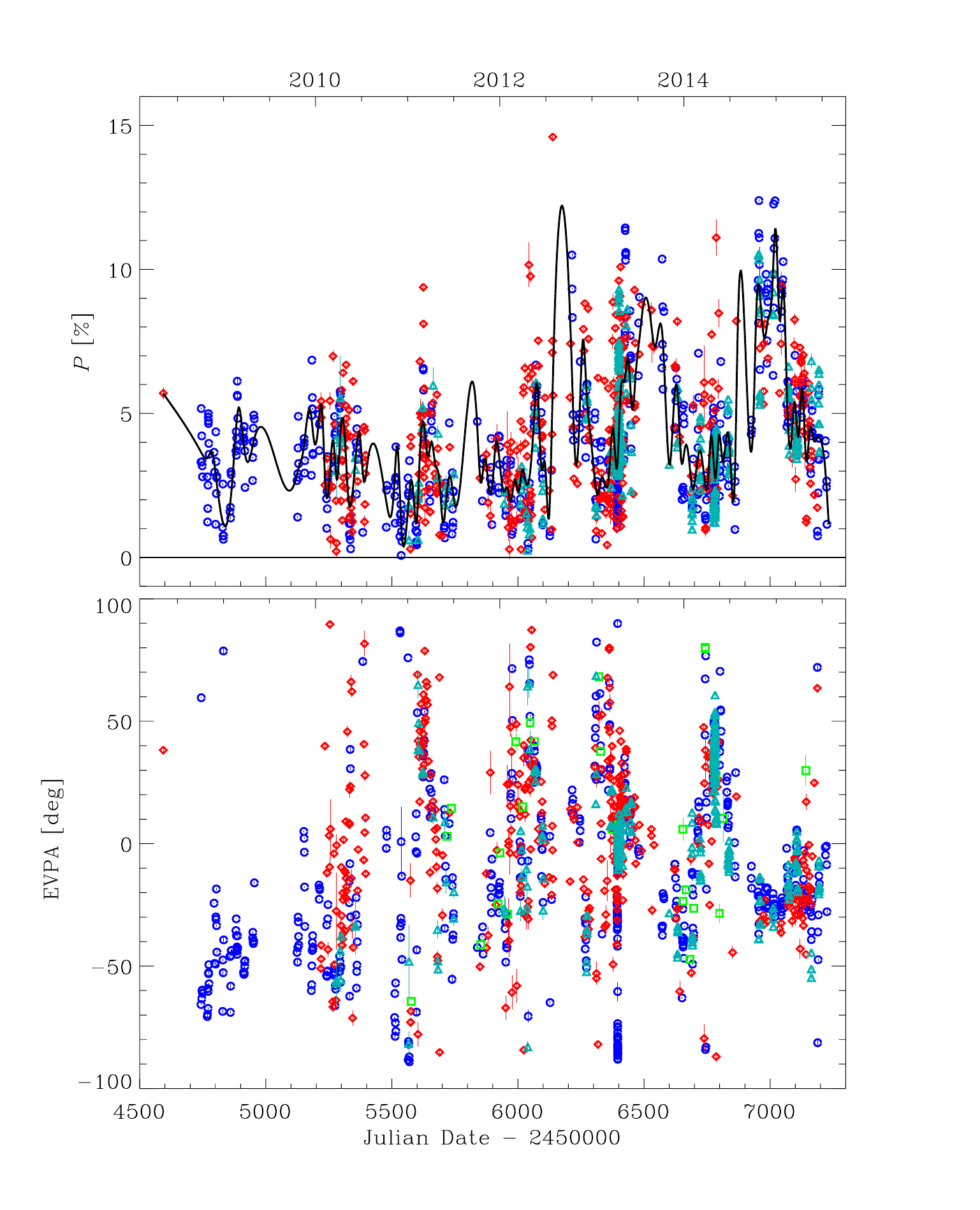}
  \caption{Top: the observed degree of polarisation as a function of time; the black line represents a cubic spline interpolation through the 15-d binned data. 
Bottom: the observed EVPAs in the $-90\degr$ and $+90\degr$ range.
Data are from the Calar Alto (green squares), Crimean (red diamonds), Lowell (cyan triangles) and Steward (blue circles) observatories.}
\label{fi:1101_pol_obs}
\end{figure}

The time evolution of the observed $P$ and EVPA is shown in Fig.\ \ref{fi:1101_pol_obs}.
A cubic spline interpolation through the 15-day binned percentage polarisation curve is drawn to highlight the long-term behaviour. 

In order to determine the degree of polarisation intrinsic to the jet, the unpolarized contribution of the galaxy must be subtracted. 
The intrinsic polarisation is computed using the following expression:
\begin{equation}\label{p_jet}
P_{\rm jet}=\frac{F_{\rm pol}}{F_{\rm jet}} = \frac{P_{\rm obs} \times F_{\rm obs}}{F_{\rm obs} - F_{\rm gal}}
\end{equation}

\begin{figure}
\centering
\includegraphics[width=8.5cm]{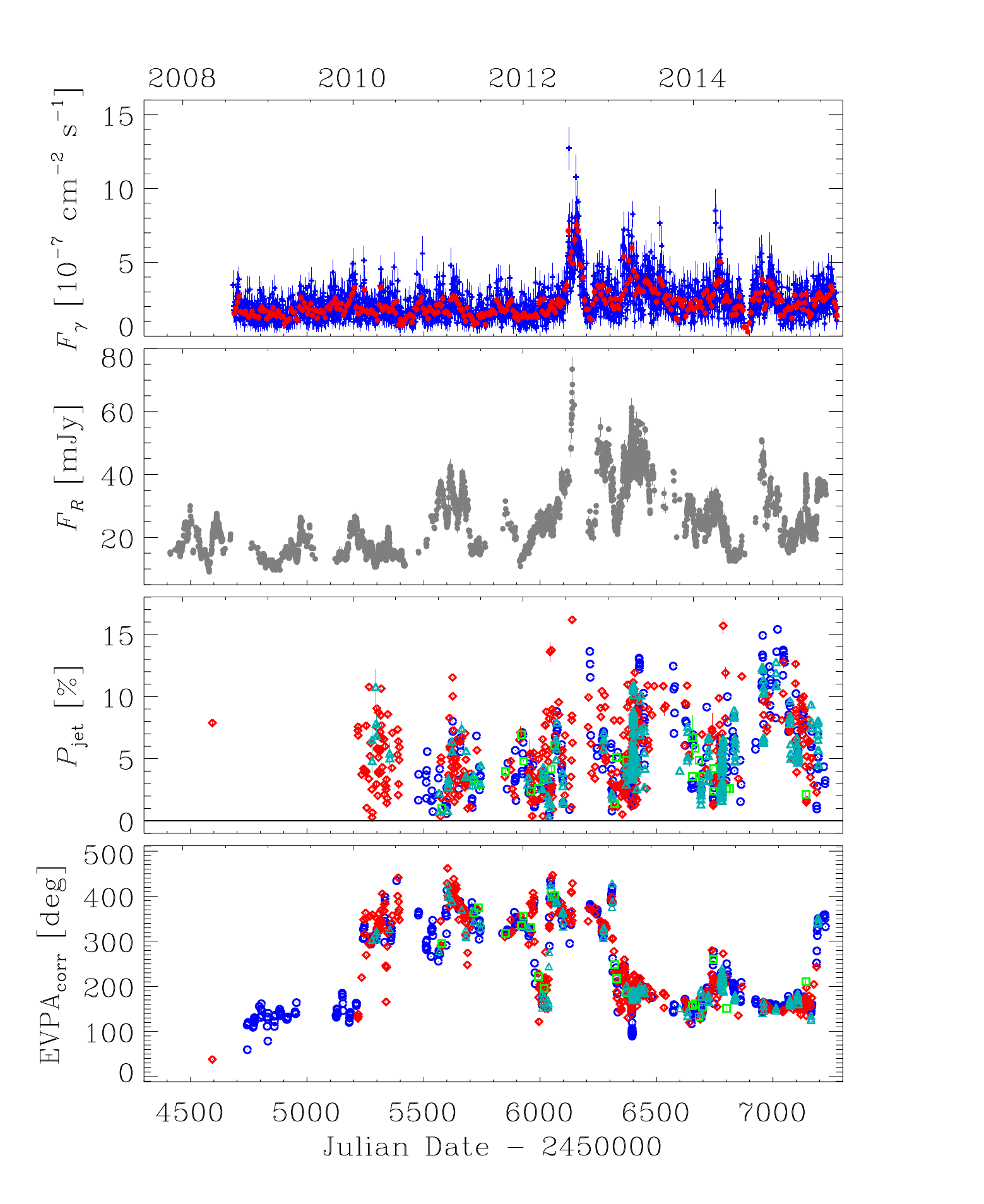}
  \caption{From top to bottom: 
the $\gamma$-ray fluxes between 100 MeV and 300 GeV; 
the $R$-band flux densities cleaned from the host-galaxy light contamination;
the degree of polarisation of the jet emission;
the EVPA after fixing the $\pm 180\degr n$ ambiguity. 
Symbols and colours as in Fig.\ \ref{fi:1101_pol_obs}.}
\label{fi:1101_pol}
\end{figure}

This is compared to the $\gamma$-ray and $R$-band light curves in Fig.\ \ref{fi:1101_pol}\footnote{The number of $P$ data points in Fig.\ \ref{fi:1101_pol} is smaller than in Fig.\ \ref{fi:1101_pol_obs} because to calculate $P_{\rm jet}$ we need $F_{\rm obs}$ (see Eq.\ \ref{p_jet}), which was not always available.}. 
The $\pm 180\degr n$ (where $n$ is an integer number) ambiguity in EVPA was treated by choosing at each epoch that angle among EVPA, EVPA+180\degr\ and EVPA-180\degr\ that leads to a minimum angle difference with respect to the previous epoch. 
In this choice we also considered the errors on the EVPAs. 

The polarisation degree ranges between 0.07 and 14.6 per cent, with a mean value of about 4 per cent. The fractional variability $F_{\rm var}$ (see Sect.\ 6.2) for $P$ is $\sim 0.56$, significantly larger than that of both optical and $\gamma$-ray fluxes, which is $\sim 0.40$. This suggests that polarisation is likely to be dominated by shorter time scale effects than the optical and $\gamma$-ray flux variability. In any case, the figure shows that there are periods where a high flux corresponds to a high polarisation degree, but this is not a general rule. 

\begin{figure}
\centering
\includegraphics[width=9cm]{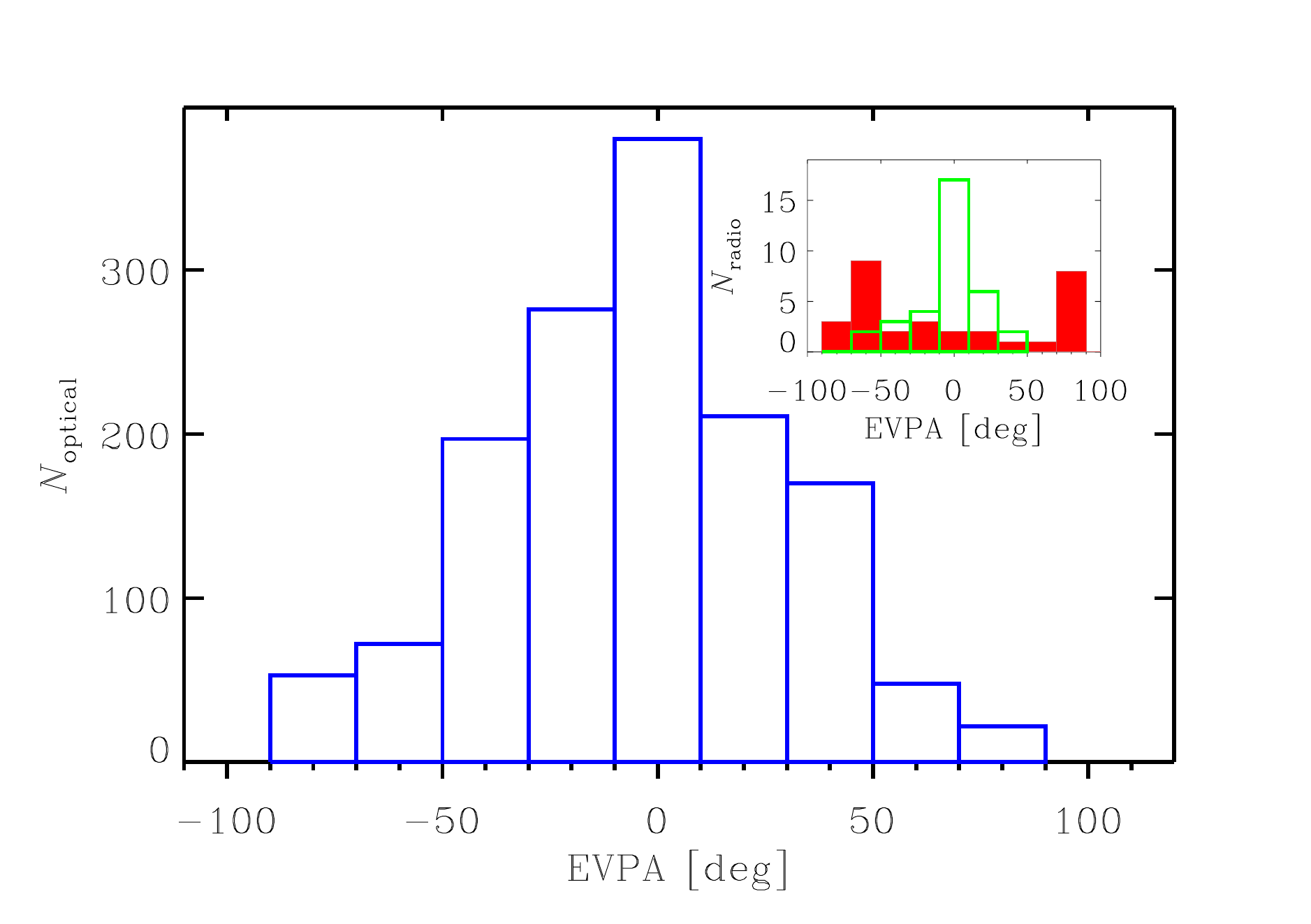}
  \caption{The distribution of the optical EVPAs; the inset displays those of the radio EVPAs at 43 GHz (green line) and at 15 GHz (red histogram).}
\label{fi:histogram_teta}
\end{figure}

The EVPA values cluster around a mean value of $\sim 4\degr$ (see Fig.\ \ref{fi:histogram_teta}), but wide rotations appear after arranging the angles to fix the $\pm 180\degr n$ ambiguity. Figure \ref{fi:histogram_teta} also shows the distribution of radio EVPAs at 43 and 15 GHz from the Boston University Blazar Group\footnote{https://www.bu.edu/blazars/research.html} and MOJAVE Project\footnote{http://www.physics.purdue.edu/MOJAVE/} \citep{lis09}, respectively. The number of radio data is small, but they suggest that the direction of the 43 GHz polarisation is aligned with the optical one, while the 15 GHz emission, likely coming from an outer jet region, has a transverse polarisation angle. A flip by 90\degr\ in EVPA usually means that there is a change from optically thin to optically thick properties of the region that most likely has happened between 43 GHz and 15 GHz. 

   \begin{figure*}
   \centering{
   %\vspace{0.0001cm}
    \epsfig{figure=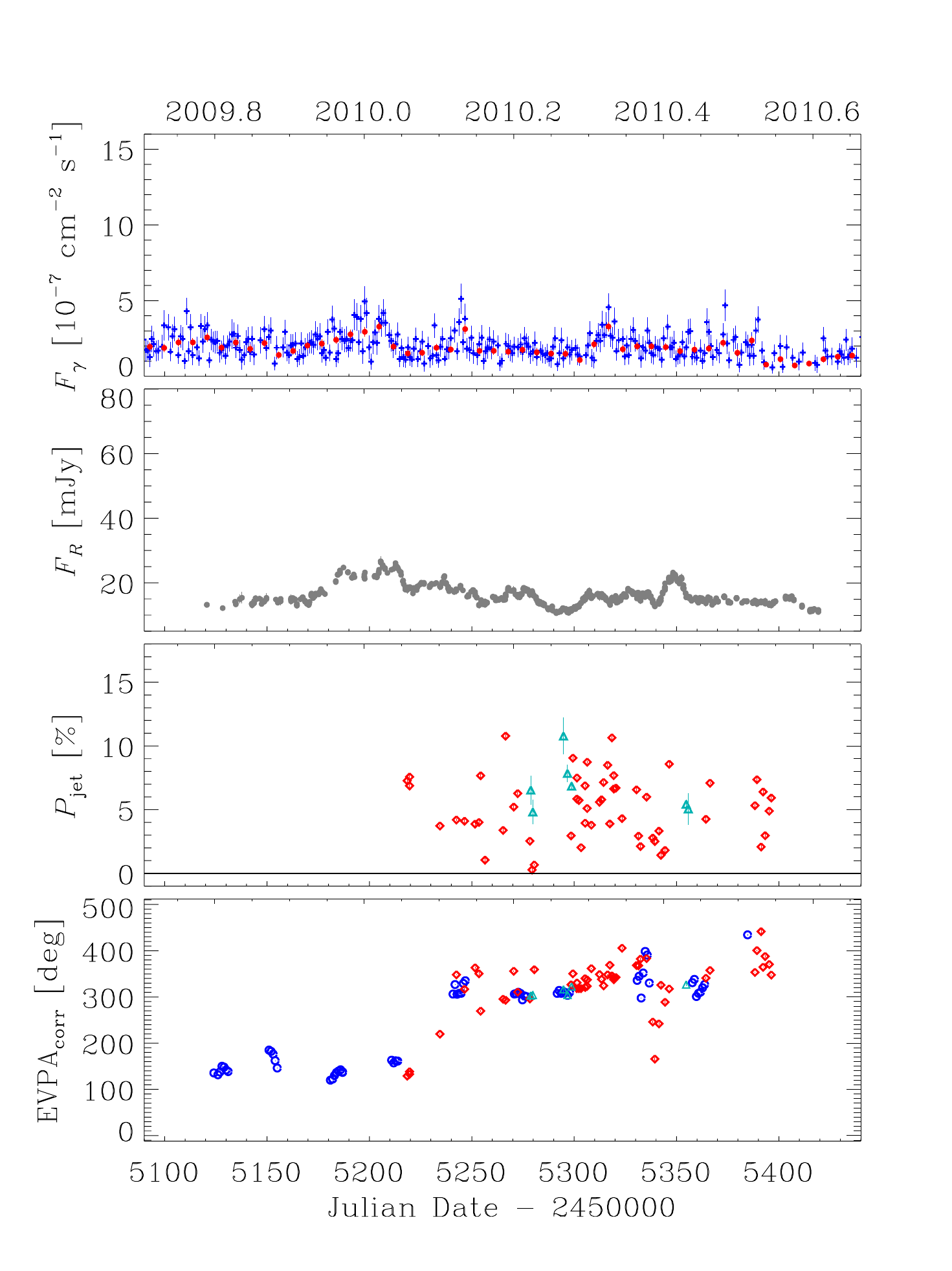,width=52.5mm}
    \epsfig{figure=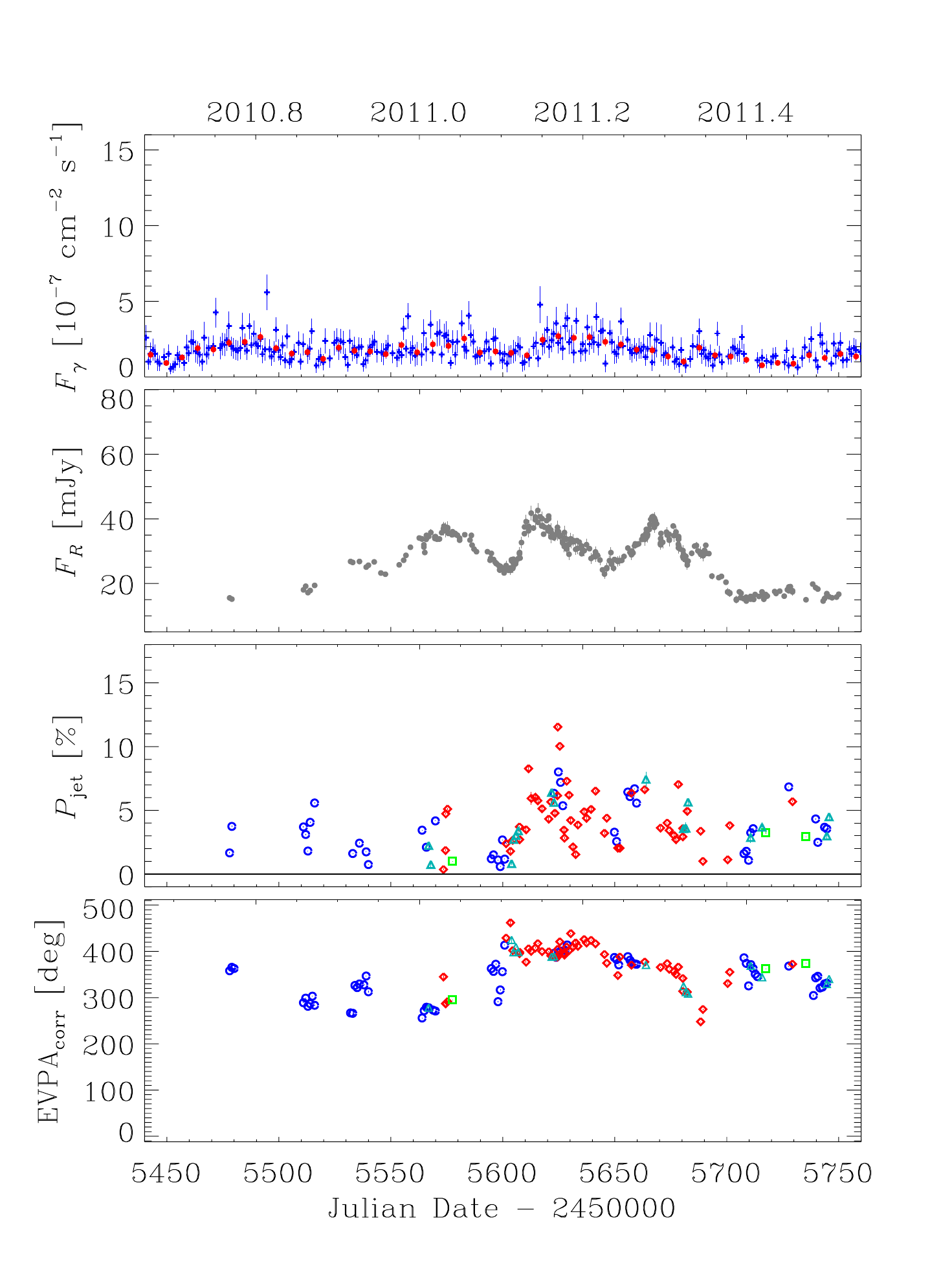,width=52.5mm}
    \epsfig{figure=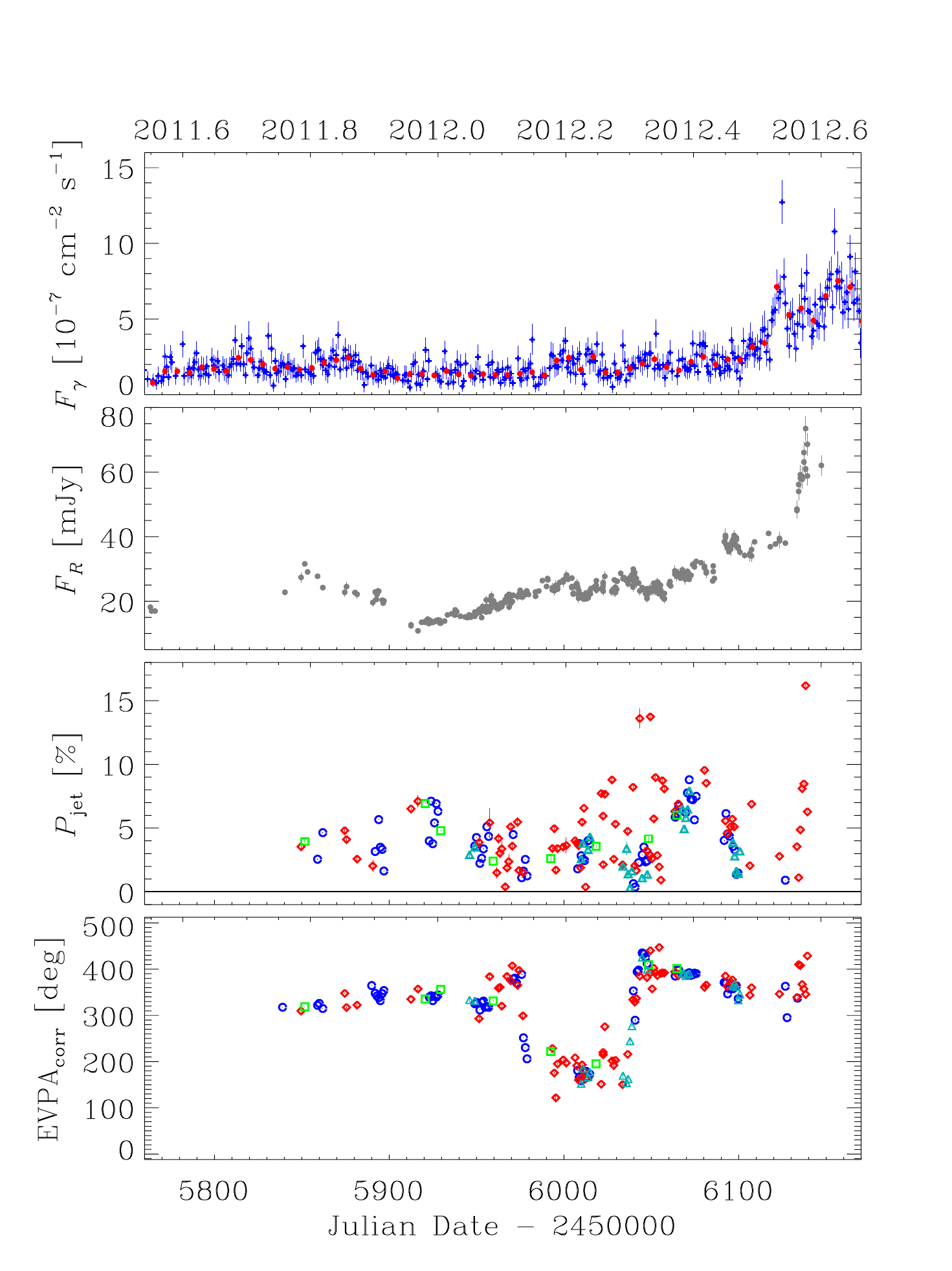,width=52.5mm}
    }
    %\vspace{0.0001cm}
 \centerline{
    \epsfig{figure=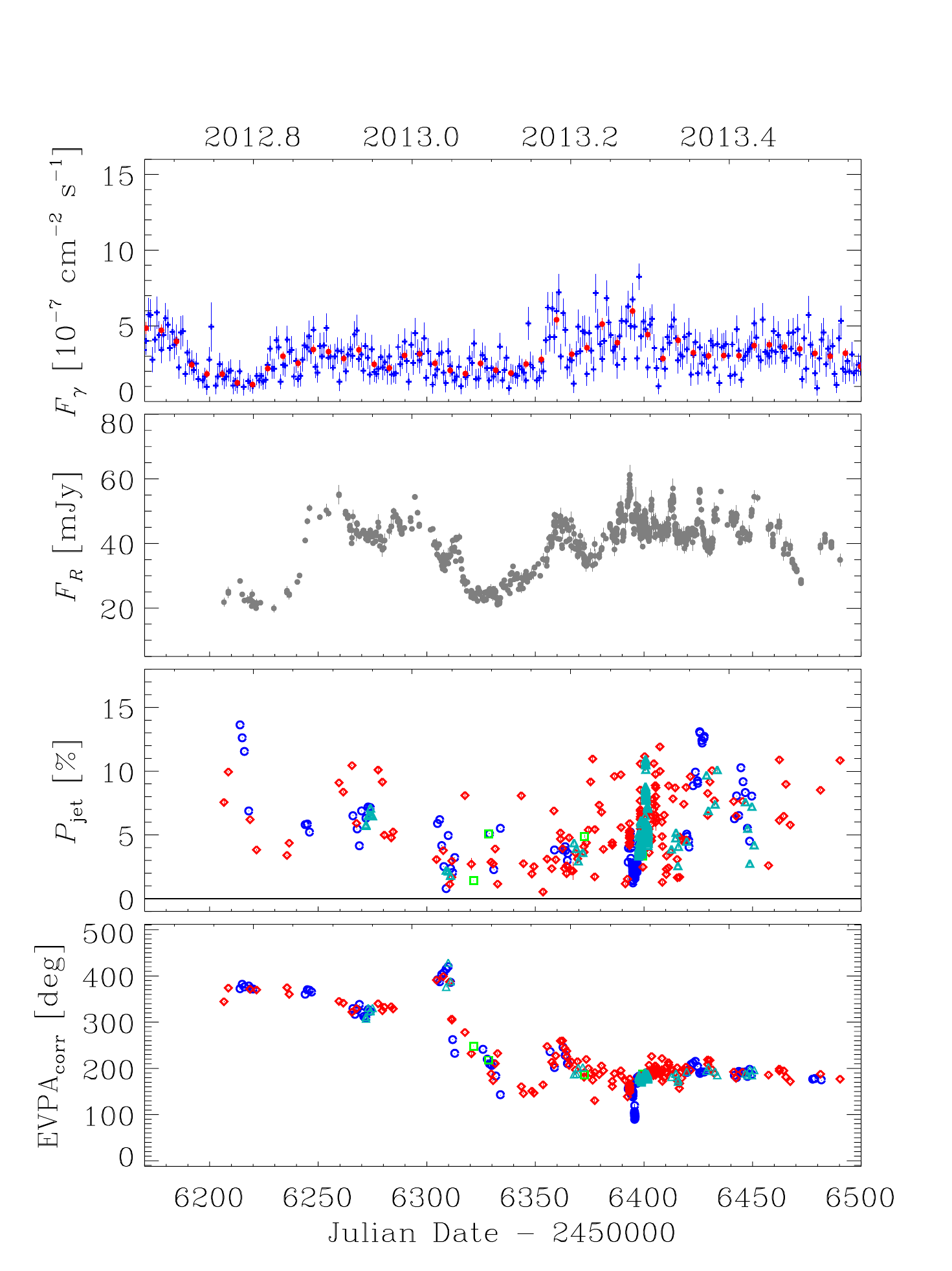,width=52.5mm}
    \epsfig{figure=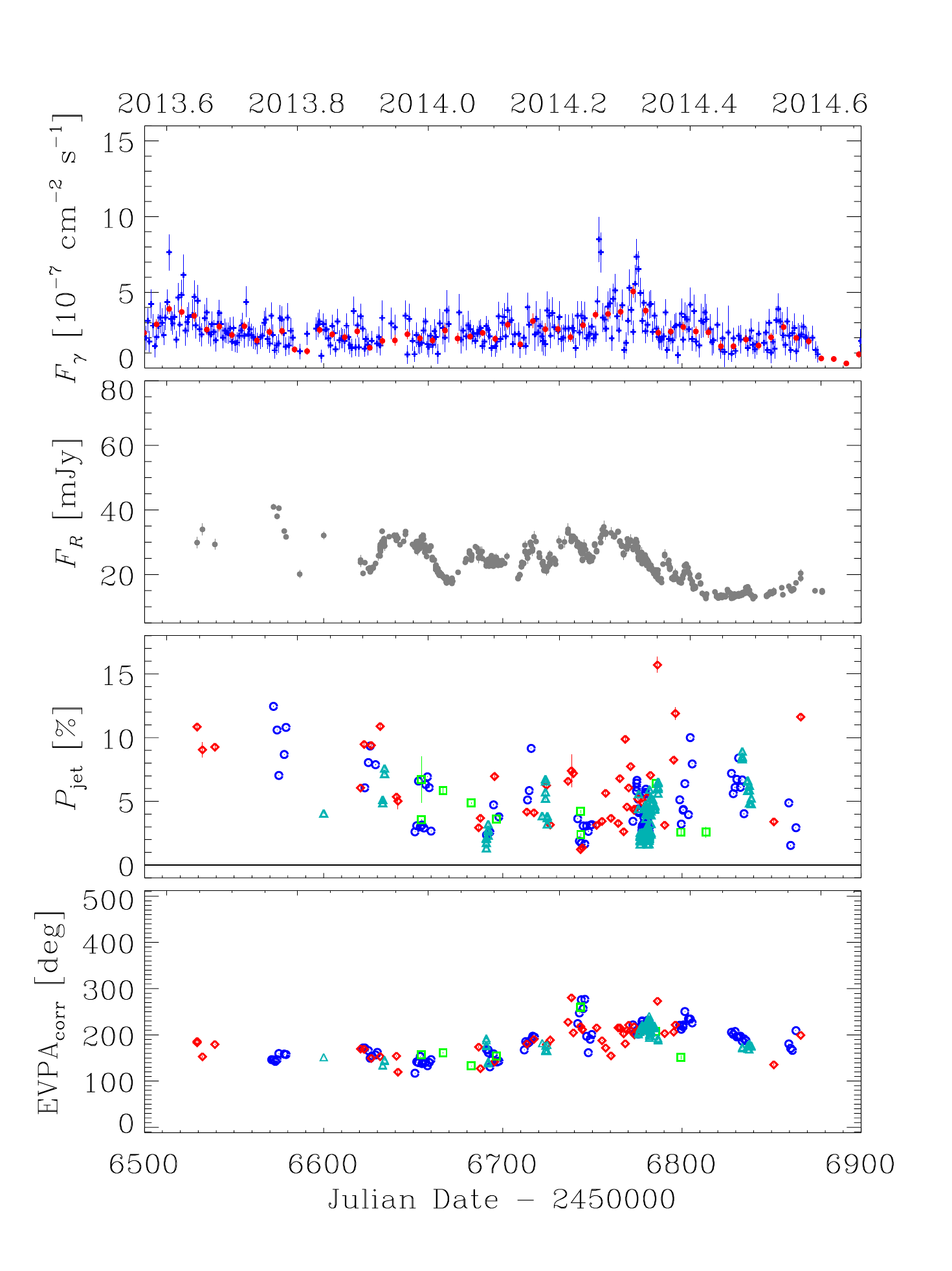,width=52.5mm}
    \epsfig{figure=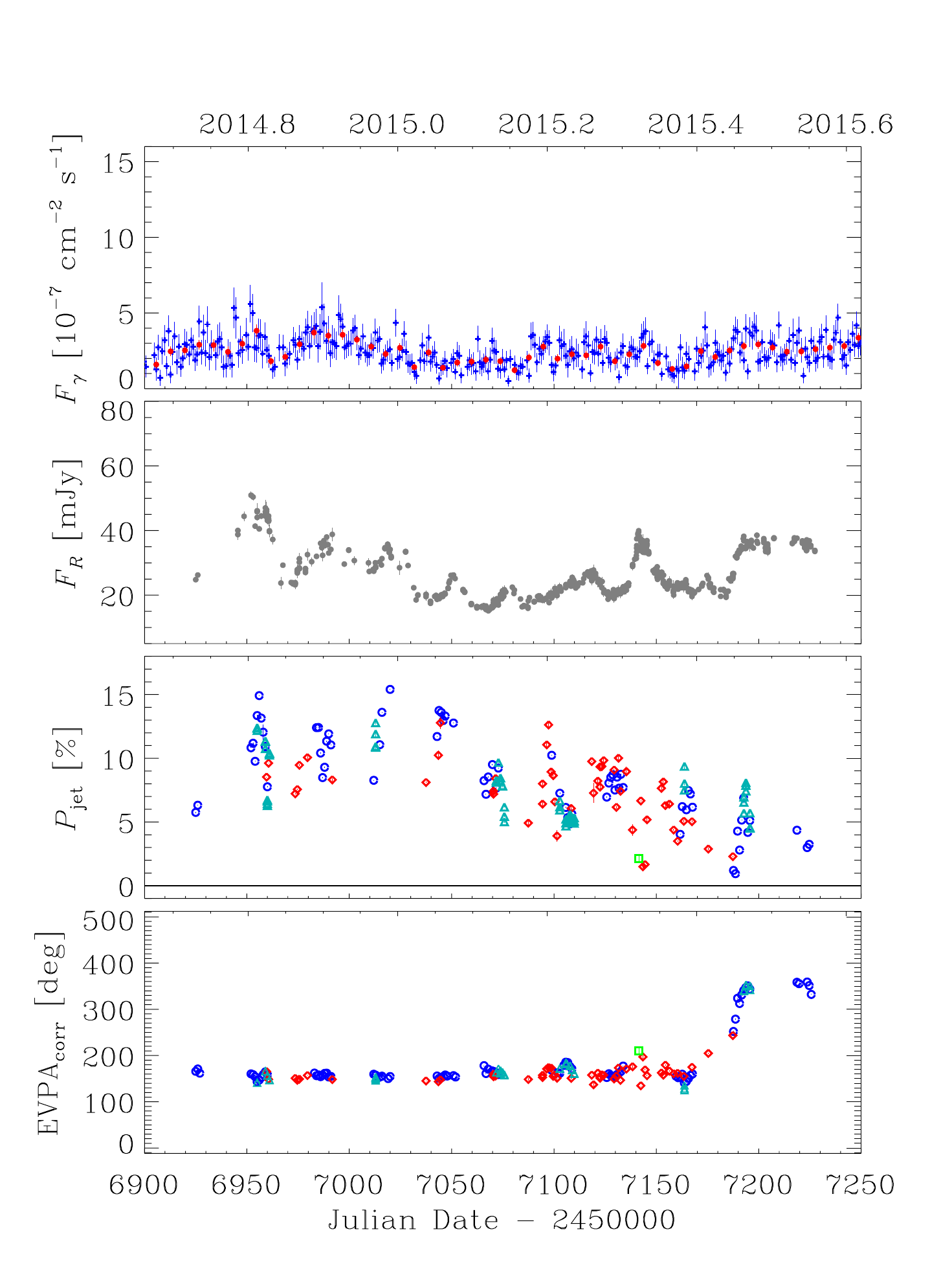,width=52.5mm}
    }
   \caption{Enlargements of Fig.\ \ref{fi:1101_pol} to appreciate short-term variability.} 
    \label{fi:1101_pol_zoom}
    \end{figure*}
In Fig.\ \ref{fi:1101_pol_zoom} we plot enlargements of Fig.\ \ref{fi:1101_pol} in different periods to better distinguish the variability properties.
We warn that in some cases wide rotations may derive from EVPA shifts performed when the angle difference between subsequent points was around $90\degr$. 
Though we have considered the angle uncertainties, it is clear that an underestimate of the error in these cases could lead to an apparent wider rotation.
This happens in the following dates: $\rm JD=2455234$, 2455340, and 2455974. 
In contrast, there are cases where the rotation appears quite robust. 
In particular, around $\rm JD=2456040$ we observe a counter-clockwise rotation of $\sim 250\degr$ in about 10 d. 
This happens when $P$ reaches a local maximum of $\sim 10$ per cent and the flux is rising toward the peak of 2012 July--August.
Another noticeable episode occurred in the last observing season. The EVPA remained stable for several months, and then rotated by $\sim 180\degr$ in a counter-clockwise direction in about one month around $\rm JD=2457180$. 
We note that $P$ experienced a local minimum at about half-way of the rotation and two symmetric maxima at the beginning and at the end of the rotation. This behaviour has already been observed in 3C 279 and was interpreted by \citet{nal10} in terms of an emitting blob encountering a major bending while travelling in the jet. If we apply that model to our case, assuming a Lorentz factor $\Gamma_{\rm jet}=\Gamma_{\rm blob}=10$ we find a minimum angle between the blob velocity vector and the line of sight of  3.4 \degr, a curvature radius of the trajectory of about $6.6 \times 10^{14} \, \rm cm$, and a distance covered by the blob between the minimum and maximum of $P$ of about $5.3 \times 10^{13} \, \rm cm$. As in the case of 3C 279, the lack of a simultaneous optical flare at the time of the minimum $P$ would mean that the blob gives only a very small contribution to the total observed flux.

\begin{figure*}
 \centerline{
    \epsfig{figure=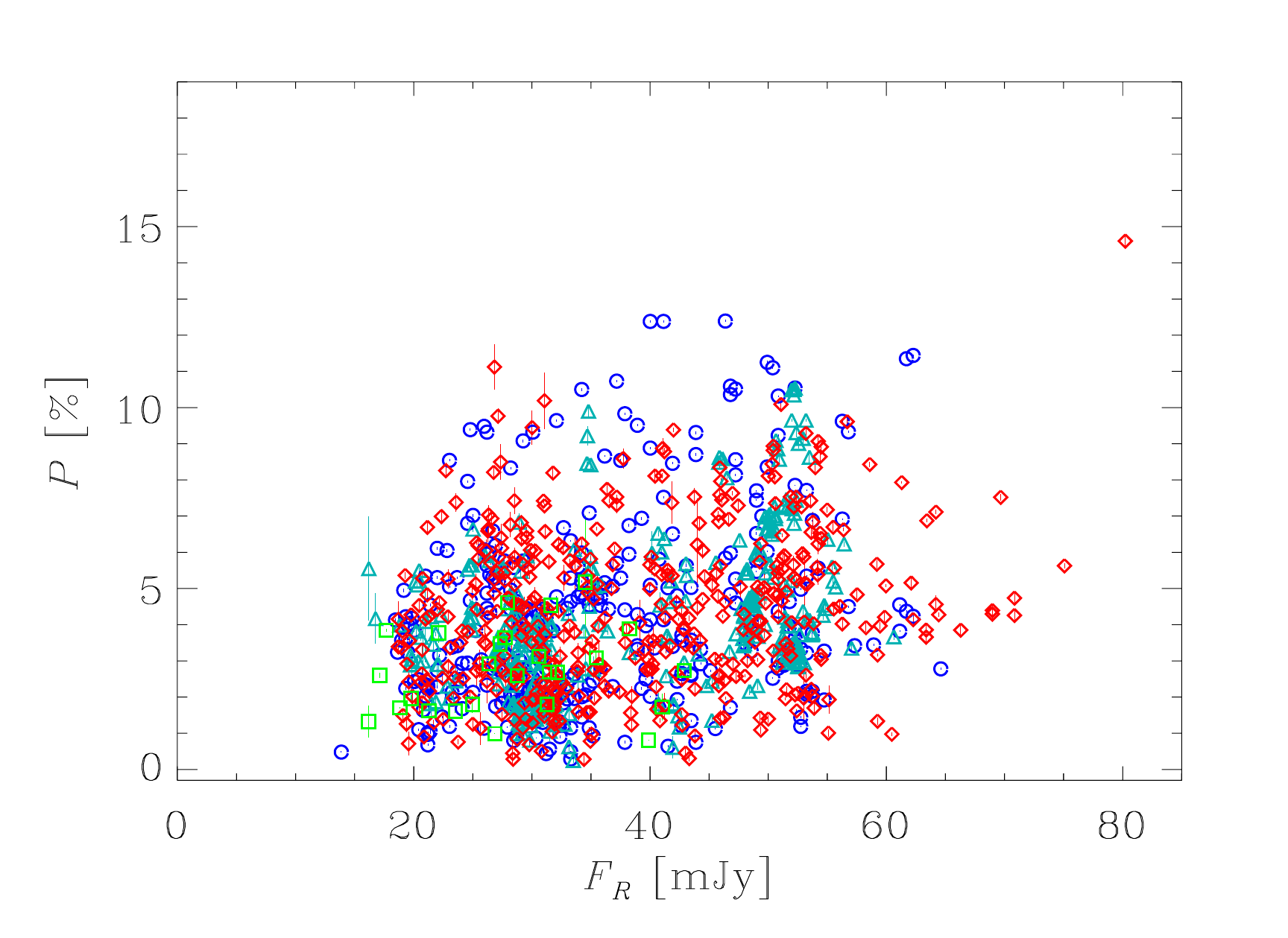,width=7cm\label{1101_opt_pol_obs}}
    \hspace{1cm}
    \epsfig{figure=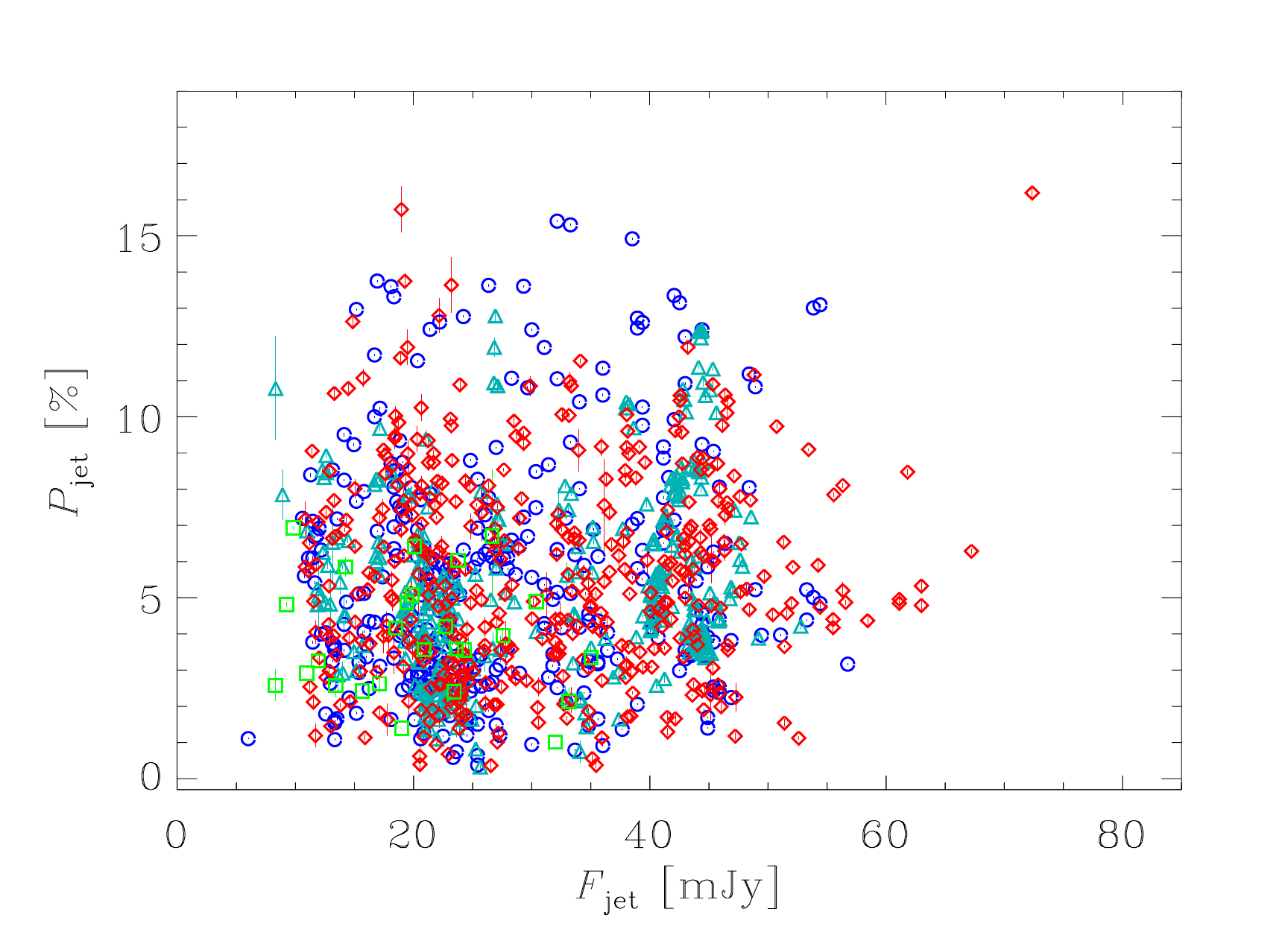,width=7cm\label{1101_opt_pol_corr}}
    }
\caption{Left: The observed degree of polarisation as a function of the observed $R$-band flux density. Right: The same plot after correcting for the host contribution.}\label{fi:1101_opt_pol}
\end{figure*}

In Fig.\ \ref{fi:1101_opt_pol} we show both $P_{\rm obs}$ versus $F_{\rm obs}$ and $P_{\rm jet}$ versus $F_{\rm jet}$. It can be noticed that in both cases no apparent correlation exists between the two quantities.

\begin{figure}
\centering
\includegraphics[width=8cm]{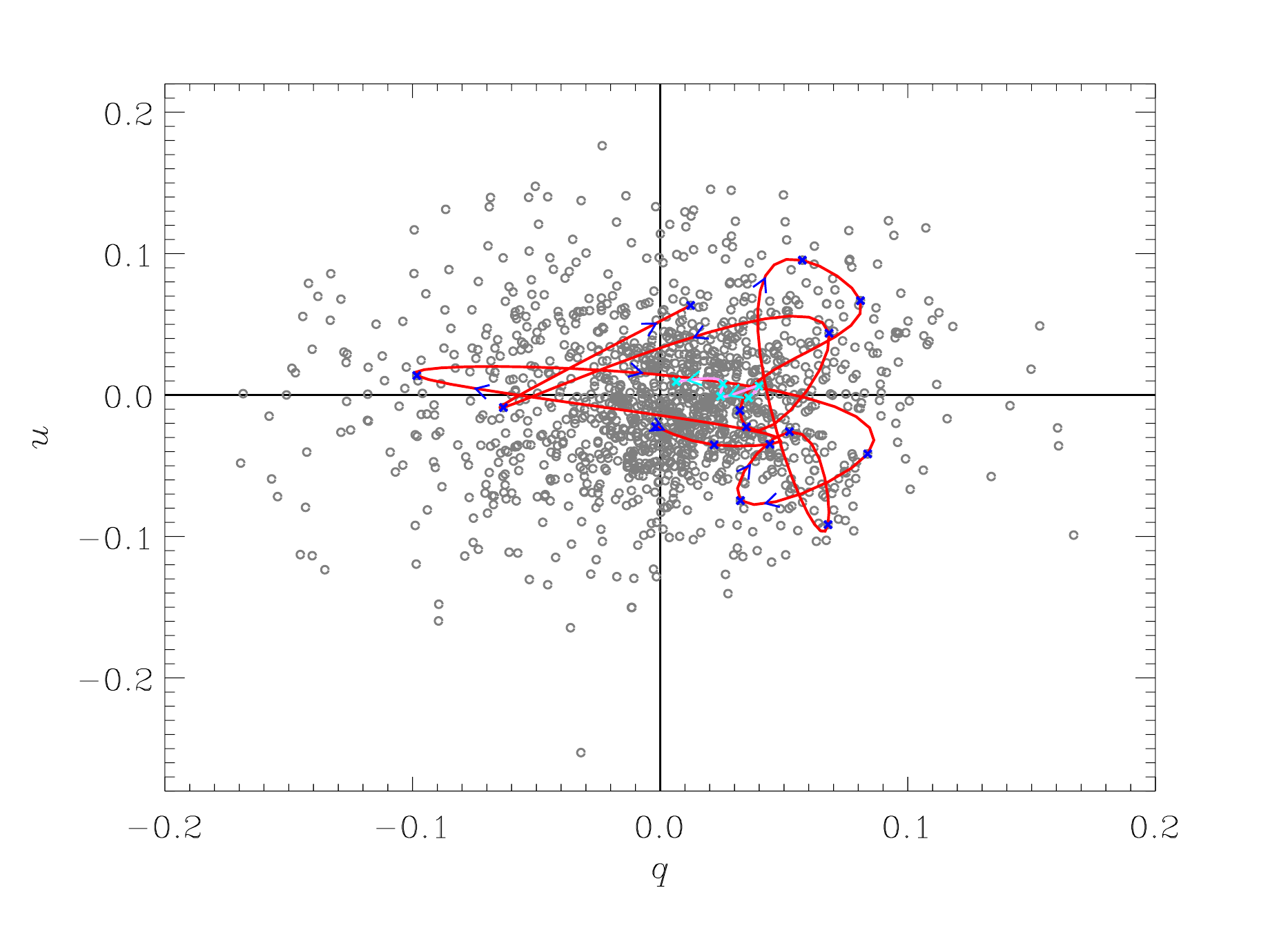}
  \caption{Distribution of the Stokes' parameters in the $u$ versus $q$ plot. The red line highlights the EVPA rotation that occurred in 2015 May--June and was obtained by connecting subsequent points with a cubic spline interpolation. The arrows indicate the time evolution.}
\label{fi:1101_q_u}
\end{figure}

Figure \ref{fi:1101_q_u} shows the distribution of the Stokes' parameters of Mrk 421 in the $u$ versus $q$ plot, highlighting the EVPA rotation that occurred in 2015 May--June. To this aim, subsequent points have been connected with a cubic spline interpolation. The distance of the spline from the origin, together with the wideness of the spline and the persistence of the direction of rotation, confirm the genuine nature of the rotation.

\section{Spectropolarimetric observations}

\begin{figure*}
\centerline{
    \epsfig{figure=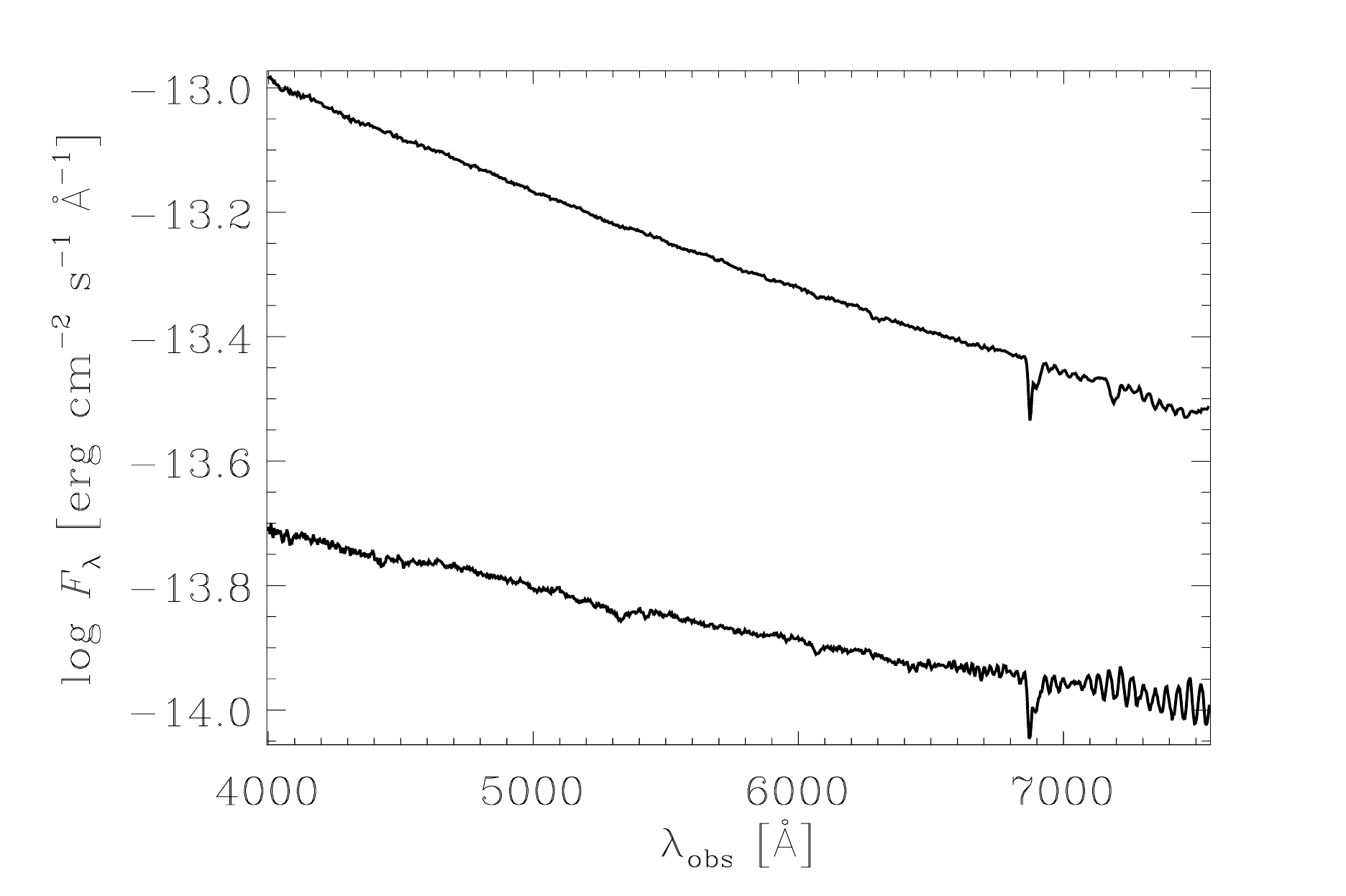,width=7cm\label{1101_maxmin_espectro}}
    \hspace{1cm}
    \epsfig{figure=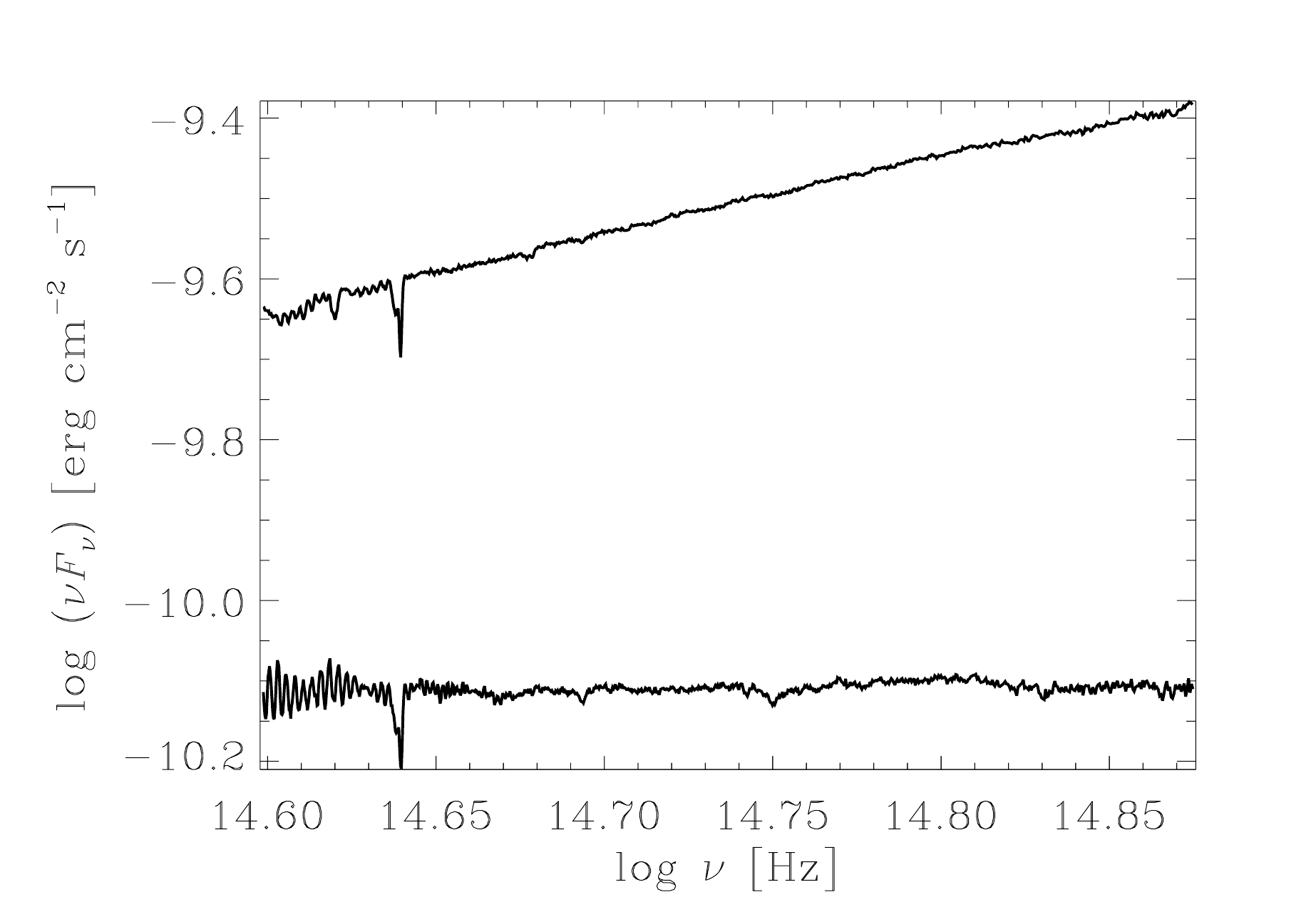,width=7cm\label{1101_sed_espectro}}
    }
\caption{Left panel: the brightest and faintest spectra in the Steward Observatory data base. 
Their flux density ratio is $\sim 5$. Right panel: the same spectra in the $\log(\nu F_\nu)$ versus $\log \nu$ representation used for the SED.}
\label{fi:espectro_maxmin}
\end{figure*}

We analysed 603 spectra from the Steward Observatory database to investigate the optical spectropolarimetric variability properties of Mrk~421.
Fig.\ \ref{fi:espectro_maxmin} shows the source spectra corresponding to the brightest and faintest states. They lack emission lines and point out a bluer-when-brighter behaviour. In the fainter spectrum, we can recognize the \ion{Mg}{I} and \ion{Na}{I} absorption lines from the host galaxy.

\begin{figure}
\centering
\includegraphics[width=6cm]{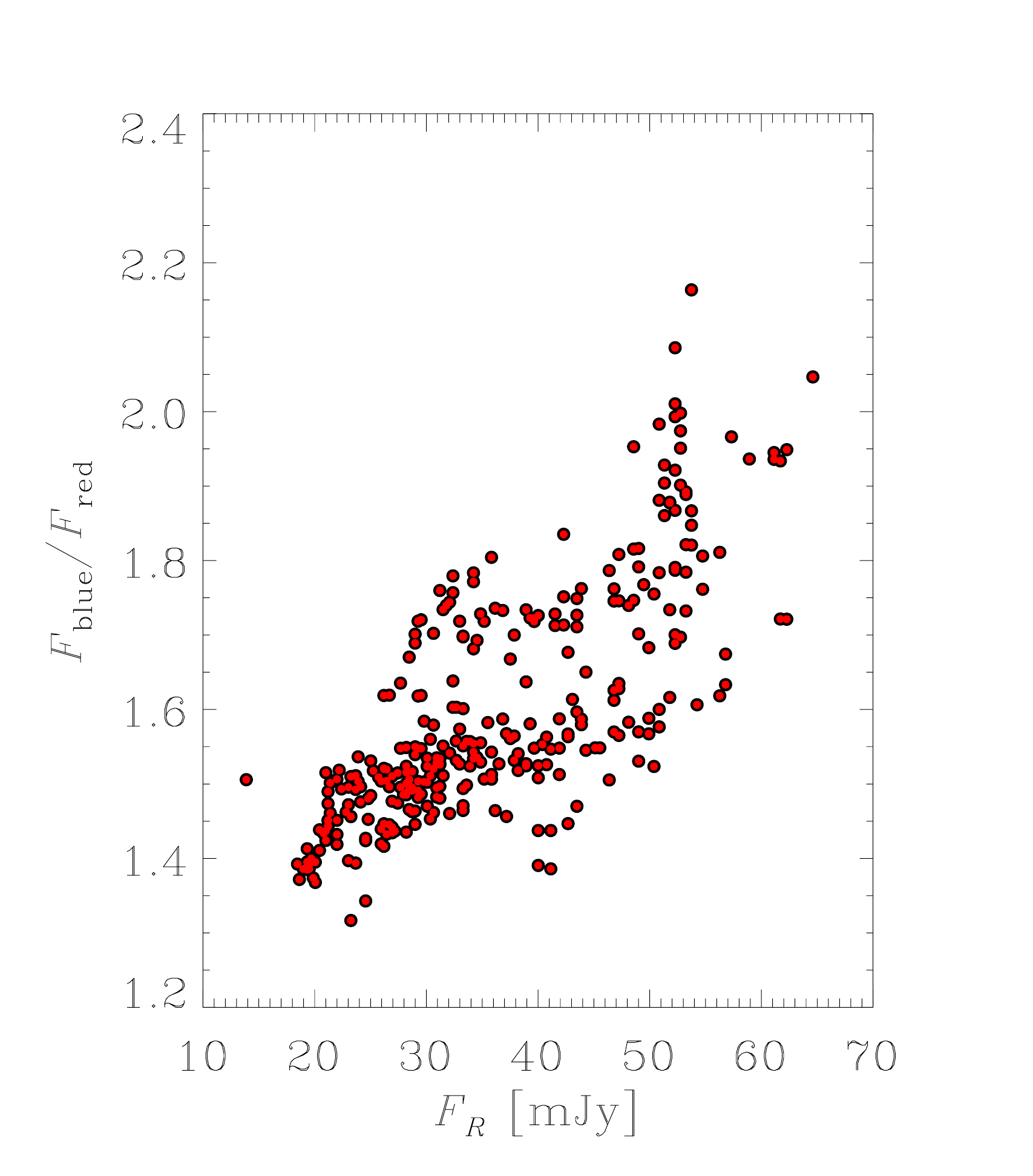}
  \caption{The Mrk~421 optical colour (``blue" to ``red" flux ratio) as a function of the source brightness (observed $R$-band flux density in mJy).} 
\label{fi:1101_color_spettri}
\end{figure}

Figure \ref{fi:1101_color_spettri} shows the optical colour of Mrk~421 as a function of the $R$-band flux density. The colour is determined as the ratio between the median flux in the range 4000--5000 \AA\ (``blue") and that in the 5800--6800 \AA\  range (``red")\footnote{We could not use the reddest part of the spectra (from 6800 to 7550 \AA) because it includes
terrestrial oxygen and water absorption features.}. Here again we note a bluer-when brighter trend.

\begin{figure}
\centering
\includegraphics[width=8cm]{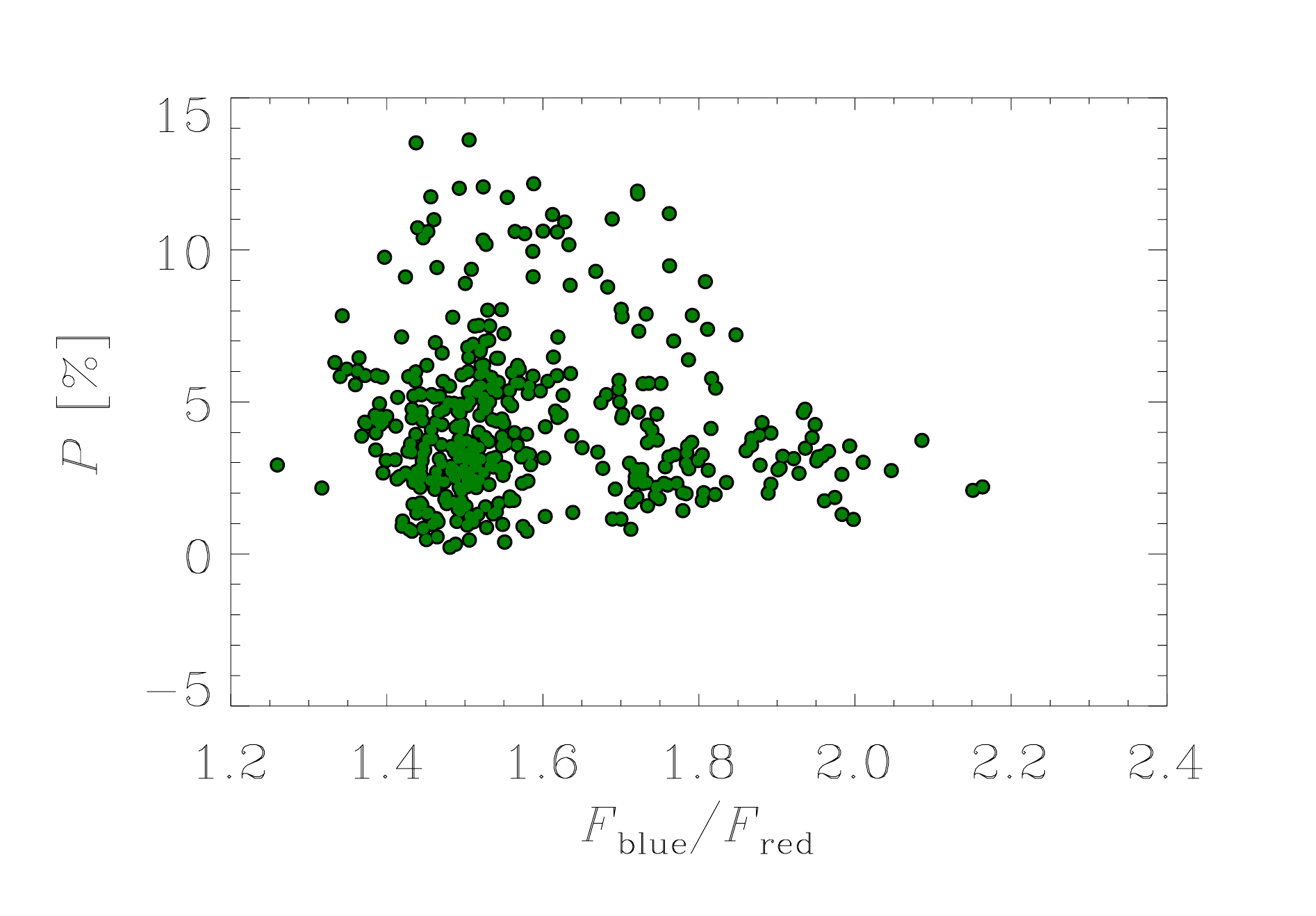}
  \caption{The observed degree of polarisation plotted against the optical colour of Mrk 421. The polarisation and flux ratio measurements are simultaneous.}
\label{fi:1101_pol_color}
\end{figure}

We then investigated the wavelength-dependence of the optical polarisation.
We did not find correlation between the degree of optical linear polarisation $P$ and the ratio between the blue and the red fluxes (see Fig.\ \ref{fi:1101_pol_color}).

\begin{figure*}
\centering
\includegraphics[width=8cm]{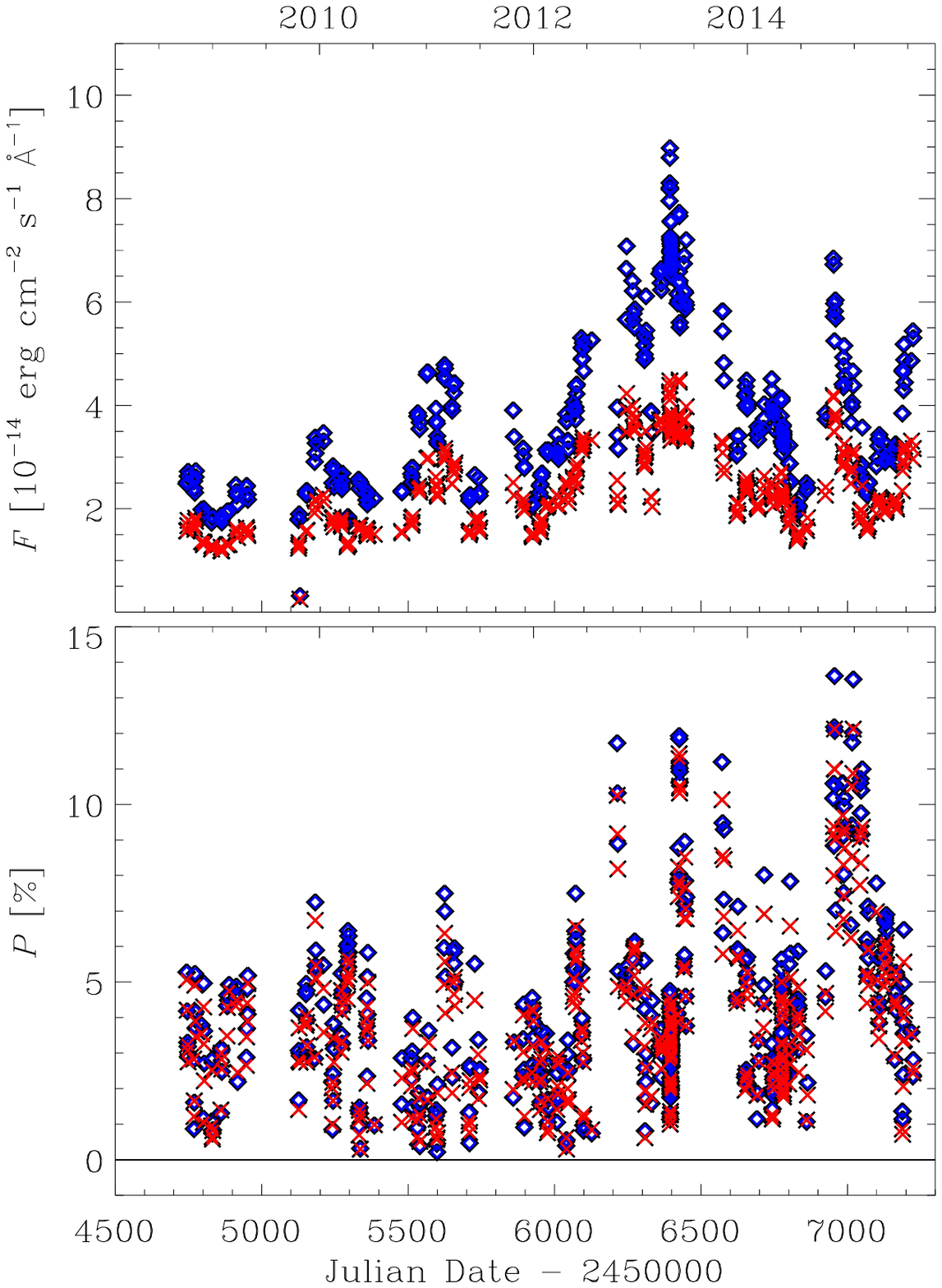}
  \caption{The observed optical flux density (top) and polarisation percentage (bottom) as a function of time in the blue and red bins defined in the text.}
\label{fi:1101_pol_rosso_blue_prueba}
\end{figure*}

Fig.\ \ref{fi:1101_pol_rosso_blue_prueba} shows flux densities and polarisation percentages for the two wavelength ranges. The blue side is characterized by higher flux variability and polarisation degree than the red one, consistently with the larger contribution of the host galaxy light in the red range.

\begin{figure*}
\centerline{
    \epsfig{figure=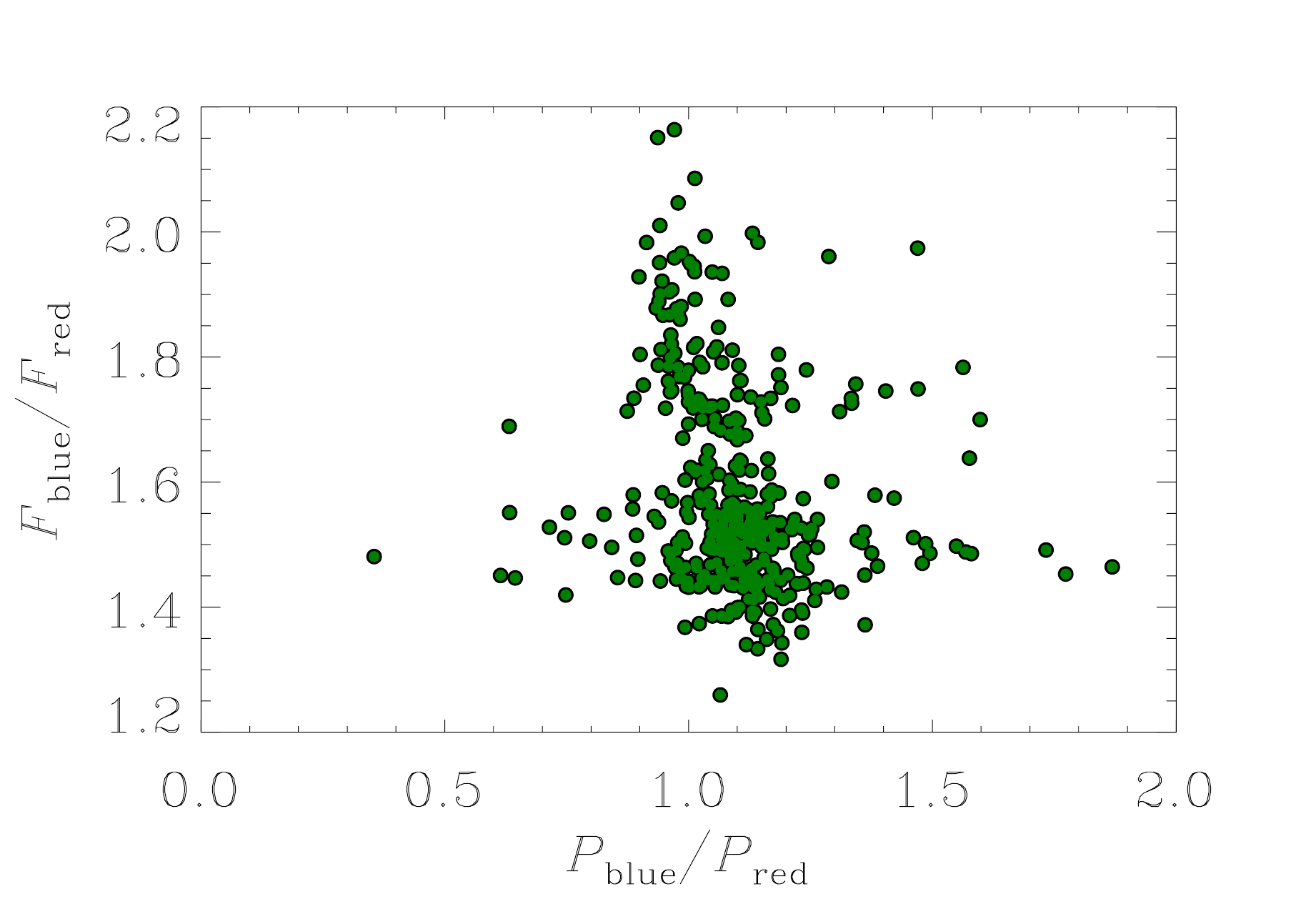,width=7cm\label{1101_flusso_pol}}
    \hspace{1cm}
    \epsfig{figure=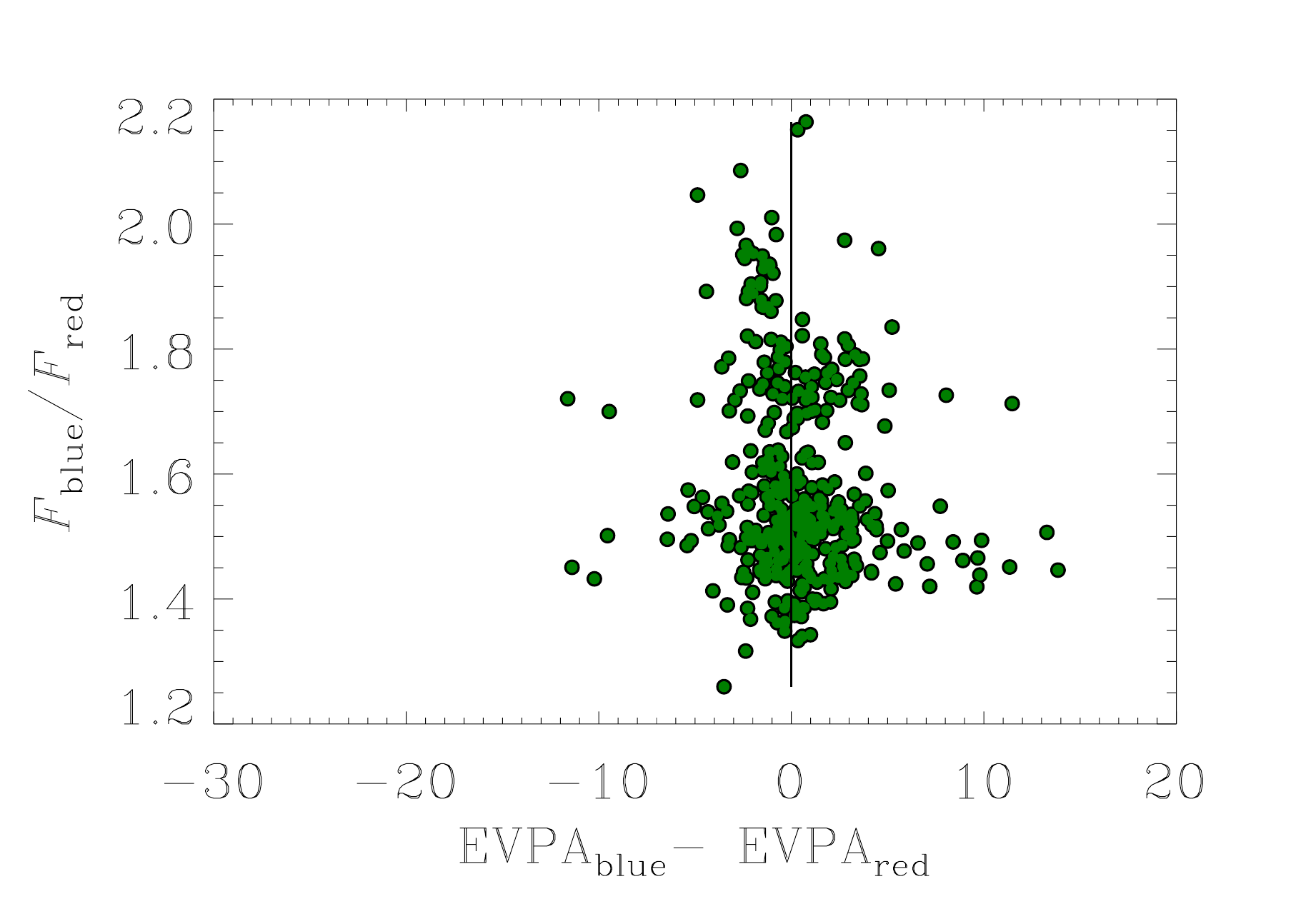,width=7cm\label{1101_flusso_teta}}
    }
\caption{Left: the relationship between the ratio of observed polarisation in the blue and red bins and the colour defined as in Fig.\ \ref{fi:1101_color_spettri}. Right: colour versus the difference between the polarisation position angles in the blue and red bins.}\label{fi:flusso_pol_teta}
\end{figure*}

Fig.\ \ref{fi:flusso_pol_teta} shows the flux ratio between the blue and the red bins plotted against the ratio of the observed polarisation in the same bins. As the blazar becomes fainter (redder), the polarisation in the blue tends to be higher than in the red because of the increasing contribution from the red and unpolarized host galaxy light. However, in the faint states the low levels of polarisation result in relatively large uncertainties in $P_{\rm blue}$/$P_{\rm red}$. Fig.\ \ref{fi:flusso_pol_teta} also shows the difference between the polarisation position angles determined in the two continuum bins. The plot suggests that there is no wavelength dependence in EVPA. 

\section{Discrete Correlation Function}

We apply the discrete correlation function (DCF) analysis to the data shown in Fig.\ \ref{fi:1101_multi} in order to investigate the existence of characteristic time scales of variability and the correlation between the $\gamma$-ray, X-ray and optical fluxes. This method is suitable to treat unevenly sampled datasets \citep{ede88, huf92}.
Correlation/anticorrelation produces a positive/negative peak of the DCF. The correlation is strong if the peak value approaches or even exceeds one.  

\begin{figure*}
\centerline{
    \epsfig{figure=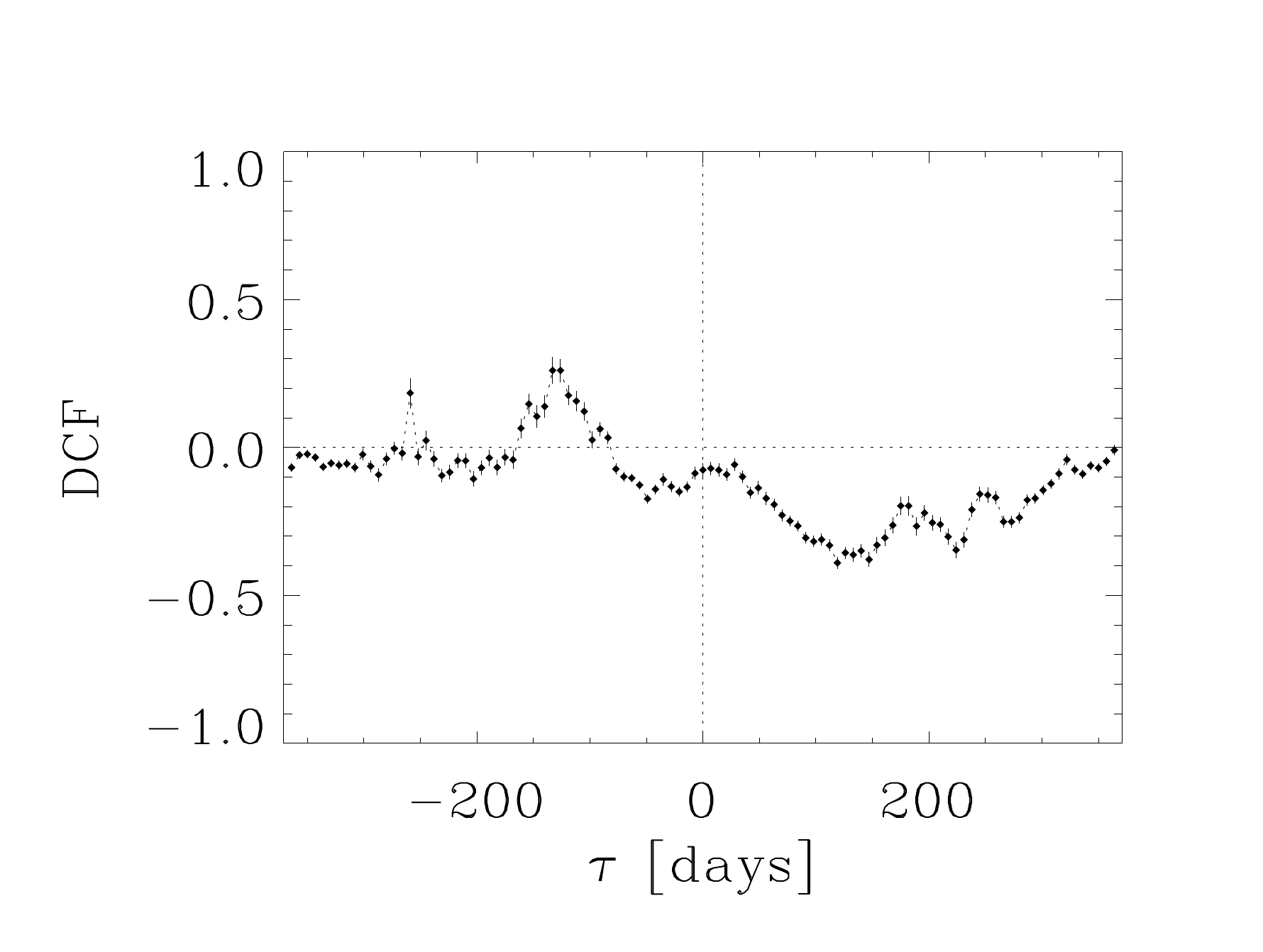,width=5.2cm}
    \hspace{0.1cm}
    \epsfig{figure=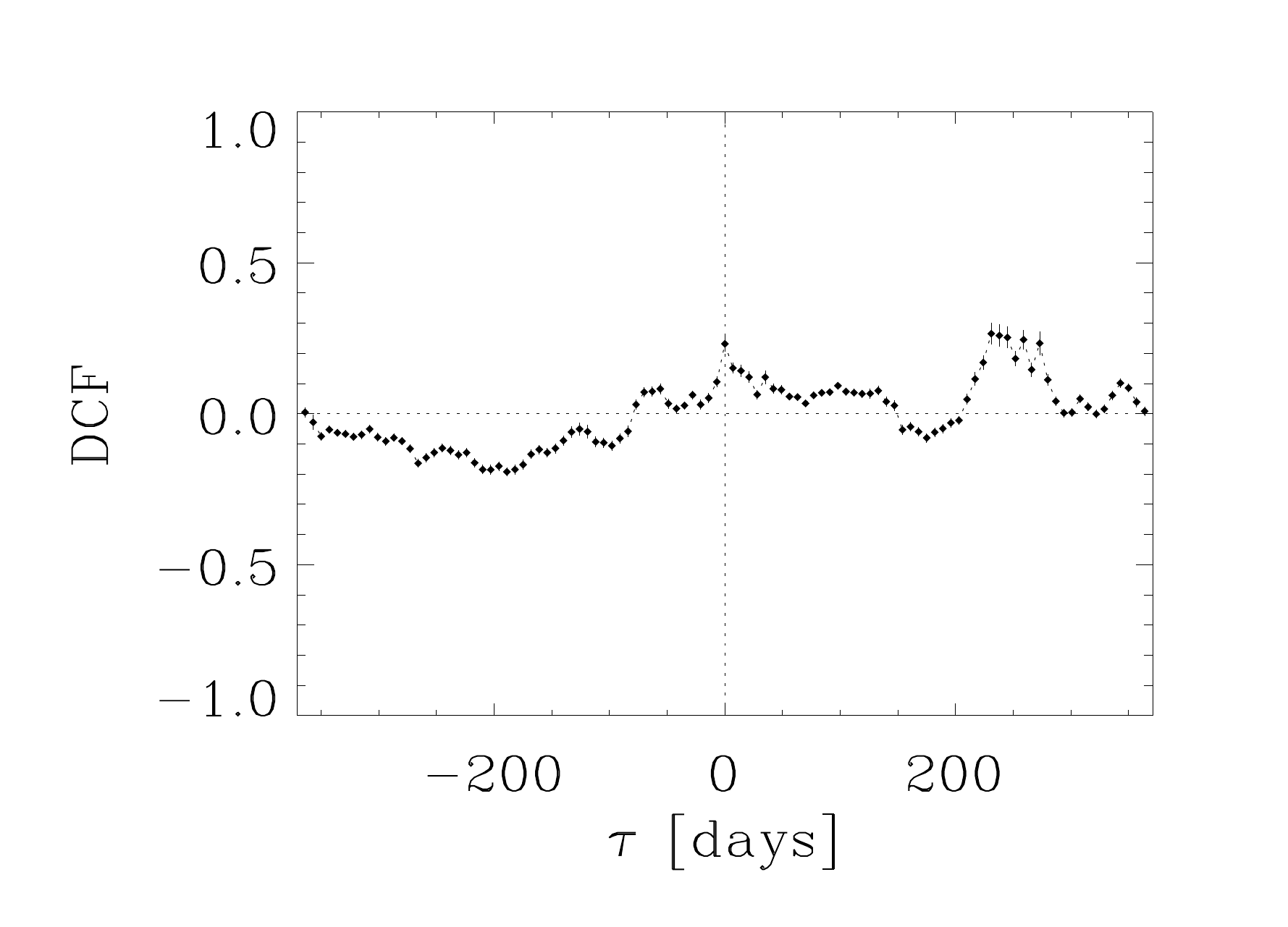,width=5.2cm}
    \hspace{0.1cm}
    \epsfig{figure=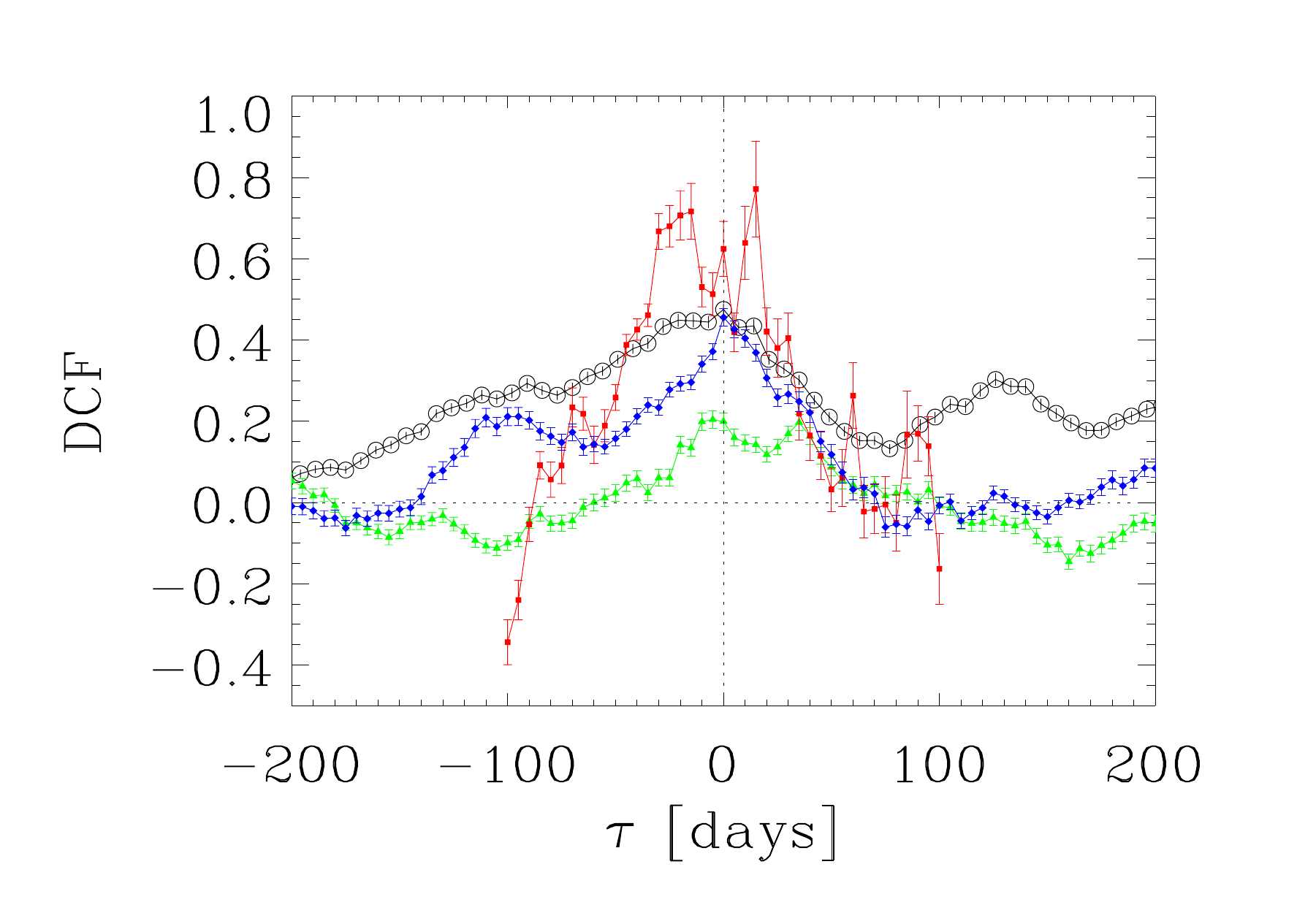,width=5.2cm}
    }
  \caption{Discrete correlation function between the X-ray and $R$-band light curves (left); the daily-binned $\gamma$-ray and X-ray light curves (middle); the daily-binned $\gamma$-ray and $R$-band light curves (right, empty black circles) over the whole 2007--2015 period. In the last panel we also show the results of the DCF run on the three subperiods corresponding to the time intervals before the 2012 outburst (green triangles), after the outburst (blue diamonds) and around the outburst (red squares).}
\label{fi:1101_dcf}
\end{figure*}

The DCF between the X-ray and $R$-band light curves over the whole 2007--2015 period is shown in Fig.\ \ref{fi:1101_dcf}. The lack of a strong signal suggests that the X-ray and optical variations are in general not correlated. The two low peaks at about $-130$ and $-260$ d indicate optical events preceding X-ray ones and likely refer to the major flares in 2012--2015.

In the same figure we show the DCF between the daily-binned $\gamma$-ray and the X-ray light curves. The central low peak is likely produced by the match of the major X-ray flare of 2013 with a contemporaneous $\gamma$-ray flare. Another low peak at a time lag of $\sim 250 \rm \, d$ comes from the correspondence between the major 2012 optical and 2013 X-ray flares. However, the value of the DCF always maintains low, implying that the correlation is weak. 

Finally, Fig.\ \ref{fi:1101_dcf} displays the DCF between the daily-binned $\gamma$-ray and the optical fluxes over the whole data trains. The value of the DCF at the central maximum is 0.48 and indicates fair correlation between the flux variations in the $\gamma$-ray and optical bands. The central maximum is broad, possibly because of the superposition of different signatures. To check this, we divided the considered period in three subperiods: the time interval before the big 2012 $\gamma$-optical outburst ($\rm JD<2456000$), the time interval after the outburst ($\rm JD>2456200$), and the time interval including the outburst ($\rm 2456000<JD<2456200$). We then calculated the $\gamma$-optical cross-correlation for the three periods separately. 
Before the big outburst the DCF value is always very low, indicating poor correlation; after the outburst the strength of the correlation increases and the central peak ($\rm DCF=0.46$) suggests fair correlation with essentially no time delay. Around the outburst the correlation is strong, as expected, but the timing is badly defined, with $\gamma$ variations that can either precede or follow the optical ones. This ambiguity is essentially due to the lack of optical data during the 2012 solar conjunction, when the $\gamma$-ray outburst was still at its highest levels.

\begin{figure*}
 \centerline{
    \epsfig{figure=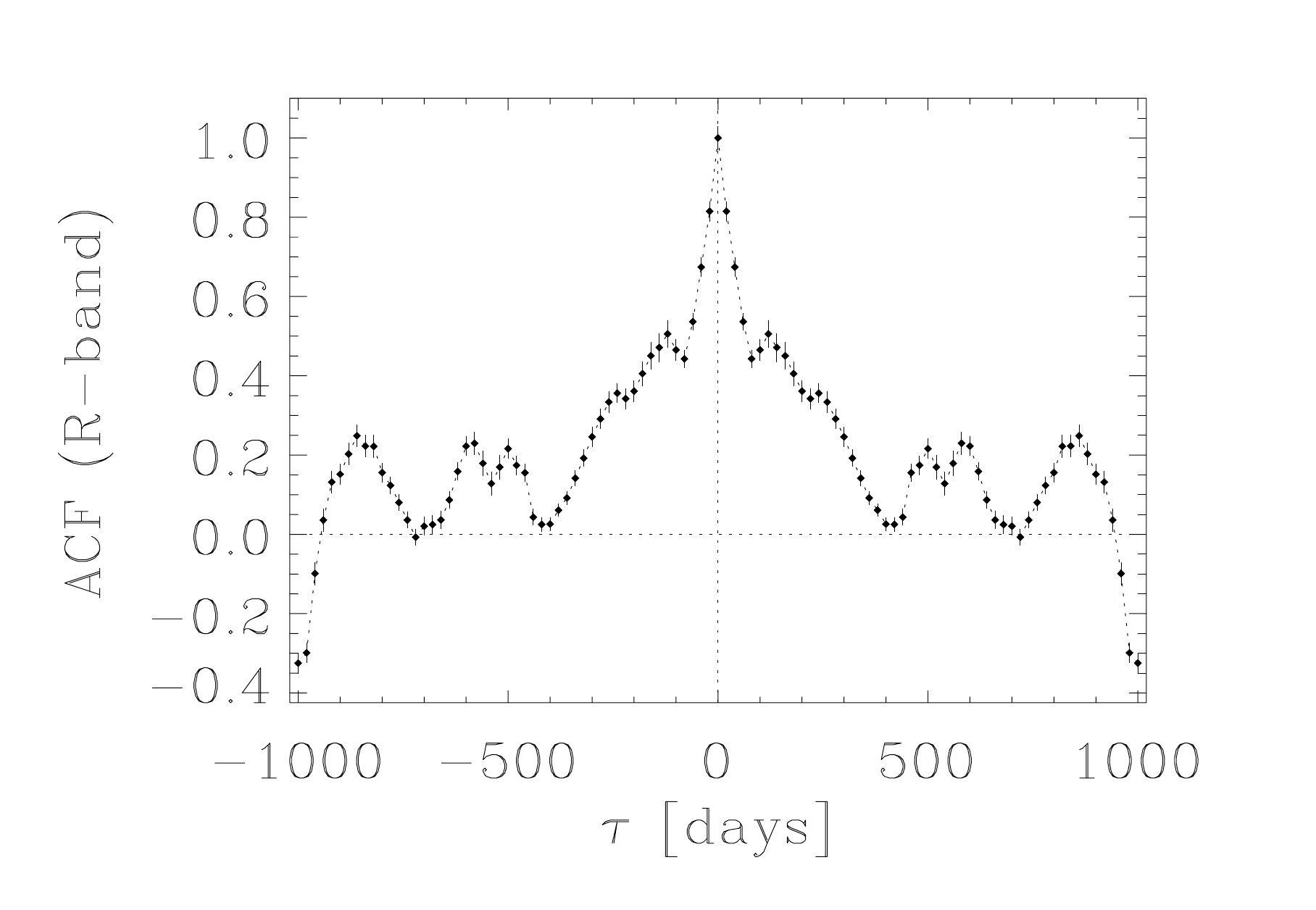,width=5.2cm\label{1101_acf_op}}
    \hspace{0.1cm}
    \epsfig{figure=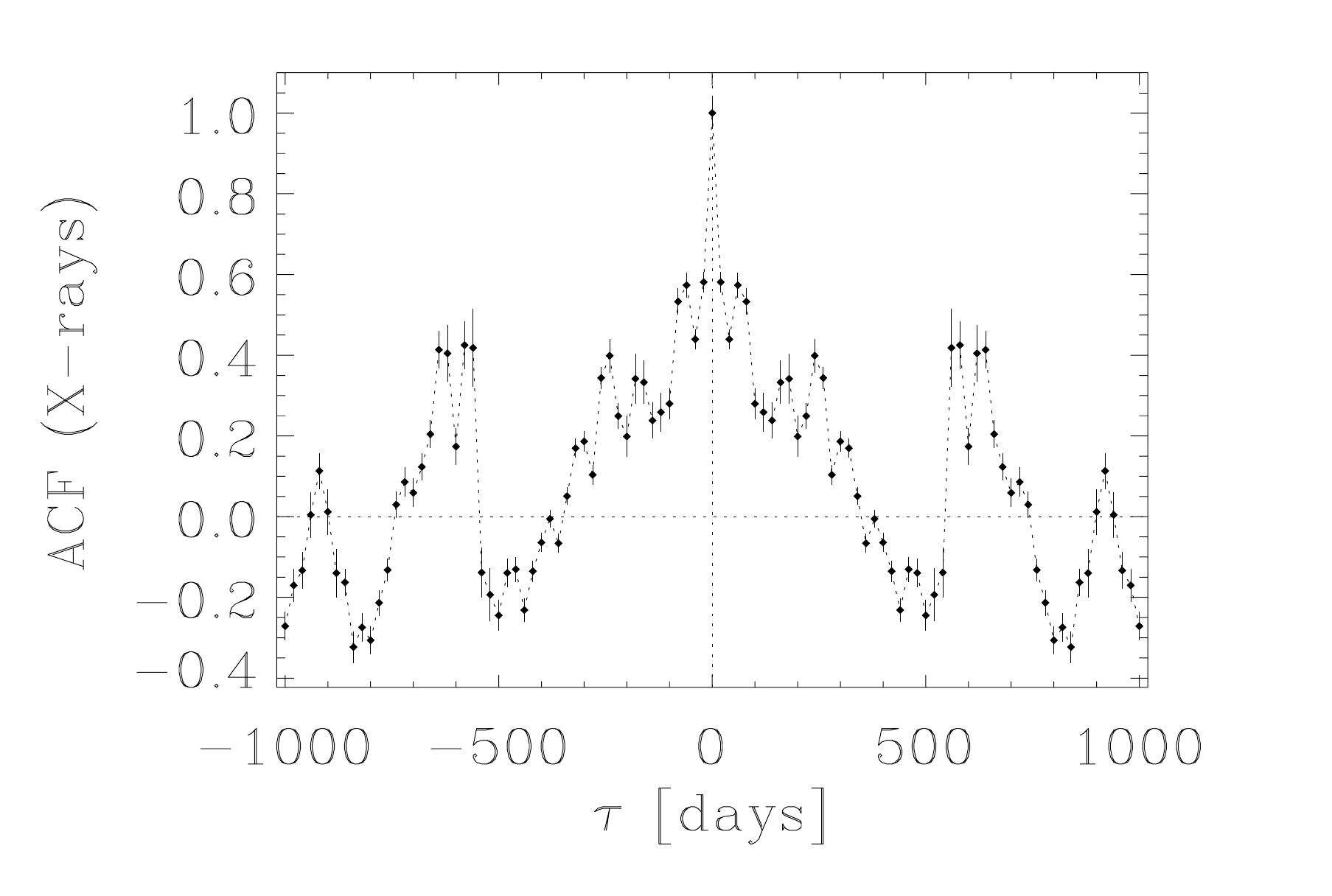,width=5.2cm\label{1101_acf_xray}}
    \hspace{0.1cm}
    \epsfig{figure=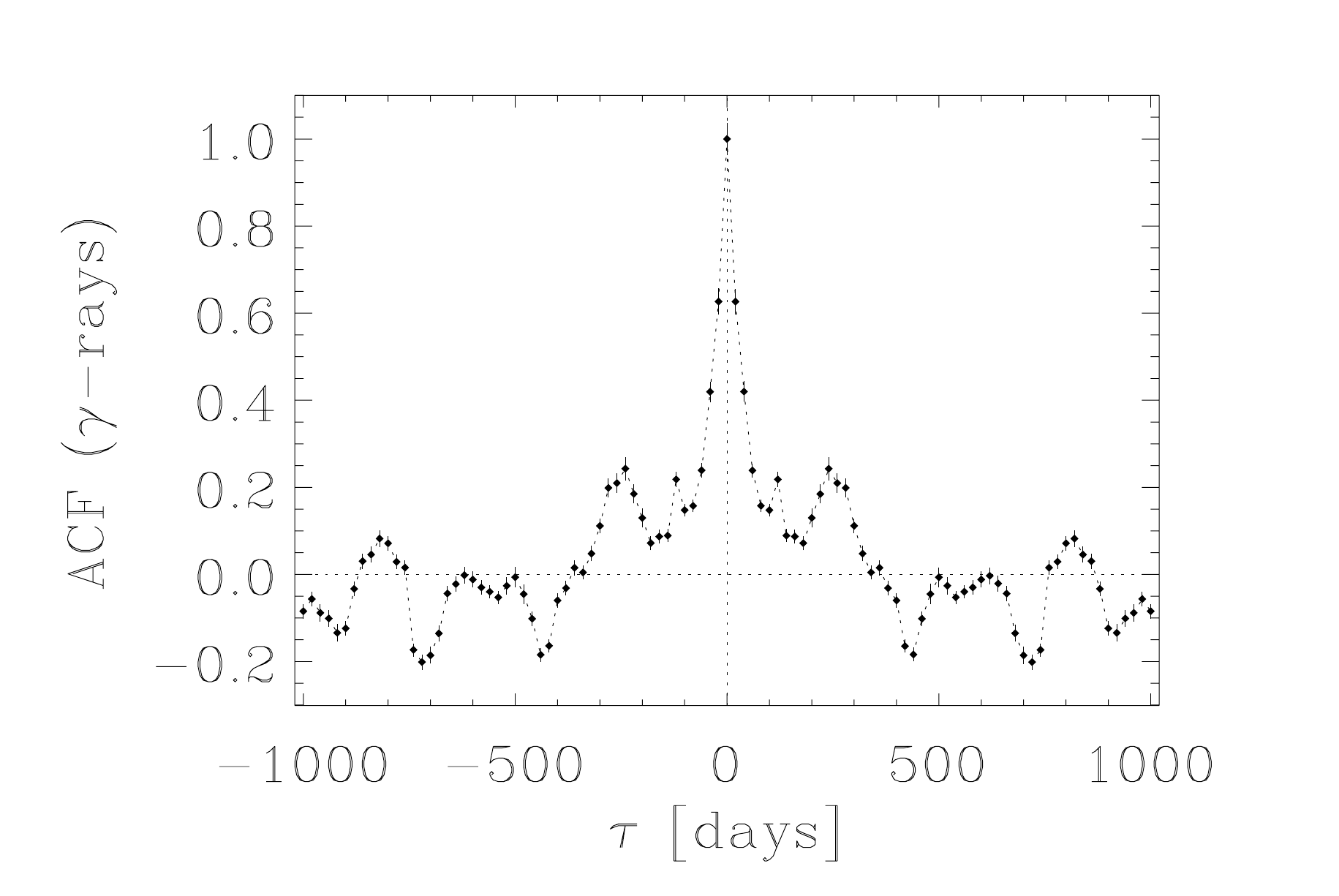,width=5.2cm\label{acf_gammad}}
    }
\caption{Auto-correlation function for the galaxy-subtracted optical flux densities (left), X-ray count rates (middle), and $\gamma$-ray fluxes (right).}\label{fi:acf_all}
\end{figure*}

The optical, X-ray, and $\gamma$-ray auto correlation functions (ACFs) are plotted in Fig.\ \ref{fi:acf_all}. 
The peaks are low and their lags reflect the time separation between flares in the corresponding light curves. None of these time scales can be considered as a periodicity. A recent detailed periodicity analysis on the $\gamma$ and optical light curves of Mrk~421 by \citet{san17} found no periodic signals.

\section{Broad-band spectral variability}

\begin{figure*}
\centering
\includegraphics[width=13cm]{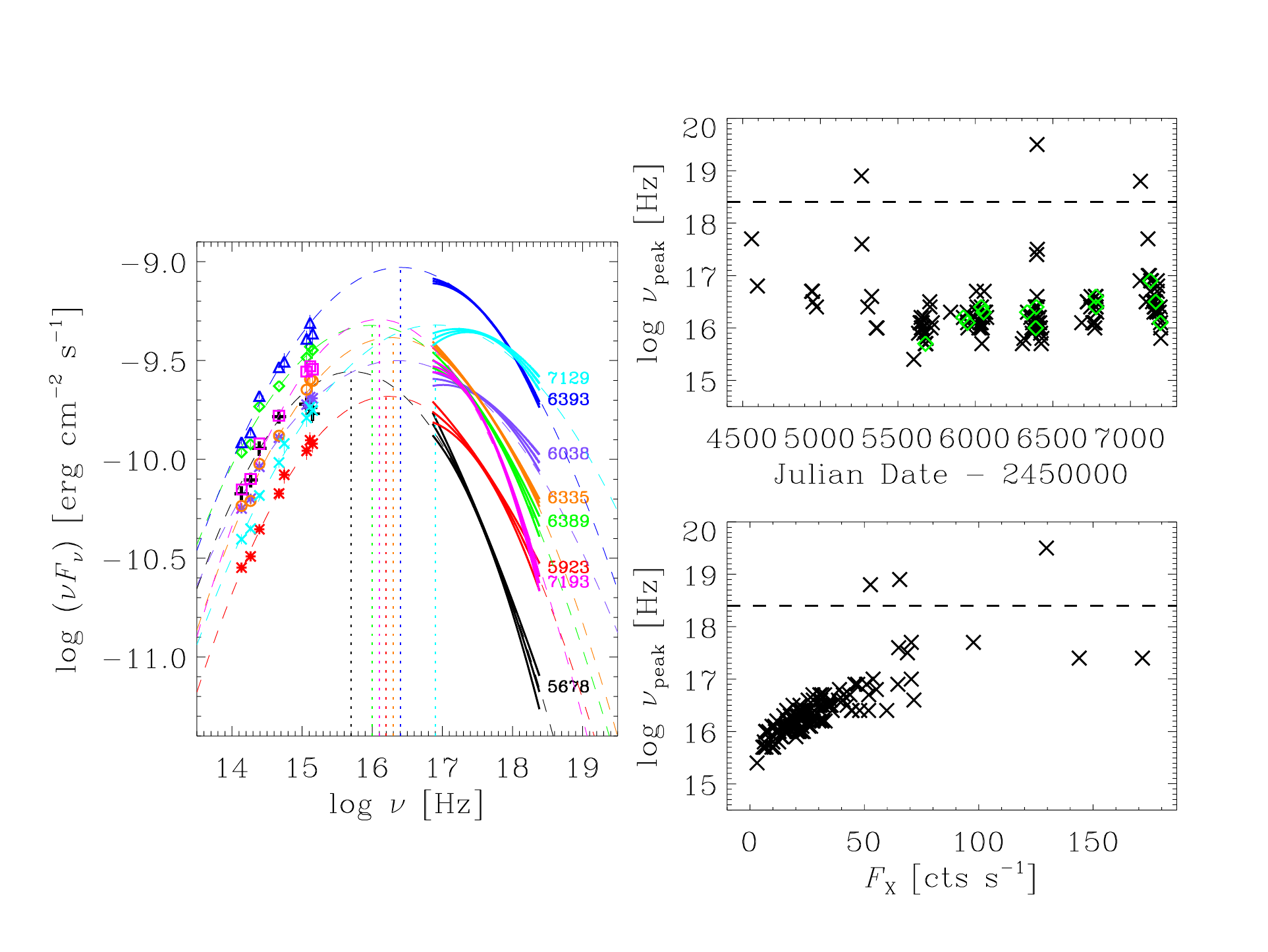}
  \caption{Left: broad-band SEDs of Mrk~421 at eight different epochs (identified on the right with their $\rm JD-2450000$) where simultaneous near-IR, optical, UV and X-ray data are available. Right: The synchrotron peak frequency versus time (top) and X-ray flux (bottom). The horizontal lines correspond to 10 keV, the upper limit of the energy range covered by XRT. In the top panel the diamonds represent the cases where also near-IR data were available and included in the SED fit. }
\label{fi:1101_sed}
\end{figure*}

We built infrared-to-X-ray SEDs to investigate the spectral variability of Mrk 421 around the synchrotron peak. 
In order to be able to reliably characterize and model the region of the SED around the synchrotron peak, we considered all the observing epochs where strictly simultaneous data in the $K, H, J, R$ bands from the WEBT Collaboration and at UV and X-ray frequencies from the \textit{Swift} satellite were available. Eight of these SEDs\footnote{The other five SEDs overlap with them and were omitted for sake of clarity.} are shown in Fig.\ \ref{fi:1101_sed}, covering a wide range of brightness states. In three cases we also found UVOT data in the $v$ band. 
The errors in the near-IR and optical flux densities are typically around three per cent and are included in the symbol size, while we conservatively assumed a 10 per cent error on the UVOT data. The X-ray spectra are the result of a log-parabola model fitting, including the uncertainties on the flux normalization and $\alpha$ and $\beta$ parameters.

We also show log-parabolic fits to the broad-band SEDs and the position of the synchrotron peak they identify.
The X-ray spectral variability is quite strong and suggests that the frequency of the synchrotron peak can shift over a large range, from $\log \nu \sim 15.7$ to 16.9 [Hz]. 
Taking into account the curvature of the X-ray spectra alone, the range could be even broader, up to $\log \nu \sim 17.4$ of the $\rm JD=2457129$ SED, i.e.\ a factor 50 in frequency.
From these few SEDs the relationship between the brightness state and the synchrotron peak frequency appears confused. In particular, the highest peak frequency ($\rm JD=2457129$) does not correspond to the highest X-ray flux ($\rm JD=2456393$). 
To better investigate this point, we built SEDs of all epochs with contemporaneous {\em Swift} and $R$-band data and fit them with a log-parabolic model. The results are shown in Fig.\ \ref{fi:1101_sed}. There is a general indication that the synchrotron peak frequency follows the X-ray activity, shifting toward higher values when the X-ray flux increases. The increase is fast below $\sim 30 \rm \, cts \, s^{-1}$, then the curve flattens. 
Values of $\log \nu_{\rm peak}$ larger than 17 are reached only if the X-ray count rate is greater than $50 \rm \, cts \, s^{-1}$. Beyond this general behaviour there are specific cases with peculiar spectral shapes. There are three cases where the X-ray spectrum is hard so that $\nu_{\rm peak}$ falls beyond the 10 keV upper limit of the XRT energy range.

The only steep UV spectrum is shown by the SED on $\rm JD=2455678$, corresponding to the faintest and softest X-ray spectrum. We also note that its near-IR and optical part fairly matches that observed at $\rm JD=2457193$, but the latter has much brighter UV and X-ray states. If the UV spectrum steepness on $\rm JD=2455678$ were real and not due to data uncertainties, its extrapolation to the higher energies would not meet the X-ray spectrum, rising the question whether more than one component is contributing to the source synchrotron flux. The same issue was encountered when comparing the optical and X-ray light curves in Section 8.
Similar mismatches in the SED have already been found for this source \citep{mas04} and were sometimes observed in other HBLs, like PG~1553+113 \citep{rai15} and H1722+119 \citep{ahn16a}. In these two cases, they have been interpreted in terms of orientation variations in an inhomogeneous helical jet.

Another pair of SEDs having the same near-IR and optical flux, but a different UV flux and very different X-ray spectra are those on $\rm JD=2456038$ and 2456335. Here a higher UV state corresponds to a softer and less curved X-ray spectrum.

\section{Summary and Conclusions}

In this paper we have analysed 8 yr of multiwavelength data on the HBL Mrk~421. 
The GASP-WEBT Collaboration and Steward Observatory provided the near-infrared and optical photometric and polarimetric data. Information in the UV and X-ray bands was acquired from the \textit{Swift} satellite and in $\gamma$-rays from the LAT instrument onboard {\em Fermi}. We have also exploited the spectropolarimetric data taken at the Steward Observatory to investigate a possible wavelength-dependence of the polarisation degree $P$ and angle EVPA. We have calculated the host galaxy contribution to the source photometry in the near-IR, optical, and UV bands and subtracted it to analyse the jet emission. 

The source showed unceasing activity at all frequencies over all the 2007--2015 period we considered. Variability in the near-IR, UV, and $\gamma$-ray flux appears well correlated with the optical one, showing prominent flares in 2012--2013. In contrast, the behaviour in X-rays does not follow the same path and X-ray flares occurred in 2008, 2010 and 2013.
In $\gamma$-rays, the most noticeable event occurred in 2012, while significant flares were also observed in 2013--2014.
Discrete correlation function analysis suggests the lack of a persistent X-ray--optical correlation while indicates fair correlation between the $\gamma$-ray and optical flux changes since 2012.

The spectral variability increases with frequency: broad-band SEDs show that the near-IR to optical spectral shape is rather stable for different brightness states, while the upturn toward the UV can be more or less pronounced, and this corresponds to a different X-ray spectral form. Indeed, X-ray spectra exhibit a large variety of slopes and curvatures and suggest that the synchrotron peak may cover a large range of frequencies and in general shifts toward higher energies when the X-ray flux increases.

The energy output from Mrk~421 has often been interpreted with one-zone SSC models \citep[e.g.][]{ale12} or with lepto-hadronic models involving proton synchrotron radiation and/or photopion interactions \citep[e.g.][]{bot13}. 
In these models, the variations of the SED shape are obtained by changing the parameters determining the jet physics. In particular, these changes can produce a shift of the synchrotron peak frequency and in turn a lack of correlation between the optical and X-ray emission, even if these two emissions are physically connected.
Alternatively, we propose the existence of at least two emitting regions in the jet to explain the different behaviour of the optical and X-ray fluxes. These regions are disjointed and experience independent short-term variability. Moreover, their orientation can likely change in time, so that periods of high X-ray activity would be observed when the part of the jet producing this radiation is more closely aligned with the line of sight. Similarly, when the viewing angle of the optical emitting region decreases, the optical flux would be more Doppler enhanced. 

We have analysed polarimetric data to look for episodes where 
the behaviour of $P$ and EVPA can suggest that the source variability is due to geometrical effects, like a curved motion of a blob in a helical jet \citep{nal10}. We found a possible candidate in the last part of the campaign, and derived the minimum angle between the blob velocity vector and the line of sight, the curvature radius of the trajectory and the distance covered by the blob during that event.
Apart from this case, we have not found correlation between the source flux and the $P$ and EVPA changes, which strongly suggests that turbulence may play a major role in determining the source polarimetric properties \citep{mar14,kie16,bli16,rai17}.

\section*{Acknowledgements}
We are grateful to an anonymous referee for useful comments and suggestions.
The data collected by the WEBT collaboration are stored in the WEBT archive at the Osservatorio Astrofisico di Torino - INAF (http://www.oato.inaf.it/blazars/webt/); for questions regarding their availability, please contact the WEBT President Massimo Villata ({\tt villata@oato.inaf.it}).
This article is partly based on observations made with the telescopes IAC80 and TCS operated by the Instituto de Astrof\'isica de Canarias in the Spanish Observatorio del Teide on
the island of Tenerife. The IAC team acknowledges the support from the group of support astronomers and telescope operators of the Observatorio del Teide.
Based (partly) on data obtained with the STELLA robotic telescopes in Tenerife, an AIP facility jointly operated by AIP and IAC.
The Steward Observatory blazar monitoring program is funded by NASA
through Fermi Guest Investigator grants NNX08AW56G, NNX09AU10G,
NNX12AO93G, and NNX15AU81G.
The Abastumani team acknowledges financial support by Shota Rustaveli NSF under contract FR/577/6-320/13.
The research at BU was supported in part by US National Science Foundation grant AST-1615796 and NASA Fermi Guest Investigator grant NNX14AQ58G. This study makes use of 43 GHz VLBA data from the VLBA-BU Blazar Monitoring Program (VLBA-BU-BLAZAR; http://www.bu.edu/blazars/VLBAproject.html), funded by NASA through the Fermi Guest Investigator Program.
The PRISM camera at Lowell Observatory was developed by K.\ Janes et al. at BU and Lowell Observatory, with funding from the NSF, BU, and Lowell Observatory.
This research has made use of data from the MOJAVE database that is maintained by the MOJAVE team \citep{lis09}. St.Petersburg University team acknowledges support from Russian RFBR
grant 15-02-00949 and St.Petersburg University research grant 6.38.335.2015.
This paper is partly based on observations carried out at the German-Spanish Calar Alto Observatory, which is jointly operated by the MPIA and the IAA-CSIC. IA research is supported by a Ram\'on y Cajal grant of the Spanish Ministerio de Econom\'ia y Competitividad (MINECO). 
Acquisition of the MAPCAT data was supported in part by MINECO through grants AYA2010-14844, AYA2013-40825-P, and AYA2016-80889-P, and by the Regional Government of Andaluc\'ia through grant P09-FQM-4784. 
This research was partially supported by Scientific Research Fund of the
Bulgarian Ministry of Education and Sciences under grants DO 02-137
(BIn-13/09) and DN 08/1. 
The Skinakas Observatory is a collaborative project of the
University of Crete, the Foundation for Research and Technology -- Hellas,
and the Max-Planck-Institut f\"ur Extraterrestrische Physik.
G.Damljanovic and O.Vince gratefully acknowledge the
observing grant support from the Institute of Astronomy and Rozhen National
Astronomical Observatory, Bulgaria Academy of Sciences, via bilateral joint
research project "Observations of ICRF radio-sources visible in optical domain"
(the head is Dr. G.Damljanovic). This work is a part of the Projects No 176011
("Dynamics and kinematics of celestial bodies and systems"), No 176004
("Stellar physics") and No 176021 ("Visible and invisible matter in nearby
galaxies: theory and observations") supported by the Ministry of Education,
Science and Technological Development of the Republic of Serbia.
The Serbian station is the Astronomical Station Vidojevica (ASV) with
the 60cm ASV telescope (and from this year, the 1.4m ASV one). 
This research was supported partly by funds of the project RD-08-81 of
Shumen University.

%%%%%%%%%%%%%%%%%%%%%%%%%%%%%%%%%%%%%%%%%%%%%%%%%%



%\bibliographystyle{mnras}

\begin{thebibliography}{99}
\bibitem[Abdo et al.(2011)]{abd11} Abdo, A.~A., Ackermann, M., Ajello, M., et al.\ 2011, \apj, 736, 131 
\bibitem[Acero et al.(2015)]{ace15} Acero, F., Ackermann, M., Ajello, M., et al.\ 2015, \apjs, 218, 23 
\bibitem[Aharonian et al.(2005)]{aha05} Aharonian, F., Akhperjanian, A.~G., Aye, K.-M., et al.\ 2005, \aap, 437, 95 
\bibitem[Ahnen et al.(2016a)]{ahn16a} Ahnen, M.~L., Ansoldi, S., Antonelli, L.~A., et al.\ 2016, \mnras, 459, 3271
\bibitem[Ahnen et al.(2016b)]{ahn16b} Ahnen, M.~L., Ansoldi, S., Antonelli, L.~A., et al.\ 2016, \aap, 593, A91 
\bibitem[Albert et al.(2007)]{alb07} Albert, J., Aliu, E., Anderhub, H., et al.\ 2007, \apj, 663, 125 
\bibitem[Aleksi{\'c} et al.(2012)]{ale12} Aleksi{\'c}, J., Alvarez, E.~A., Antonelli, L.~A., et al.\ 2012, \aap, 542, A100
\bibitem[Aleksi{\'c} et al.(2015a)]{ale15a} Aleksi{\'c}, J., Ansoldi, S., Antonelli, L.~A., et al.\ 2015, \aap, 576, A126 
\bibitem[Aleksi{\'c} et al.(2015b)]{ale15b} Aleksi{\'c}, J., Ansoldi, S., Antonelli, L.~A., et al.\ 2015, \aap, 578, A22 
\bibitem[Atwood et al.(2009)]{atw09} Atwood, W.~B., Abdo, A.~A., Ackermann, M., et al.\ 2009, \apj, 697, 1071
\bibitem[Atwood et al.(2013)]{atw13} Atwood, W., Albert, A., Baldini, L., et al.\ 2013, arXiv:1303.3514 
\bibitem[Balokovi{\'c} et al.(2016)]{bal16} Balokovi{\'c}, M., Paneque, D., Madejski, G., et al.\ 2016, \apj, 819, 156 
\bibitem[Bessell et al.(1998)]{bes98} Bessell, M.~S., Castelli, F., \& Plez, B.\ 1998, \aap, 333, 231
\bibitem[Blasi et al.(2013)]{bla13} Blasi, M.~G., Lico, R., Giroletti, M., et al.\ 2013, \aap, 559, A75 
\bibitem[{{Blinov} {et~al}\mbox{.}(2016){Blinov}, {Pavlidou}, {Papadakis},
  {Kiehlmann}, {Liodakis}, {Panopoulou}, {Pearson}, {Angelakis},
  {Balokovi{\'c}}, {Hovatta}, {Joshi}, {King}, {Kus}, {Kylafis}, {Mahabal},
  {Marecki}, {Myserlis}, {Paleologou}, {Papamastorakis}, {Pazderski},
  {Prabhudesai}, {Ramaprakash}, {Readhead}, {Reig}, {Tassis}, \&
  {Zensus}}]{bli16}
{Blinov} D. {et~al.}, 2016, \mnras, 462, 1775
\bibitem[B{\"o}ttcher et al.(2013)]{bot13} B{\"o}ttcher, M., Reimer, A., Sweeney, K., \& Prakash, A.\ 2013, \apj, 768, 54
\bibitem[Burrows et al.(2005)]{bur05} Burrows, D.~N., Hill, J.~E., Nousek, J.~A., et al.\ 2005, \ssr, 120, 165
\bibitem[Cardelli et al.(1989)]{car89} Cardelli, J.~A., Clayton, G.~C., \& Mathis, J.~S.\ 1989, \apj, 345, 245
\bibitem[Carnerero et al.(2015)]{car15} Carnerero, M.~I., Raiteri, C.~M., Villata, M., et al.\ 2015, \mnras, 450, 2677 
\bibitem[Carnerero et al.(2016)]{car16} Carnerero, M., Raiteri, C., Villata, M., et al.\ 2016, Galaxies, 4, 61 \bibitem[Donnarumma et al.(2009)]{don09} Donnarumma, I., Vittorini, V., Vercellone, S., et al.\ 2009, \apjl, 691, L13 
\bibitem[Edelson \& Krolik(1988)]{ede88} Edelson, R.~A., \& Krolik, J.~H.\ 1988, \apj, 333, 646 
\bibitem[{{Hartman} {et~al}\mbox{.}(1999){Hartman}, {Bertsch}, {Bloom}, {Chen},
  {Deines-Jones}, {Esposito}, {Fichtel}, {Friedlander}, {Hunter}, {McDonald},
  {Sreekumar}, {Thompson}, {Jones}, {Lin}, {Michelson}, {Nolan}, {Tompkins},
  {Kanbach}, {Mayer-Hasselwander}, {M{\"u}cke}, {Pohl}, {Reimer}, {Kniffen},
  {Schneid}, {von Montigny}, {Mukherjee}, \& {Dingus}}]{har99}
{Hartman} R.~C. {et~al.}, 1999, \apjs, 123, 79
\bibitem[Hovatta et al.(2015)]{hov15} Hovatta, T., Petropoulou, M., Richards, J.~L., et al.\ 2015, \mnras, 448, 3121 
\bibitem[Hufnagel \& Bregman(1992)]{huf92} Hufnagel, B.~R., \& Bregman, J.~N.\ 1992, \apj, 386, 473 
\bibitem[Ikejiri et al.(2011)]{ike11} Ikejiri, Y., Uemura, M., Sasada, M., et al.\ 2011, \pasj, 63, 639
\bibitem[Jorstad et al.(2010)]{jor10} Jorstad, S.~G., Marscher, A.~P., Larionov, V.~M., et al.\ 2010, \apj, 715, 362
\bibitem[Kapanadze et al.(2016)]{kap16} Kapanadze, B., Dorner, D., Vercellone, S., et al.\ 2016, \apj, 831, 102 
\bibitem[{{Kiehlmann} {et~al}\mbox{.}(2016){Kiehlmann}, {Savolainen},
  {Jorstad}, {Sokolovsky}, {Schinzel}, {Marscher}, {Larionov}, {Agudo},
  {Akitaya}, {Ben{\'{\i}}tez}, {Berdyugin}, {Blinov}, {Bochkarev}, {Borman},
  {Burenkov}, {Casadio}, {Doroshenko}, {Efimova}, {Fukazawa}, {G{\'o}mez},
  {Grishina}, {Hagen-Thorn}, {Heidt}, {Hiriart}, {Itoh}, {Joshi}, {Kawabata},
  {Kimeridze}, {Kopatskaya}, {Korobtsev}, {Krajci}, {Kurtanidze}, {Kurtanidze},
  {Larionova}, {Larionova}, {Lindfors}, {L{\'o}pez}, {McHardy}, {Molina},
  {Moritani}, {Morozova}, {Nazarov}, {Nikolashvili}, {Nilsson}, {Pulatova},
  {Reinthal}, {Sadun}, {Sasada}, {Savchenko}, {Sergeev}, {Sigua}, {Smith},
  {Sorcia}, {Spiridonova}, {Takaki}, {Takalo}, {Taylor}, {Troitsky}, {Uemura},
  {Ugolkova}, {Ui}, {Yoshida}, {Zensus}, \& {Zhdanova}}]{kie16}
{Kiehlmann} S. {et~al.}, 2016, \aap, 590, A10
\bibitem[Larionov et al.(2008)]{lar08} Larionov, V.~M., Jorstad, S.~G., Marscher, A.~P., et al.\ 2008, \aap, 492, 389
\bibitem[Lico et al.(2014)]{lic14} Lico, R., Giroletti, M., Orienti, M., et al.\ 2014, \aap, 571, A54 
\bibitem[Lister et al.(2009)]{lis09} Lister, M.~L., Aller, H.~D., Aller, M.~F., et al.\ 2009, \aj, 137, 3718-3729 
\bibitem[Lockman \& Savage(1995)]{loc95} Lockman, F.~J., \& Savage, B.~D.\ 1995, \apjs, 97, 1
\bibitem[Mannucci et al.(2001)]{man01} Mannucci, F., Basile, F., Poggianti, B.~M., et al.\ 2001, \mnras, 326, 745
\bibitem[Marscher(2014)]{mar14} Marscher, A.~P.\ 2014, \apj, 780, 87 
\bibitem[Massaro et al.(2004)]{mas04} Massaro, E., Perri, M., Giommi, P., \& Nesci, R.\ 2004, \aap, 413, 489 
\bibitem[Mattox et al.(1996)]{mat96} Mattox, J.~R., Bertsch, D.~L., Chiang, J., et al.\ 1996, \apj, 461, 396 
\bibitem[Miller(1975)]{mil75} Miller, H.~R.\ 1975, \apjl, 201, L109 
\bibitem[Nalewajko(2010)]{nal10} Nalewajko, K.\ 2010, International Journal of Modern Physics D, 19, 701
\bibitem[{{Nilsson} {et~al}\mbox{.}(2007){Nilsson}, {Pasanen}, {Takalo},
  {Lindfors}, {Berdyugin}, {Ciprini}, \& {Pforr}}]{nil07}
{Nilsson} K., {Pasanen} M., {Takalo} L.~O., {Lindfors} E., {Berdyugin} A.,
  {Ciprini} S., {Pforr} J., 2007, \aap, 475, 199
\bibitem[Paliya et al.(2015)]{pal15} Paliya, V.~S., B{\"o}ttcher, M., Diltz, C., et al.\ 2015, \apj, 811, 143 
\bibitem[Peterson et al.(1998)]{pet98} Peterson, B.~M., Wanders, I., Horne, K., et al.\ 1998, \pasp, 110, 660 
\bibitem[{{Peterson}(2001)}]{pet01}
{Peterson} B.~M., 2001, in Advanced Lectures on the Starburst-AGN Connection,
  {Aretxaga} I., {Kunth} D., {M{\'u}jica} R., eds., Singapore: World
  Scientific, p.~3
\bibitem[Pian et al.(2014)]{pia14} Pian, E., T{\"u}rler, M., Fiocchi, M., et al.\ 2014, \aap, 570, A77 
\bibitem[Piner et al.(2010)]{pin10} Piner, B.~G., Pant, N., \& Edwards, P.~G.\ 2010, \apj, 723, 1150 
\bibitem[{{Polletta} {et~al}\mbox{.}(2007){Polletta}, {Tajer}, {Maraschi},
  {Trinchieri}, {Lonsdale}, {Chiappetti}, {Andreon}, {Pierre}, {Le F{\`e}vre},
  \& {Zamorani}}]{polletta2007}
{Polletta} M. {et~al.}, 2007, \apj, 663, 81
\bibitem[Punch et al.(1992)]{pun92} Punch, M., Akerlof, C.~W., Cawley, M.~F., et al.\ 1992, \nat, 358, 477
\bibitem[Raiteri et al.(2003)]{rai03} Raiteri, C.~M., Villata, M., Tosti, G., et al.\ 2003, \aap, 402, 151 
\bibitem[Raiteri et al.(2010)]{rai10} Raiteri, C.~M., Villata, M., Bruschini, L., et al.\ 2010, \aap, 524, A43
\bibitem[{{Raiteri} {et~al}\mbox{.}(2013){Raiteri}, {Villata}, {D'Ammando},
  {Larionov}, {Gurwell}, {Mirzaqulov}, {Smith}, {Acosta-Pulido}, {Agudo},
  {Ar{\'e}valo}, {Bachev}, {Ben{\'{\i}}tez}, {Berdyugin}, {Blinov}, {Borman},
  {B{\"o}ttcher}, {Bozhilov}, {Carnerero}, {Carosati}, {Casadio}, {Chen},
  {Doroshenko}, {Efimov}, {Efimova}, {Ehgamberdiev}, {G{\'o}mez},
  {Gonz{\'a}lez-Morales}, {Hiriart}, {Ibryamov}, {Jadhav}, {Jorstad}, {Joshi},
  {Kadenius}, {Klimanov}, {Kohli}, {Konstantinova}, {Kopatskaya}, {Koptelova},
  {Kimeridze}, {Kurtanidze}, {Larionova}, {Larionova}, {Ligustri}, {Lindfors},
  {Marscher}, {McBreen}, {McHardy}, {Metodieva}, {Molina}, {Morozova},
  {Nazarov}, {Nikolashvili}, {Nilsson}, {Okhmat}, {Ovcharov}, {Panwar},
  {Pasanen}, {Peneva}, {Phipps}, {Pulatova}, {Reinthal}, {Ros}, {Sadun},
  {Schwartz}, {Semkov}, {Sergeev}, {Sigua}, {Sillanp{\"a}{\"a}}, {Smith},
  {Stoyanov}, {Strigachev}, {Takalo}, {Taylor}, {Thum}, {Troitsky}, {Valcheva},
  {Wehrle}, \& {Wiesemeyer}}]{rai13}
{Raiteri} C.~M. {et~al.}, 2013, \mnras, 436, 1530
\bibitem[Raiteri et al.(2014)]{rai14} Raiteri, C.~M., Villata, M., Carnerero, M.~I., et al.\ 2014, \mnras, 442, 629 
\bibitem[Raiteri et al.(2015)]{rai15} Raiteri, C.~M., Stamerra, A., Villata, M., et al.\ 2015, \mnras, 454, 353 
\bibitem[Raiteri et al.(2017)]{rai17} Raiteri, C.~M., Nicastro, F., Stamerra, A.,  et al.\ 2017, \mnras, in press 
\bibitem[{{Roming} {et~al}\mbox{.}(2005){Roming}, {Kennedy}, {Mason}, {Nousek},
  {Ahr}, {Bingham}, {Broos}, {Carter}, {Hancock}, {Huckle}, {Hunsberger},
  {Kawakami}, {Killough}, {Koch}, {McLelland}, {Smith}, {Smith}, {Soto},
  {Boyd}, {Breeveld}, {Holland}, {Ivanushkina}, {Pryzby}, {Still}, \&
  {Stock}}]{rom05}
{Roming} P.~W.~A. {et~al.}, 2005, Space Science Reviews, 120, 95
\bibitem[Sandrinelli et al.(2017)]{san17} Sandrinelli, A., Covino, S., Treves, A., et al.\ 2017, arXiv:1701.04454 
\bibitem[{{Schlegel}, {Finkbeiner} \& {Davis}(1998){Schlegel}, {Finkbeiner}, \&
  {Davis}}]{sch98}
{Schlegel} D.~J., {Finkbeiner} D.~P., {Davis} M., 1998, \apj, 500, 525
\bibitem[Sikora et al.(2001)]{sikora01} Sikora, M., B{\l}a{\.z}ejowski, M., Begelman, M.~C., \& Moderski, R.\ 2001, \apj, 554, 1 
\bibitem[Sinha et al.(2015)]{sin15} Sinha, A., Shukla, A., Misra, R., et al.\ 2015, \aap, 580, A100 
\bibitem[Skrutskie et al.(2006)]{skr06} Skrutskie, M.~F., Cutri, R.~M., Stiening, R., et al.\ 2006, \aj, 131, 1163 
\bibitem[{{Smith}(1996)}]{smi96}
{Smith} A.~G., 1996, in Astronomical Society of the Pacific Conference Series,
  Vol. 110, Blazar Continuum Variability, {Miller} H.~R., {Webb} J.~R., {Noble}
  J.~C., eds., p.~3
\bibitem[Smith et al.(2003)]{smi03} Smith, P.~S., Schmidt, G.~D., Hines, D.~C., \& Foltz, C.~B.\ 2003, \apj, 593, 676 
\bibitem[Ulrich et al.(1975)]{ulr75} Ulrich, M.-H., Kinman, T.~D., Lynds, C.~R., Rieke, G.~H., \& Ekers, R.~D.\ 1975, \apj, 198, 261
\bibitem[Villata et al.(1998)]{vil98} Villata, M., Raiteri, C.~M., Lanteri, L., Sobrito, G., \& Cavallone, M.\ 1998, \aaps, 130, 305
\bibitem[Villata et al.(2009a)]{vil09a} Villata, M., Raiteri, C.~M., Gurwell, M.~A., et al.\ 2009, \aap, 504, L9 
\bibitem[Villata et al.(2009b)]{vil09b} Villata, M., Raiteri, C.~M., Larionov, V.~M., et al.\ 2009, \aap, 501, 455 
\bibitem[Visvanathan \& Wills(1998)]{vis98} Visvanathan, N., \& Wills, B.~J.\ 1998, \aj, 116, 2119
\bibitem[Wilms et al.(2000)]{wil00} Wilms, J., Allen, A., \& McCray, R.\ 2000, \apj, 542, 914 


\end{thebibliography}


\vspace{1cm}\noindent
{\large \bf AFFILIATIONS}

\vspace{0.5cm}\noindent
{\it
$^{ 1}$INAF, Osservatorio Astrofisico di Torino, via Osservatorio 20, I-10025, Pino Torinese, Italy        \\
$^{ 2}$Instituto de Astrofisica de Canarias (IAC), La Laguna, E-38200 Tenerife, Spain                      \\
$^{ 3}$Departamento de Astrofisica, Universidad de La Laguna, La Laguna, E-38205 Tenerife, Spain           \\
$^{ 4}$Astronomical Institute, St.\ Petersburg State University, Universitetsky pr. 28, Petrodvoretz, 198504 St.\ Petersburg, Russia                                                  \\
$^{ 5}$Pulkovo Observatory, 196140 St.\ Petersburg, Russia                                                         \\
$^{ 6}$Steward Observatory, University of Arizona, Tucson, AZ 85721, USA                                         \\
$^{ 7}$INAF, Istituto di Radioastronomia, via Gobetti 101, I-40129 Bologna, Italy                                            \\
$^{ 8}$DIFA, Universit\'a di Bologna, Viale B. Pichat 6/2, I-40127 Bologna, Italy                          \\
$^{ 9}$Instituto de Astrof\'{i}sica de Andaluc\'{i}a, CSIC, Apartado 3004, E-18080, Granada, Spain                                 \\
$^{10}$Institute of Astronomy and NAO, Bulgarian Academy of Sciences, 72 Tsarigradsko shosse Blvd., 1784 Sofia, Bulgaria                                  \\
$^{11}$Department of Physics, Salt Lake Community College, Salt Lake City, Utah 84070, USA                 \\
$^{12}$Dept.\ of Astronomy, Faculty of Physics, Sofia University,BG-1164 Sofia, Bulgaria                                 \\
$^{13}$EPT Observatories, Tijarafe, E-38780 La Palma, Spain                                                       \\
$^{14}$INAF, TNG Fundaci\'on Galileo Galilei, E-38712 La Palma, Spain                                               \\
$^{15}$Max-Planck-Institut f\"ur Radioastronomie, Bonn, Germany                                            \\
$^{16}$Graduate Institute of Astronomy, National Central University, 300 Jhongda Rd, Jhongli City, Taoyuan County 32001, Taiwan (R.O.C.)                               \\
$^{17}$Astronomical Observatory, Volgina 7, 11060 Belgrade, Serbia                                         \\
$^{18}$Agrupaci\'o Astron\`omica de Sabadell, Spain                                                        \\
$^{19}$Department of Physics and Astronomy, Brigham Young University, Provo, UT 84602, USA               \\
$^{20}$Department of Theoretical and Applied Physics, University of Shumen, 9712 Shumen, Bulgaria               \\
$^{21}$School of Cosmic Physics, Dublin Institute For Advanced Studies, Ireland                            \\
$^{22}$Institute for Astrophysical Research, Boston University, 725 Commonwealth Avenue, Boston, MA 02215, USA                                    \\
$^{23}$Abastumani Observatory, Mt. Kanobili, 0301 Abastumani, Georgia                                      \\
$^{24}$Engelhardt Astronomical Observatory, Kazan Federal University, Tatarstan, Russia                    \\
$^{25}$Center for Astrophysics, Guangzhou University, Guangzhou 510006, China \\
$^{26}$Circolo Astrofili Talmassons, Italy                                                                 \\
$^{27}$UCD School of Physics, University College Dublin, Dublin, Ireland                                   \\
$^{28}$Finnish Centre for Astronomy with ESO (FINCA), University of Turku, Piikki\"o, Finland              \\
$^{29}$Department of Physical Science, Southern Utah University, Cedar City, UT 84721, USA                 \\
$^{30}$Osservatorio Astronomico Sirio, Piazzale Anelli, I-70013 Castellana Grotte, Italy                                       \\
$^{31}$Department of Physics, University of Colorado Denver, CO, 80217-3364 USA                            \\
$^{32}$Cork Institute of Technology, Cork, Ireland                                                         \\
 }

% Don't change these lines
\bsp	% typesetting comment
\label{lastpage}
\end{document}